\documentclass[12pt,natbib,iop]{emulateapj}

\def \HST{{\emph{HST}}}

\def \Spitzer{{\emph{Spitzer}}}

\def \um {{$\mu$m}}

\def \msun {{M$_\odot$}}

\begin{document}
\slugcomment{01/09/12}

\title{The Contribution of TP-AGB and RHeB Stars to the Near-IR Luminosity of Local Galaxies: Implications for Stellar Mass Measurements of High Redshift Galaxies}

\author{J. Melbourne\altaffilmark{1}, Benjamin F. Williams\altaffilmark{2}, Julianne J. Dalcanton\altaffilmark{2}, Philip Rosenfield\altaffilmark{2}, L\'eo Girardi\altaffilmark{3}, P. Marigo\altaffilmark{4}, D. Weisz\altaffilmark{2}, A. Dolphin\altaffilmark{5}, Martha L. Boyer\altaffilmark{6}, Knut Olsen\altaffilmark{7}, E. Skillman\altaffilmark{8}, Anil C. Seth\altaffilmark{9}}
\altaffiltext{1}{Caltech Optical Observatories, Division of Physics, Mathematics and Astronomy, Mail Stop 301-17, California Institute of Technology, Pasadena, CA 91125, jmel@caltech.edu}
\altaffiltext{2}{Department of Astronomy, Box 351580, University of Washington, Seattle, WA 98195, USA; ben@astro.washington.edu, jd@astro.washington.edu, philrose@astro.washington.edu, dweisz@astro.washington.edu}
\altaffiltext{3}{Osservatorio Astronomico di Padova---INAF, Padova, Italy; leo.girardi@oapd.inaf.it}
\altaffiltext{4}{Dipartimento di Astronomia, Universit\`a di Padova
Vicolo dell'Osservatorio 3, I-35122 Padova Italy; paola.marigo@unipd.it}
\altaffiltext{5}{Raytheon, 1151 E. Hermans Road, Tucson, AZ 85706; adolphin@raytheon.com}
\altaffiltext{6}{Space Telescope Science Institute, 3700 San Martin Dr., Baltimore, MD 21218; mboyer@stsci.edu}
\altaffiltext{7}{National Optical Astronomical Observatories, 950 N. Cherry Ave.,Tucson, AZ 85719; kolsen@noao.edu} 
\altaffiltext{8}{University of Minnesota, Deptartment of Astronomy, 116 Church St. SE, Minneapolis, MN, 55455; skillman@astro.umn.edu}
\altaffiltext{9}{University of Utah, Salt Lake City, UT 84112; aseth@astro.utah.edu}

\begin{abstract}
Using high spatial resolution Hubble Space Telescope Wide Field Camera 3 and Advance Camera for Surverys imaging of resolved stellar populations, we constrain the contribution of thermally-pulsing asymptotic giant branch (TP-AGB) stars and red helium burning (RHeB) stars to the 1.6 \um\ near-infrared (NIR) luminosities of 23 nearby galaxies,  including dwarfs and spirals.  The TP-AGB phase contributes as much as 17\% of the integrated $F160W$ flux, even when the red giant branch is well populated.  The RHeB population contribution can match or even exceed the TP-AGB contribution, providing as much as 21\% (18\% after a statistical correction for foreground) of the integrated $F160W$ light.  We estimate that these two short lived phases may account for up to 70\% of the rest-frame NIR flux at higher redshift.  The NIR mass-to-light (M/L) ratio should therefore be expected to vary significantly due to fluctuations in the star formation rate over timescales from 25 Myr to several Gyr, an effect that may be responsible for some of the lingering scatter in NIR galaxy scaling relations such as the Tully-Fisher and metallicity-luminosity relations.  We compare our observational results to predictions based on optically derived star formation histories and stellar population synthesis (SPS) models, including models based on the 2008 Padova isochrones (used in popular SPS programs) and the updated 2010 Padova isochrones, which shorten the lifetimes of low-mass (old) low-metallicity TP-AGB populations. The updated (2010) SPS models generally reproduce the expected numbers of TP-AGB stars in the sample; indeed, for 65\% of the galaxies, the discrepancy between modeled and observed numbers is smaller than the measurement uncertainties.  The weighted mean model/data number ratio for TP-AGB stars is 1.5 (1.4 with outliers removed) with a standard deviation of 0.5. The same SPS models, however, give a larger discrepancy in the $F160W$ flux contribution from the TP-AGB stars, over-predicting the flux by a weighted mean factor of 2.3 (2.2 with outliers removed) with a standard deviation of 0.8.  This larger offset is driven by the prediction of modest numbers of high luminosity TP-AGB stars at young ($<300$ Myrs) ages.  The best-fit SPS models simultaneously tend to under-predict the numbers and fluxes of stars on the RHeB sequence, typically by a factor of $2.0 \pm0.6$ for galaxies with significant numbers of RHeBs.  Possible explanations for both the TP-AGB and RHeB model results include: (1) difficulties with measuring the SFHs of galaxies especially on the short timescales over which these stars evolve (several Myrs); (2)  issues with the way the SPS codes populate the CMDs (e.g. how they handle pulsations or self extinction), and/or (3) lingering issues with the lifetimes of these stars in the stellar evolution codes.  Coincidentally these two competing discrepancies --- over-prediction of the TP-AGB and under-prediction of the RHeBs --- result in a predicted NIR M/L ratio largely unchanged for a rapid star formation rate, after correcting for these effects.  However, the NIR-to-optical flux ratio of galaxies could be significantly smaller than AGB-rich models would predict, an outcome that has been observed in some intermediate redshift post-starburst galaxies.
\end{abstract}


\section{Introduction}
One of the primary objectives of extragalactic observational astronomy is to measure and track the growth of stellar mass in galaxies across cosmic time \citep[e.g.][]{Bundy05,Fontana06,Ilbert10,Pozzetti10,Vulcani10}.  To accomplish this task, rest-frame ultra-violet (UV) through near-infrared (NIR) observations have been obtained for hundreds of thousands of galaxies \citep[e.g.][]{Giavalisco04,Davis07,Sanders07}.   However, the interpretation of these observations requires stellar population synthesis codes that incorporate models of the initial mass function, star formation histories, and stellar evolution tracks \citep[e.g.][]{BC03}.  The detailed prescriptions for these inputs can affect the resulting estimates of stellar population age, and total stellar mass \citep{Maraston06,Ilbert10}.

Until recently, NIR passbands were assumed to provide an ideal window on the stellar masses of galaxies \citep[e.g.][]{Bundy05}.  Compared to optical and UV passbands, NIR wavelengths are significantly less affected by massive main sequence stars formed in bursts of star formation, which can decrease the mass-to-light (M/L) ratio in the optical passbands.   NIR observations are also less affected by dust obscuration, which can increase the M/L ratio at shorter wavelengths.  In addition, deep Spitzer IRAC observations provide an ideal window on the rest-frame NIR fluxes of high redshift galaxies, and have been used extensively for estimating stellar masses.

Unfortunately, while massive main sequence stars do not have a large impact on the NIR luminosities of galaxies, intermediate-mass ($2-10\; M_{\odot}$) evolved stars have been shown to contribute significantly to integrated NIR fluxes, even when they represent a negligible contribution to the stellar mass \citep{Persson83,Frogel90}.  In recent years, renewed effort has been given to understanding the contribution of thermally-pulsing asymptotic giant branch (TP-AGB) stars to the NIR M/L ratios of galaxies \citep{Maraston06}. The TP-AGB represents a brief period ($1-2$ Myr) of double shell burning at the end of stellar evolution. During this phase, a star swells, undergoes pulsations, and ultimately loses as much as 80\% of its stellar mass before fading to a white dwarf \citep{Iben83,Vassiliadis93,Kennicutt94}.   The most massive of these TP-AGB stars can be very luminous in the NIR, exceeding the luminosity of the tip of the red giant branch (TRGB) by several magnitudes. Models  that neglect TP-AGB stars have been shown to over-estimate the masses of distant galaxies by factors of two or more in comparison to models that include them \citep{Ilbert10}.  

TP-AGB stars are now routinely included in population synthesis models of galaxies, although in different proportions depending on the technique adopted \citep[see][]{Charlot91,Bressan94,Maraston06, Bruzual07,Conroy09}. Many issues still remain, primarily because late stage stellar evolution is difficult to follow from first principles. Stellar evolution codes require knowledge of hard to model processes, such as:  recurrent third dredge-up events, hot-bottom burning, long period variability, and mass loss  \citep{Marigo07}.  To account for these processes,  modelers often resort to simplified TP-AGB stellar evolution codes with parameters tuned to observational data sets \citep{Marigo08}.   Currently, the most complete data sets of evolved stars come from studies of the Large and Small Magellanic Clouds \citep{Frogel90,Cioni99, Blum06, Boyer11}.  The TP-AGB evolution code used to build the Padova isochrones (Girardi \& Marigo 2007; Marigo et al. 2008) has been successfully tuned to reproduce the numbers and optical luminosities of TP-AGB stars in these systems.

Unfortunately, the stellar populations of the Magellanic Clouds only span a narrow region of age and metallicity.  Codes tuned to the Magellanic clouds can fail dramatically when used to predict the resolved stellar populations of other nearby galaxies.    For instance, \citet{Gullieuszik08} showed that these codes over-predicted the numbers of carbon-rich AGB stars in the Leo~II dSph by a factor of six.  Similarly, in nearby (2.5 Mpc) dwarf irregular galaxy KKH 98, the models have been found to over-predict  the numbers of TP-AGB stars by factors of  2-3 compared to observations \citep{Melbourne10}.  These codes also had difficulty modelling more massive metal-rich galaxies at larger distances.  For instance, the Virgo Cluster shows a deficiency of AGB stars compared to model predictions \citep{Williams07}, and a sample of intermediate redshift post-starburst galaxies shows spectral energy distributions that rule out large flux contributions from TP-AGB stars \citep{Kriek10}. %In both cases, the metallicity of the system being studied was significantly lower than the Magellanic Clouds. 

Star count analysis has been extended to  galaxies beyond the Local Group, where resolved stellar populations are best studied with \HST\ \citep[e.g. the ACS Nearby Galaxy Survey Treasury, ANGST][]{Dalcanton09}, or with adaptive optics on large ground based telescopes \citep{Gullieuszik08a,Melbourne10,Davidge10}. In one of the largest such studies to date, \citet{Girardi10} found that that the 2008 Padova isochrones were over-predicting the TP-AGB in optical \HST\ observations of 10 old, metal-poor galaxies from the ANGST sample.   However, the models could be brought into agreement with the data by lowering the estimated lifetimes of low-mass (old), low-metallicity TP-AGB stars, making them roughly equivalent to the lifetimes of higher mass (younger) TP-AGB stars.  These revisions have been incorporated into the 2010 versions of Padova stellar evolution codes and isochrones \footnote{http:/stev.oapd.inaf.it/cmd}.     
 
 \begin{figure*}[ht]
\centering
\includegraphics[scale=0.5]{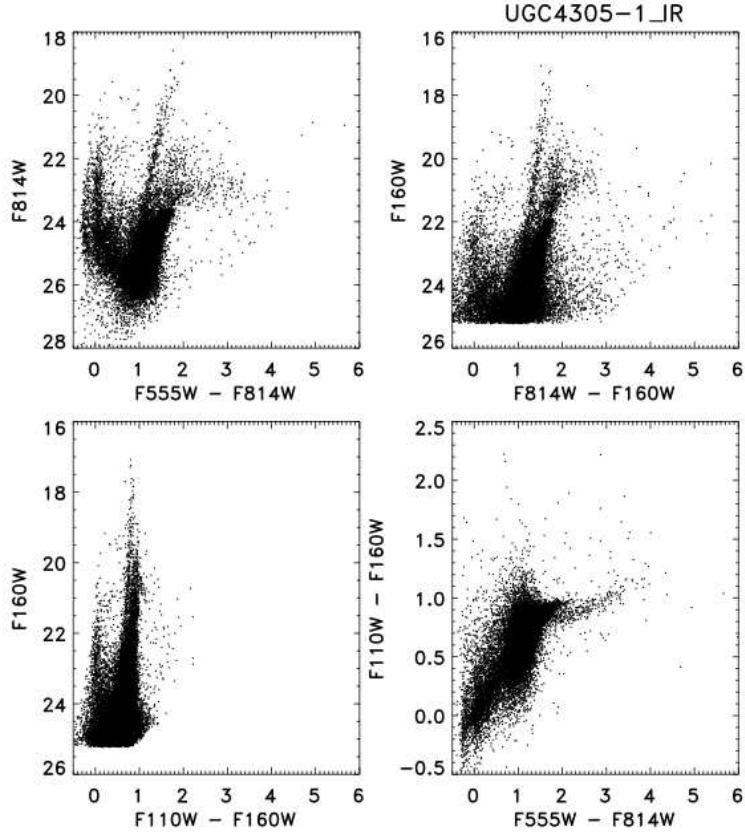}
\caption{\label{fig:AllCMD} Example CMDs of the stars in program galaxy UGC-4305-1.  These CMDs contain only stars that are matched across all four optical and NIR bands; hence, stars that are blue and faint tend to be missing from the optical CMD (upper left).  Note, the optical and optical/NIR hybrid (upper right) CMDs provide a much larger color spread for IR luminous stars, compared with the NIR only CMD (bottom left).  This color spread makes it easier to distinguish between RHeB stars and TP-AGB stars.  We therefore use the optical/NIR hybrid CMD to identify sequences of evolved stars (see Figure \ref{fig:CMD}).}
\end{figure*}

While significant effort has recently been given to the TP-AGB phase, another phase may be equally important for understanding NIR M/L ratios of galaxies. Luminous red helium burning stars (RHeBs) are massive (i.e. $> 3.5 M_{\odot}$) core helium burning stars that form a tight sequence at the luminous end of optical and NIR CMDs \citep{Dohm-Palmer02,McQuinn11}.  These stars have even shorter lifetimes than TP-AGB stars, and the RHeB sequence is only obvious in CMDs with a burst of star formation younger than $\sim300$ Myrs.  As with TP-AGB stars, these stars are difficult to model,  with significant uncertainties associated with convective/mixing processes (overshooting and rotation) and mass loss.   Unfortunately, less attention has been given to this phase of stellar evolution for stellar population synthesis models, even though at high redshift where there is significant ongoing star formation, RHeBs will be one of the dominant contributors to the NIR luminosity \citep[see][]{Dalcanton12}.
 
\begin{deluxetable*}{lccccccc}
\tabletypesize{\small}
\tablecaption{Properties of the Sample Galaxies Measured from the SFH Fitting Routine CalcSFH\label{tab:results}}
\tablehead{\colhead{Galaxy} & \colhead{Distance} & \colhead{Metallicity} & \colhead{Fraction of Mass} & \colhead{Fraction of Mass} \\
& \colhead{Modulus}& \colhead{[M/H]} & \colhead{Younger than 2 Gyr} & \colhead{Younger than 0.3 Gyrs} }
\startdata
                DDO71&     27.67&     -1.09 $\pm$       0.31&  1.43e-02 $\pm$    6.0e-03&  0.00e-00 $\pm$    0.0e-00\\
               DDO78&     27.84&     -1.15 $\pm$       0.16&  1.88e-02 $\pm$    3.9e-03&  1.00e-04 $\pm$    1.0e-04\\
               DDO82&     28.04&     -1.11 $\pm$       0.16&  1.75e-02 $\pm$    3.6e-03&  7.00e-04 $\pm$    1.0e-04\\
          ESO540-030&     27.76&     -1.06 $\pm$       0.24&  6.15e-02 $\pm$    2.4e-02&  1.30e-03 $\pm$    6.0e-04\\
               HS117&     27.93&     -0.61 $\pm$       0.30&  5.30e-02 $\pm$    5.5e-03&  1.60e-03 $\pm$    1.0e-04\\
          IC2574-SGS&     27.98&     -0.97 $\pm$       0.43&  9.27e-02 $\pm$    8.7e-03&  1.44e-02 $\pm$    5.1e-03\\
               KDG73&     27.90&     -1.28 $\pm$       0.14&  6.36e-02 $\pm$    1.5e-02&  9.70e-03 $\pm$    4.5e-03\\
               KKH37&     27.66&     -1.04 $\pm$       0.11&  2.55e-02 $\pm$    3.7e-03&  2.10e-03 $\pm$    6.0e-04\\
            M81-DEEP&     27.78&     -0.41 $\pm$       0.28&  2.09e-02 $\pm$    1.1e-02&  2.90e-03 $\pm$    1.8e-03\\
       NGC0300-WIDE1&     26.55&     -0.75 $\pm$       0.36&  8.83e-02 $\pm$    2.0e-02&  1.96e-02 $\pm$    6.5e-03\\
      NGC2403-HALO-6&     27.51&     -0.65 $\pm$       0.51&  9.88e-02 $\pm$    6.5e-03&  1.17e-02 $\pm$    4.1e-03\\
        NGC2976-DEEP&     27.73&     -0.78 $\pm$       0.40&  1.80e-03 $\pm$    4.7e-03&  2.00e-04 $\pm$    1.0e-04\\
     NGC3077-PHOENIX&     27.95&     -0.94 $\pm$       0.11&  5.10e-03 $\pm$    6.0e-04&  3.80e-03 $\pm$    6.0e-04\\
             NGC3741&     27.49&     -1.27 $\pm$       0.16&  1.11e-01 $\pm$    1.1e-02&  2.06e-02 $\pm$    4.6e-03\\
             NGC4163&     27.36&     -1.19 $\pm$       0.12&  2.49e-02 $\pm$    4.3e-03&  3.50e-03 $\pm$    7.0e-04\\
      NGC7793-HALO-6&     27.91&     -0.69 $\pm$       0.34&  6.07e-02 $\pm$    9.1e-03&  1.19e-02 $\pm$    1.8e-03\\
             SCL-DE1&     28.22&     -1.19 $\pm$       0.17&  2.36e-02 $\pm$    1.1e-02&  4.00e-04 $\pm$    3.0e-04\\
           UGC4305-1&     27.64&     -1.15 $\pm$       0.12&  1.03e-01 $\pm$    9.2e-03&  2.39e-02 $\pm$    5.1e-03\\
           UGC4305-2&     27.64&     -1.14 $\pm$       0.19&  1.11e-01 $\pm$    1.1e-02&  2.58e-02 $\pm$    6.4e-03\\
             UGC4459&     27.79&     -1.16 $\pm$       0.16&  6.60e-02 $\pm$    2.1e-02&  9.00e-03 $\pm$    1.3e-03\\
             UGC5139&     27.91&     -0.78 $\pm$       0.36&  1.03e-01 $\pm$    1.4e-02&  1.97e-02 $\pm$    5.6e-03\\
             UGC8508&     27.04&     -1.25 $\pm$       0.14&  5.62e-02 $\pm$    1.2e-02&  1.33e-02 $\pm$    2.9e-03\\
             UGCA292&     27.54&     -1.51 $\pm$       0.06&  1.97e-01 $\pm$    3.5e-02&  4.96e-02 $\pm$    1.7e-02\\

\enddata
\end{deluxetable*}

In this paper, we build on \citet{Girardi10}, now examining the TP-AGB and RHeB stars within a diverse sample of 23 dwarf and spiral galaxies in the nearby universe, many with significant on-going star formation.   Individual stars within these galaxies are resolved with high spatial resolution Hubble Space Telescope (\HST) observations in the optical (ACS) and NIR (WFC3).  \HST\ crowded-field photometry techniques provide the distributions of stars in color-magnitude space.  The ACS observations are deep enough to constrain the star formation histories (SFHs) of these galaxies \citep{Williams09a,Williams10,Weisz11}, while the WFC3 data provide constraints on the NIR luminosities of the TP-AGB, RHeB, and red giant branch (RGB) stars \citep{Dalcanton12}.   First we calculate the fraction of the 1.6 \um\ galaxy flux contributed by TP-AGB and RHeB stars as a function of the population age, a number that can be used to correct the NIR M/L ratios of galaxies for evolved stellar populations.  Then we compare the observations to the numbers and fluxes of TP-AGB and RHeB stars predicted by  stellar population synthesis (SPS) models and the 2008 and 2010 Padova isochrones.  Finally we discuss the implications of IR luminous stars for models of high redshift galaxies.  The next paper in this series Rosenfield et al. (in preparation) will explore these results further and provide updates to the Padova stellar evolution codes where needed.  

%Our results continue to show that current models over-predict the AGB as expected, and that corrections to these models such as those supplied by \citet{Girardi10} are necessary to match observations.    
  
%In addition, we show that RHeB stars can contribute as much or more to the NIR luminosities of galaxies as the AGB itself.  The RHeB populations are also not well produced by stellar evolution codes.  However, unlike AGB stars, the current codes tend to under-predict the numbers of RHeB stars in most galaxies.  This result again has significant ramifications for estimating accurate stellar masses at high redshift, where there is expected to be huge amounts of ongoing star formation.

% In section XXX, we provide a tool for predicting the correction to current models for AGB stars, as a function of stellar age and metallicity.   For high redshift observations where all of the stellar mass is younger than 1-2 Gyrs, these corrections could be even larger.  

\begin{figure*}[t]
\centering
\includegraphics[scale=0.35]{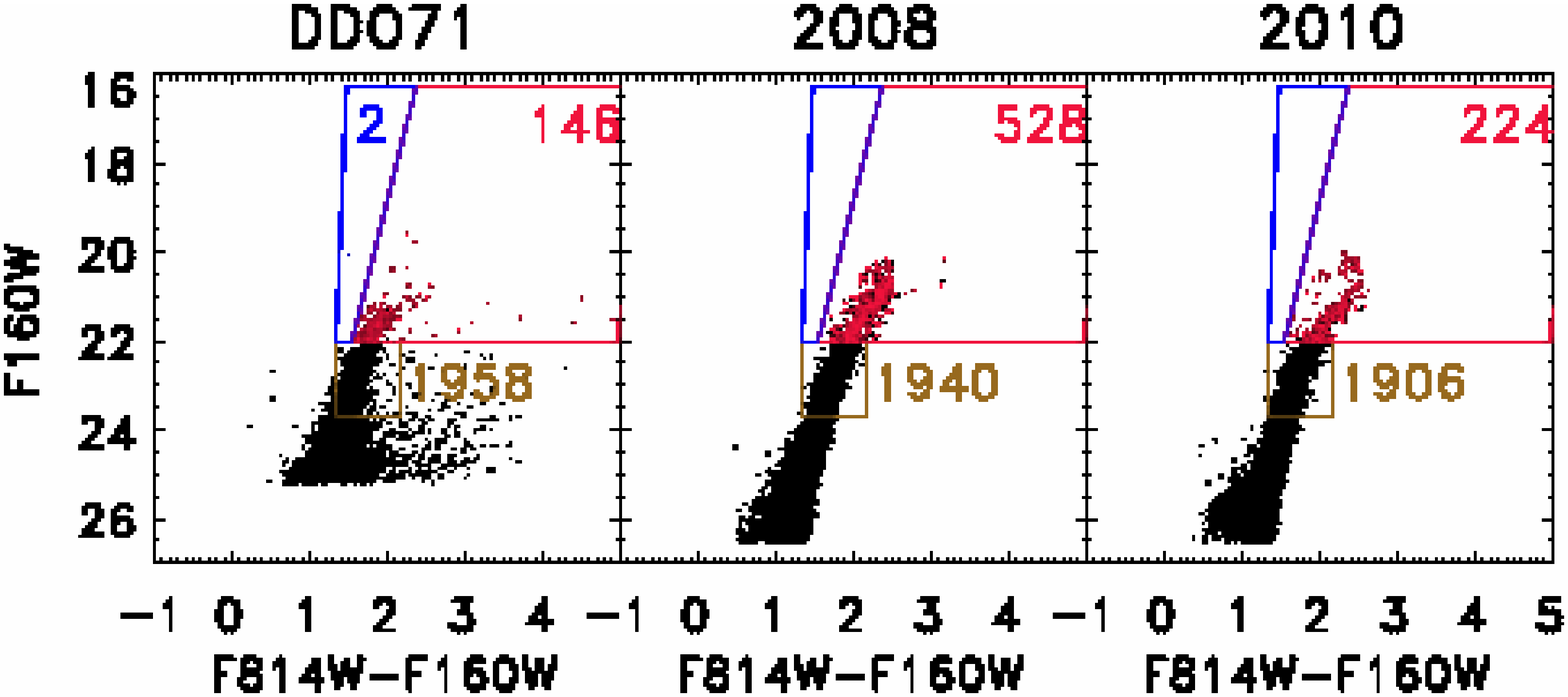}
\vspace{1mm}
\includegraphics[scale=0.35]{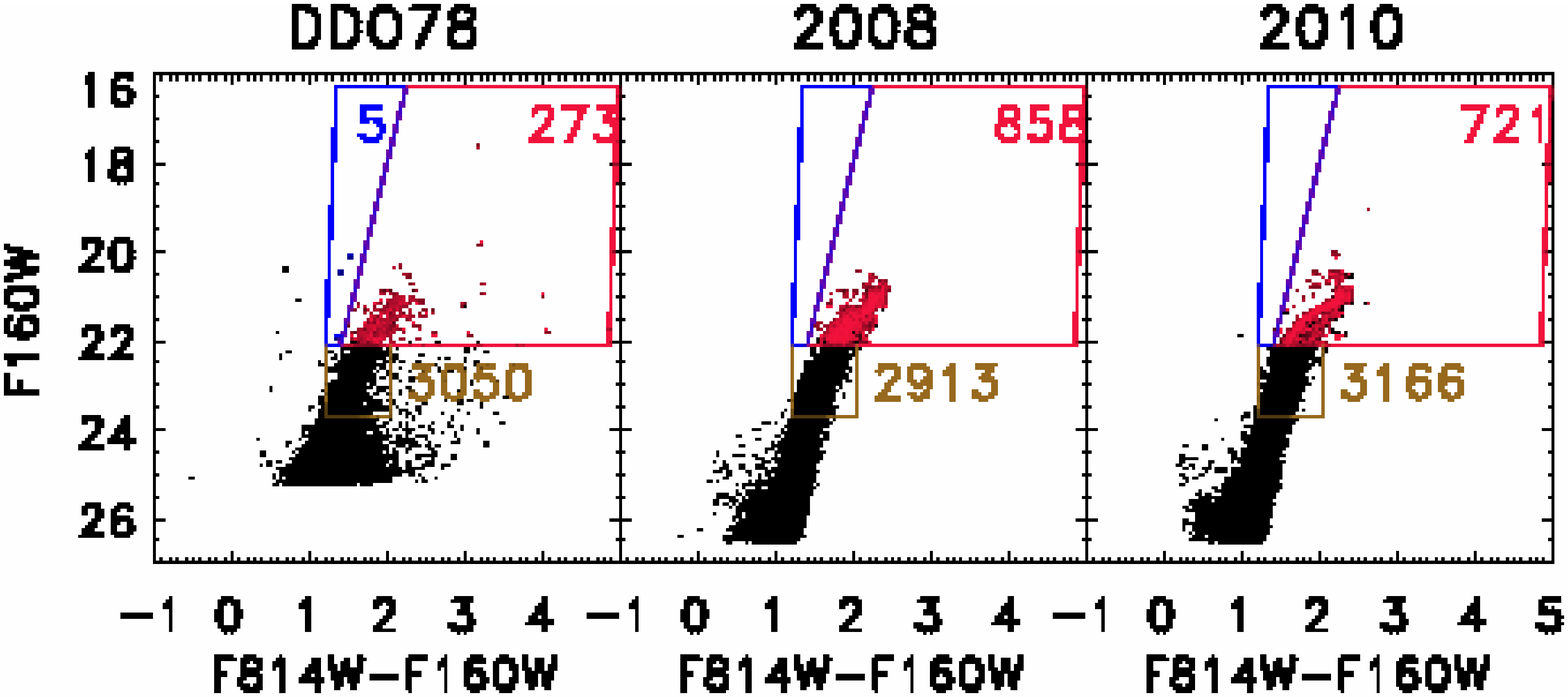}
\vspace{1mm}
\includegraphics[scale=0.35]{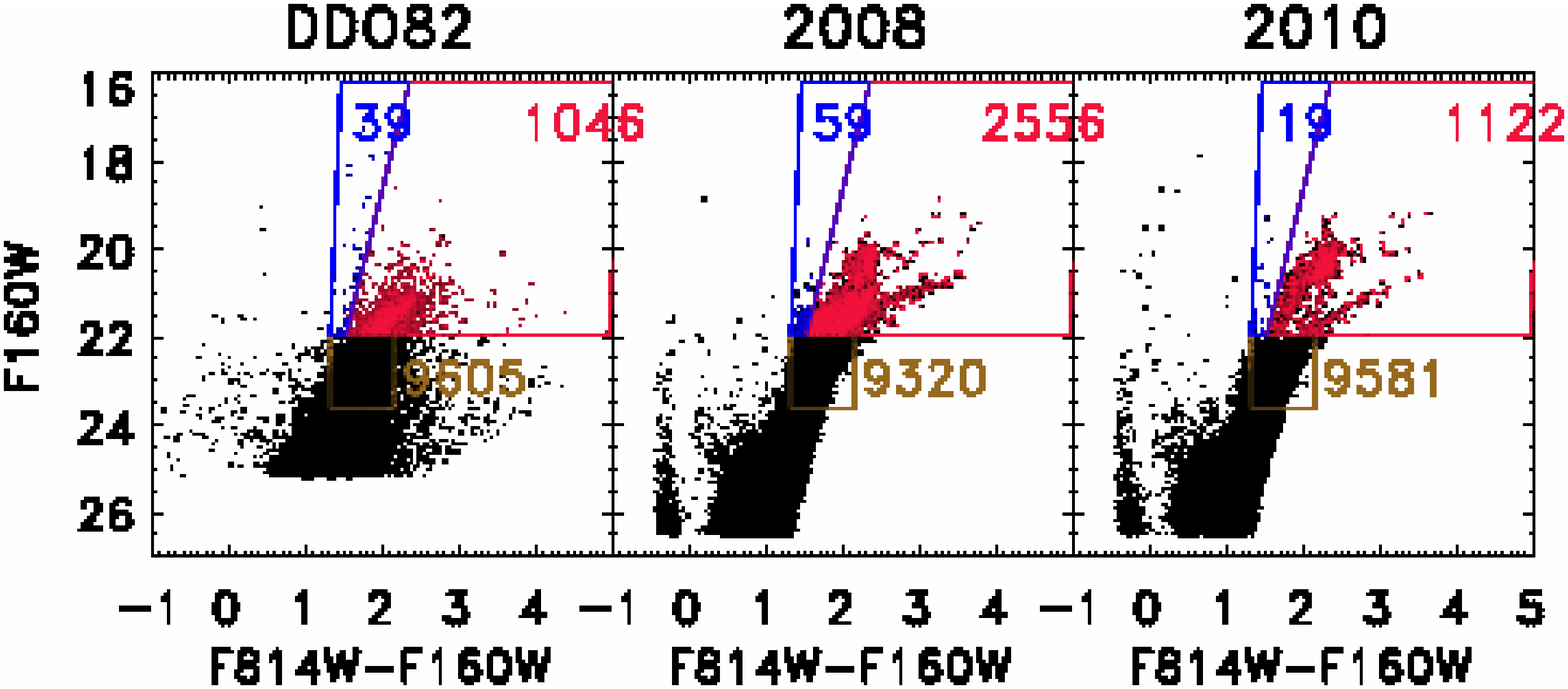}
\caption{\label{fig:CMD} CMDs of the program galaxies in the $(F814W-F160W)$ bands (top left).  Regions that identify different features in the CMDs are shown: RHeB (blue), TP-AGB (red), and RGB (brown).  Model CMDs created from the best-fit star formation histories are also shown for the SPS codes based on the 2008 Padova isochrones (top middle) and 2010 isochrones (top right).   The models do a good job of reproducing the RGB.  As expected the 2008 models significantly over-predict the numbers of TP-AGB stars (red points), while the 2010 models are a much better match.  However, both models tend to over-predict the luminous TP-AGB populations.  The SPS models also tend to under-predict the numbers of RHeB  stars (blue points), especially at the luminous end. }
\end{figure*}

This paper is arranged as follows.  Section 2 describes the optical and NIR \HST\ observations of the sample galaxies.  Section 3 provides the flux fractions contributed by TP-AGB and RHeB stars from both our data and simulations. Section 4 examines the results in more detail and discusses the implications of studies at high redshift. Section 5 summarizes our conclusions. Magnitudes are reported in the Vega system, and we assume the canonical $\Lambda$CDM cosmology with $\Omega_M=0.3$ and $\Omega_{\Lambda}=0.7$ \citep{Spergel07}.

\section{The Data: Resolved Stellar Populations from \HST}
Studies of resolved stellar populations require very high spatial resolution ($<0.1\arcsec$) imaging for galaxies outside the Local Group . Even at the resolution of \HST, these studies are only possible within roughly the local 4 Mpc volume.   ANGST provided the first uniform observational data-set of optical \HST\ imaging of galaxies within the local volume.  A subset of the ANGST  sample has subsequently been observed with high spatial resolution NIR imaging with \HST\ WFC3 \citep{Dalcanton12}.  In this paper, we use the optical and NIR \HST\ ANGST observations of 23 nearby galaxies to study luminous TP-AGB and RHeB stars.

\begin{figure*}
\centering
\figurenum{2}
\includegraphics[scale=0.35]{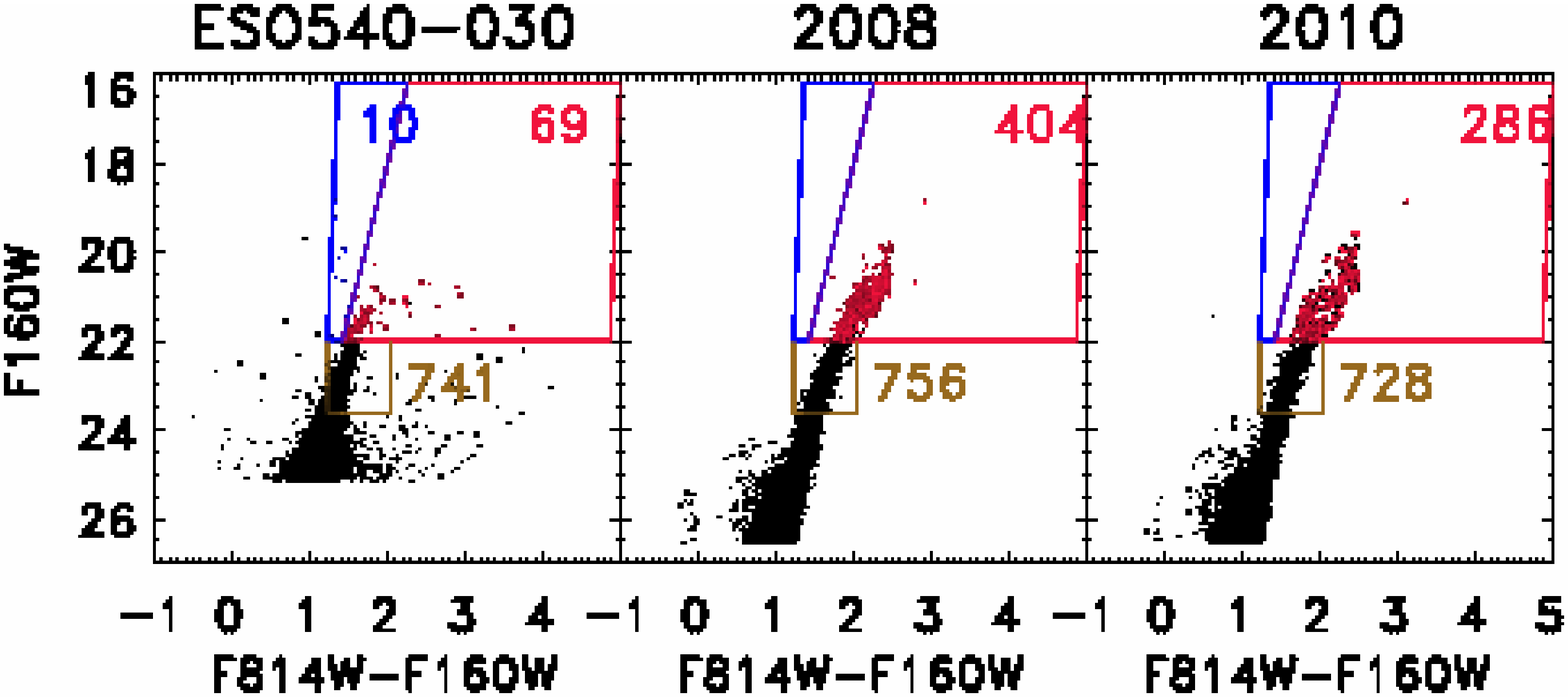}
\vspace{1mm}
\includegraphics[scale=0.35]{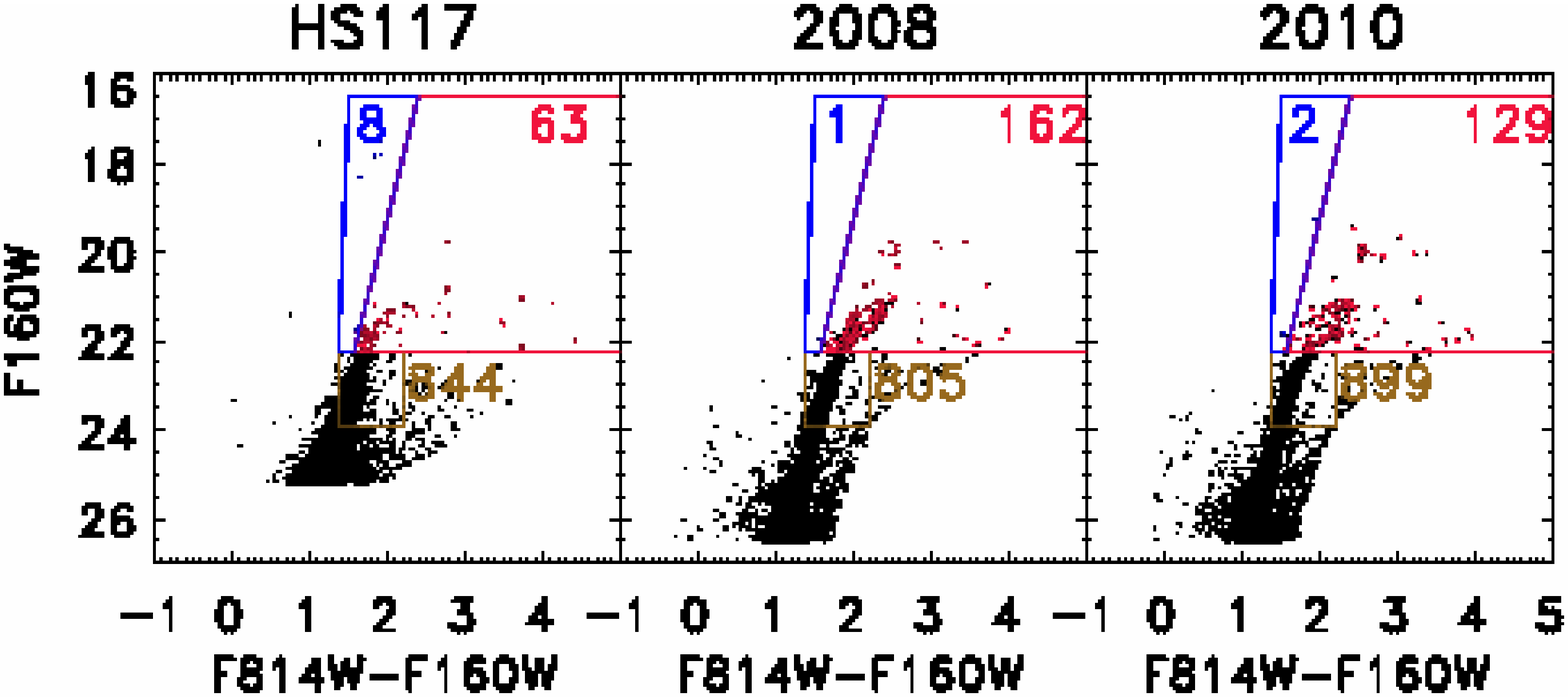}
\vspace{1mm}
\includegraphics[scale=0.35]{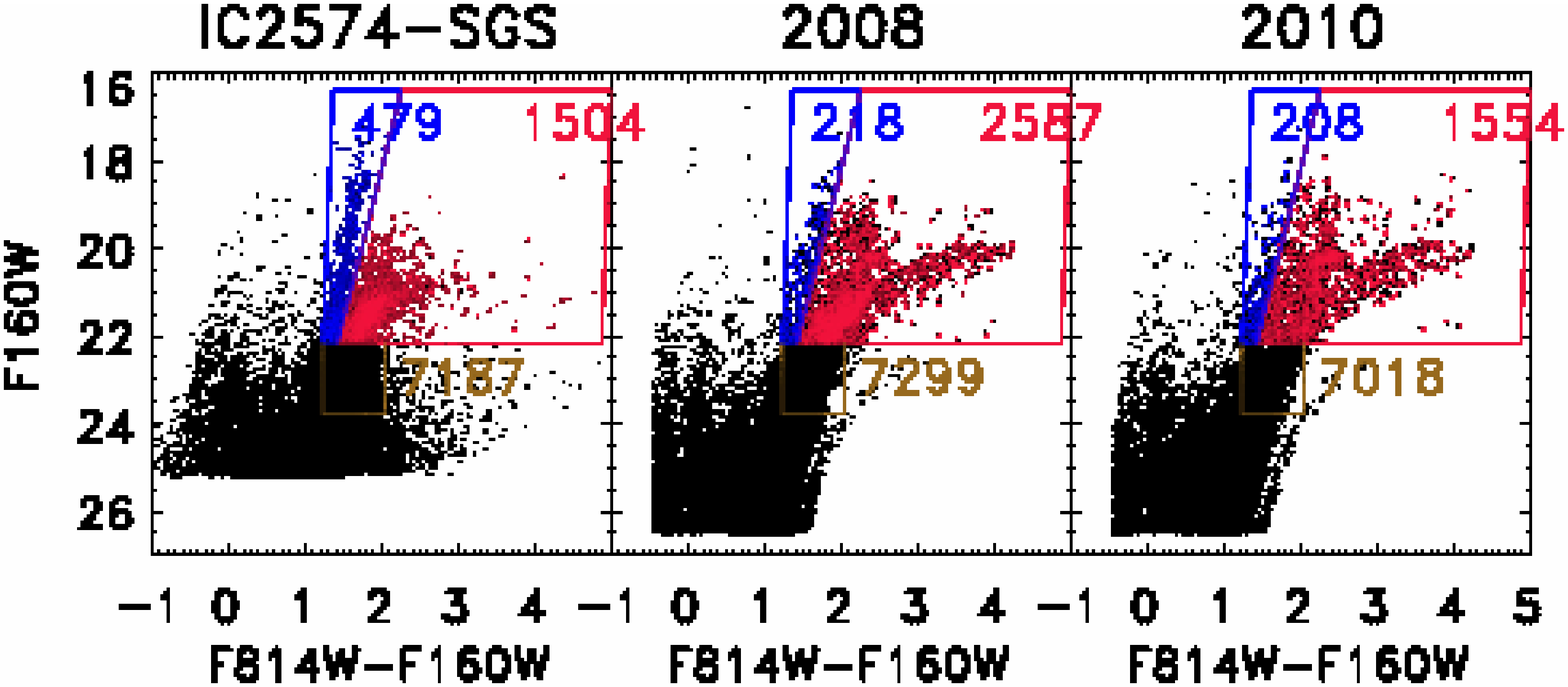}
\caption{continued}
\end{figure*}

\subsection{Optical ACS Imaging and Star Formation Histories \label{sec:sfh}}
Deep multi-band optical \HST\ imaging of a complete set of galaxies within the local 4 Mpc volume was either obtained by the ANGST team or culled from the archive and included in the ANGST program.  ANGST galaxies were observed in at least two filters to provide color and luminosity information for the stars.  Each galaxy was observed in the red $F814W$ filter.  At least one bluer band was also obtained, usually in the $F475W$, $F555W$, or $F606W$ bands. 

The basic image reductions were described in \citet{Dalcanton09}.  Photometry of these fields were obtained with the \HST\ crowded-field photometry package DOLPHOT, a version of HSTPHOT \citep{Dolphin00}, which has been optimized for use with ACS and WFC3. The sensitivities of these photometric data-sets are provided in \citet{Dalcanton09}, but typically were deep enough to reach the red clump and the main sequence turn-off for populations younger than 1 Gyr.  Figure \ref{fig:AllCMD} shows an example optical CMD.

The optical multi-band photometry was used to constrain the SFH of each galaxy.  The numbers and positions of stars across color-magnitude space are set by stellar evolution and the SFH of each galaxy.  Both the youngest and oldest stellar populations are thought to be well constrained by the CMDs in the ANGST sample \citep[see tests in][]{Weisz11}.  The youngest populations are constrained by luminous main sequence stars and evolving supergiant stars.  Older populations are well constrained by the RGB, which becomes well-populated for galaxies older than $\sim2$ Gyrs.    

The global SFHs of the ANGST dwarf galaxies are described in \citet{Weisz11}, and were determined with the 2-D CMD fitting routine CalcSFH \citep[part of the MATCH package][]{Dolphin02}. We constructed the SFH of each galaxy based on the numbers of stars within color magnitude bins on the observed CMD, with color bins of size 0.05 mags, and magnitude bins of 0.1 mags.  We used 71 logarithmic time bins from 4 Myrs to 12 Gyrs old, and 24 different metallicity bins ranging from metallicities of $[M/H] = -2.25$ to 0.05.  We assumed a single-slope power-law IMF with a spectral index of -1.30 over a mass range of $0.1-120$ \msun, and a binary fraction of 0.35 with a flat secondary mass distribution.   The difference between our selected IMF and a Kroupa IMF \citep{Kroupa01} is negligible, as the ANGST CMDs are limited to stellar masses $>0.8$ \msun. 

The methods for estimating uncertainties in the SFHs are also described in \citet{Weisz11}.  We used the standard MATCH routine Monte Carlo approach to estimate both the random and systematic uncertainties. For each Monte Carlo run, the observed CMD was randomly resampled and refit with CalcSFH. Additive errors in $M_{bol}$ and log($T_{eff}$) were introduced when generating the model CMDs for these solutions. This method was developed to account for the full range of systematic differences between isochrone sets that use different prescriptions for various phenomena such as stellar rotation and convective overshooting. 

\begin{figure*}
\figurenum{2}
\centering
\includegraphics[scale=0.35]{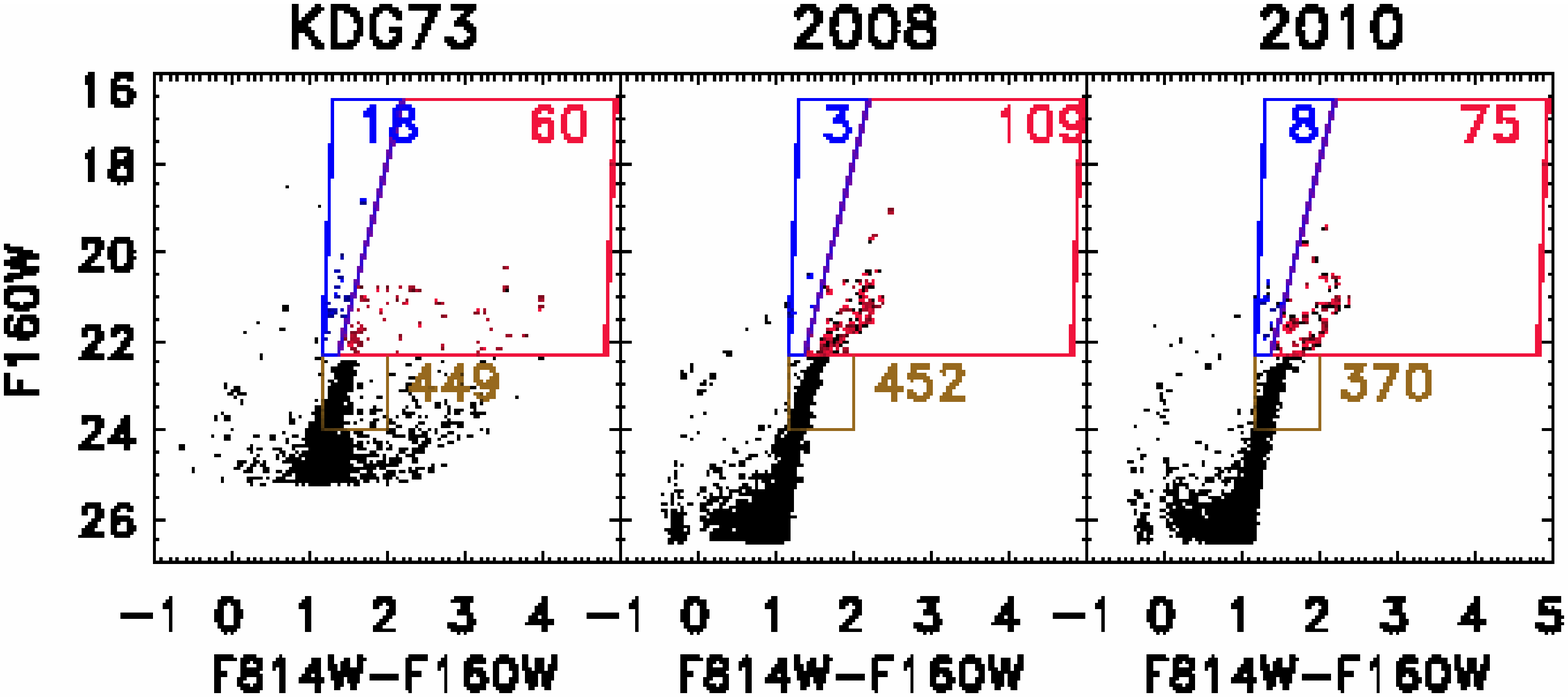}
\vspace{1mm}
\includegraphics[scale=0.35]{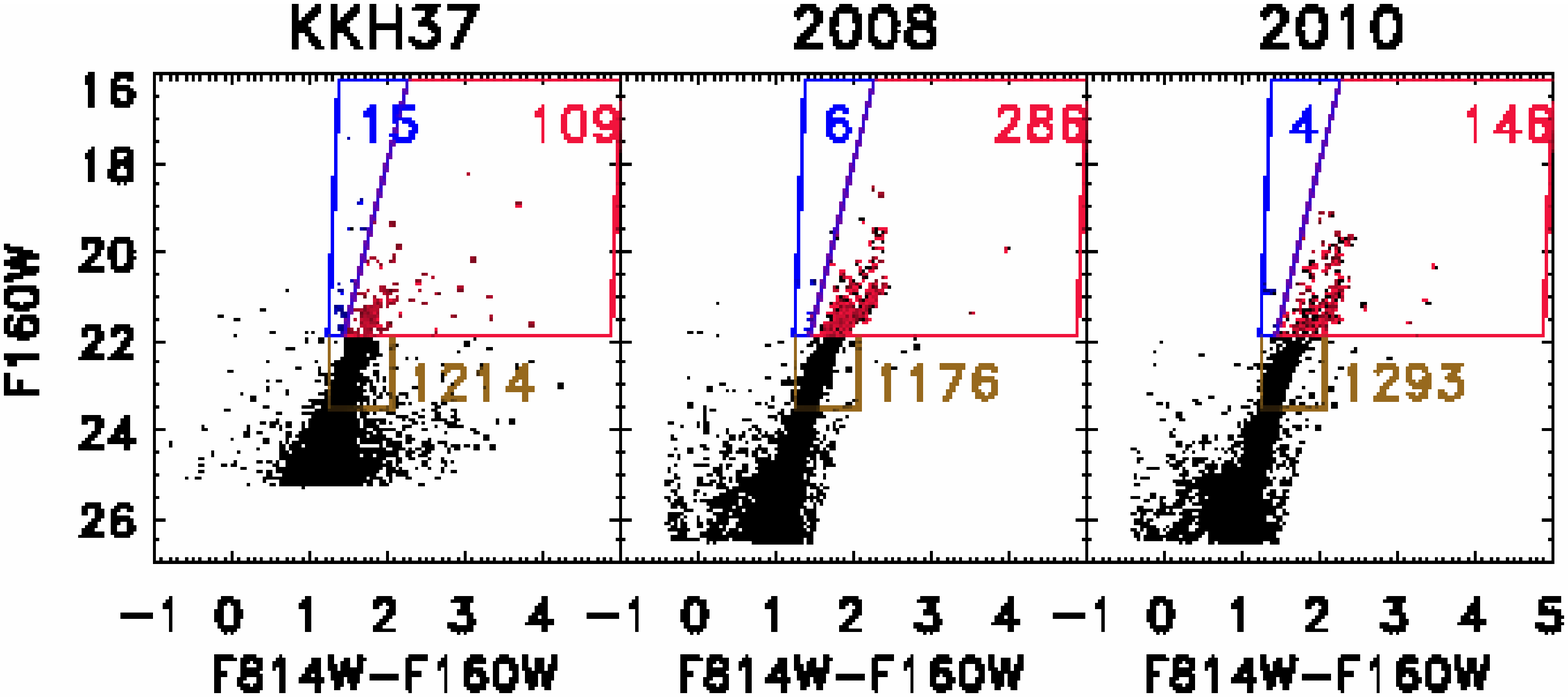}
\vspace{1mm}
\includegraphics[scale=0.35]{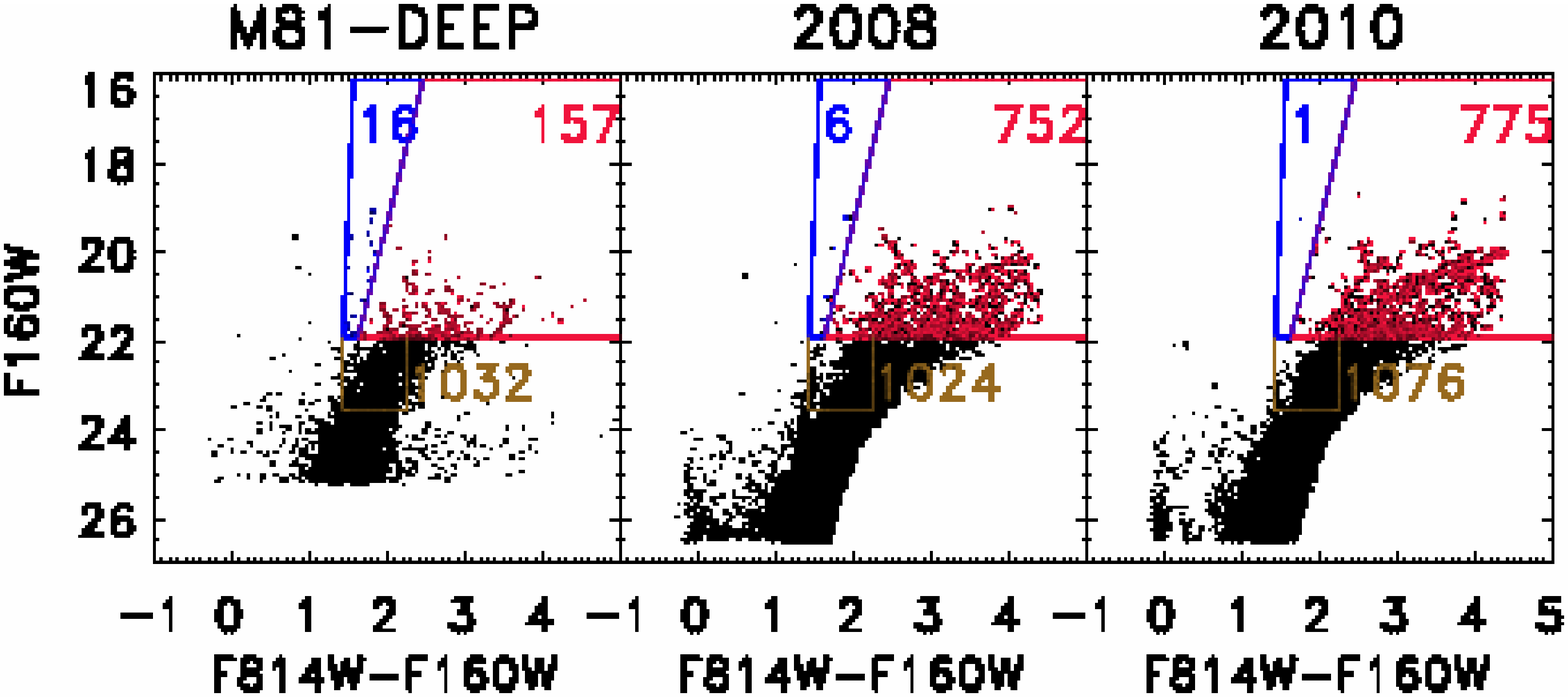}
\caption{continued}
\end{figure*}

For this paper, we focus on the sub-regions of the ANGST galaxies that were observed in both the optical with ACS and the NIR with WFC3.  These sub-regions are smaller than the full ACS fields, and may have different SFHs than the full fields.  We therefore re-ran the SFH modeling codes using the optical photometry of just the overlap region, but following the procedure and binning used in \citet{Weisz11}.  This reanalysis typically did not change the relative amount of star formation in each time bin but only the overall scaling. Table \ref{tab:results} gives the distance modulus, metallicity, and fractions of young stars for each galaxy as measured by the SFH routine CalcSFH.  Table \ref{tab:results} also provides uncertainties for these parameters. Both metallicity and fractions of young stars are characterized by relatively large uncertainties compared to Local Group galaxies for which the SFH has been determined with the same method \citep{Dolphin05, Holtzman06,  Gallart08, Williams09b,Weisz11b}. As discussed in \citet{Girardi10}, larger SFH uncertainties are the price to pay for observing large samples of TP-AGB stars with only a single \HST\ pointing per galaxy, and with minimal contamination from foreground stars.

\subsection{\HST\ NIR WFC3 Imaging}
\HST\ NIR images of a subset of the ANGST sample were obtained during Cycle 17, in program SNAP-11719.    Imaging was obtained in both the $F110W$ and $F160W$ filters, with total exposure times of 597.7\,s and 897.7\,s respectively.  The observations and image reduction of the WFC3 data are described in \citet{Dalcanton12}. As with the optical data, photometry of the WFC3 observations was done with the DOLPHOT package, which has been updated to include a module for the processing of WFC3 data.  

Figure \ref{fig:AllCMD} shows an example NIR CMD produced from these data.  Typical uncertainties for the stellar photometry range from 0.01 mags at the $F160W\sim18$ to 0.10 mags at $F160W\sim24$.  Each galaxy is observed to several magnitudes below the TRGB.  These limits are not typically faint enough to detect the main sequence turn off for stars older than a couple hundred Myrs, or the well-populated red clump.  However, the CMDs do show AGB, RGB, and RHeB populations (Figure 1).  

\subsection{Catalogue Matching}

By design there is significant overlap between the optical and NIR images of each galaxy.  We generate optical through NIR matched catalogues to identify TP-AGB and RHeB stars. 
To do a proper transformation between the two coordinate systems, we select $\sim150$ stars that are bright in both the optical and NIR data sets and that spatially span the entire overlap region between the WFC3 and ACS images.   Starting with the optical and NIR catalogues from DOLPHOT, we cull the lists to only include stars that are in the spatial overlap region.  Then we select all of the reasonably bright red stars in each dataset, with optical color $> 0.7$ mags and $F814W < 26$ mags, and IR color $(F110W - F160W) > 0.5$ mags and $F160W < 24$ mags.  We sort these two lists by luminosity and select 150 stars roughly evenly spaced across the image choosing the more luminous stars first.  Note that the final list is not the 150 brightest, because these are often spatially clustered and do not span the full area.  

\begin{figure*}
\figurenum{2}
\centering
\includegraphics[scale=0.35]{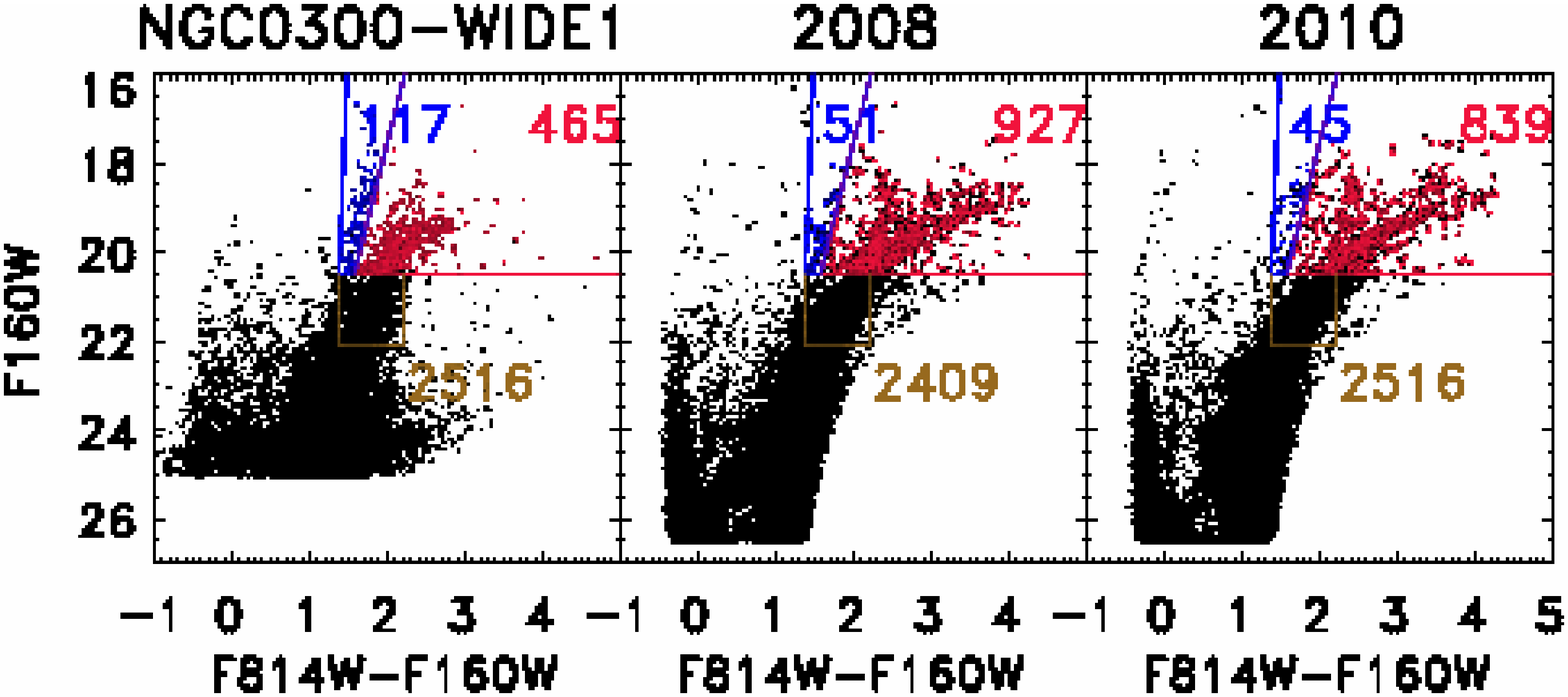}
\vspace{1mm}
\includegraphics[scale=0.35]{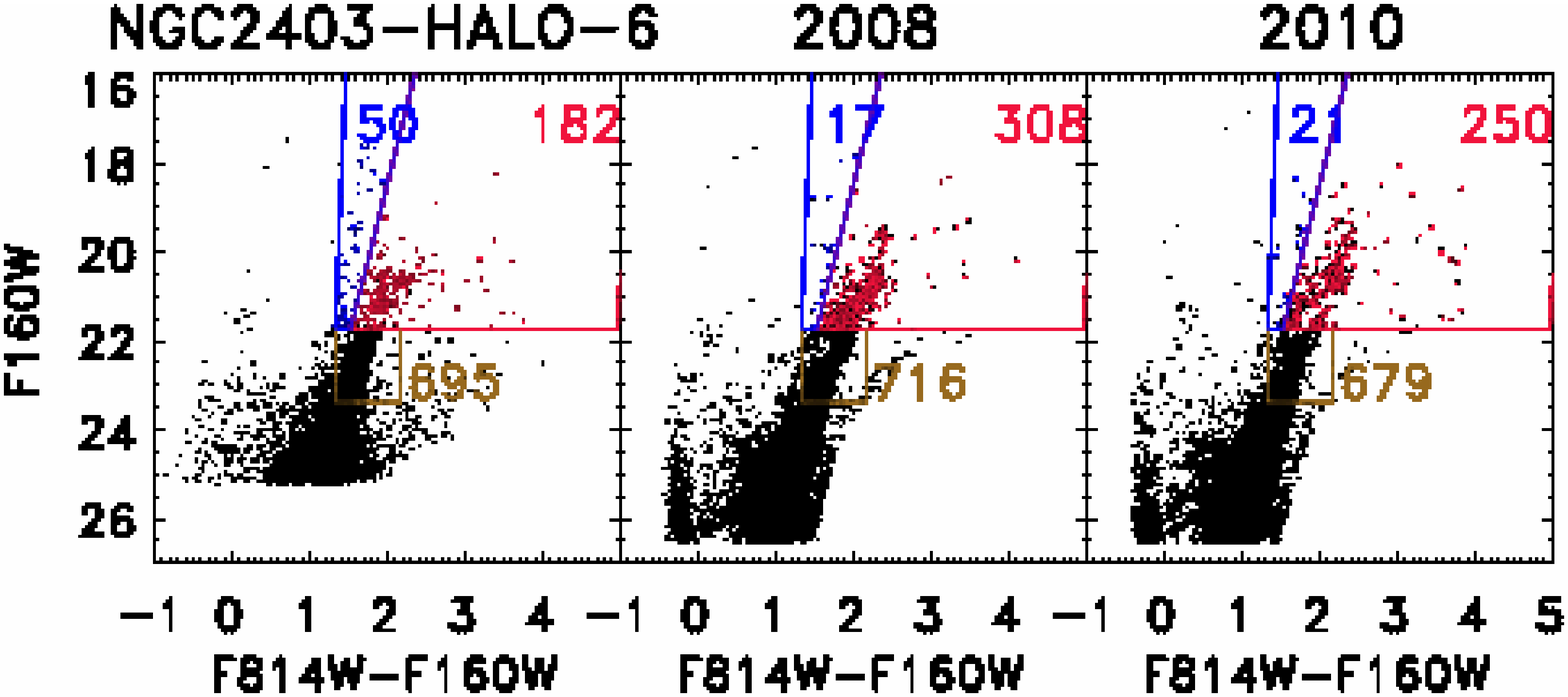}
\vspace{1mm}
\includegraphics[scale=0.35]{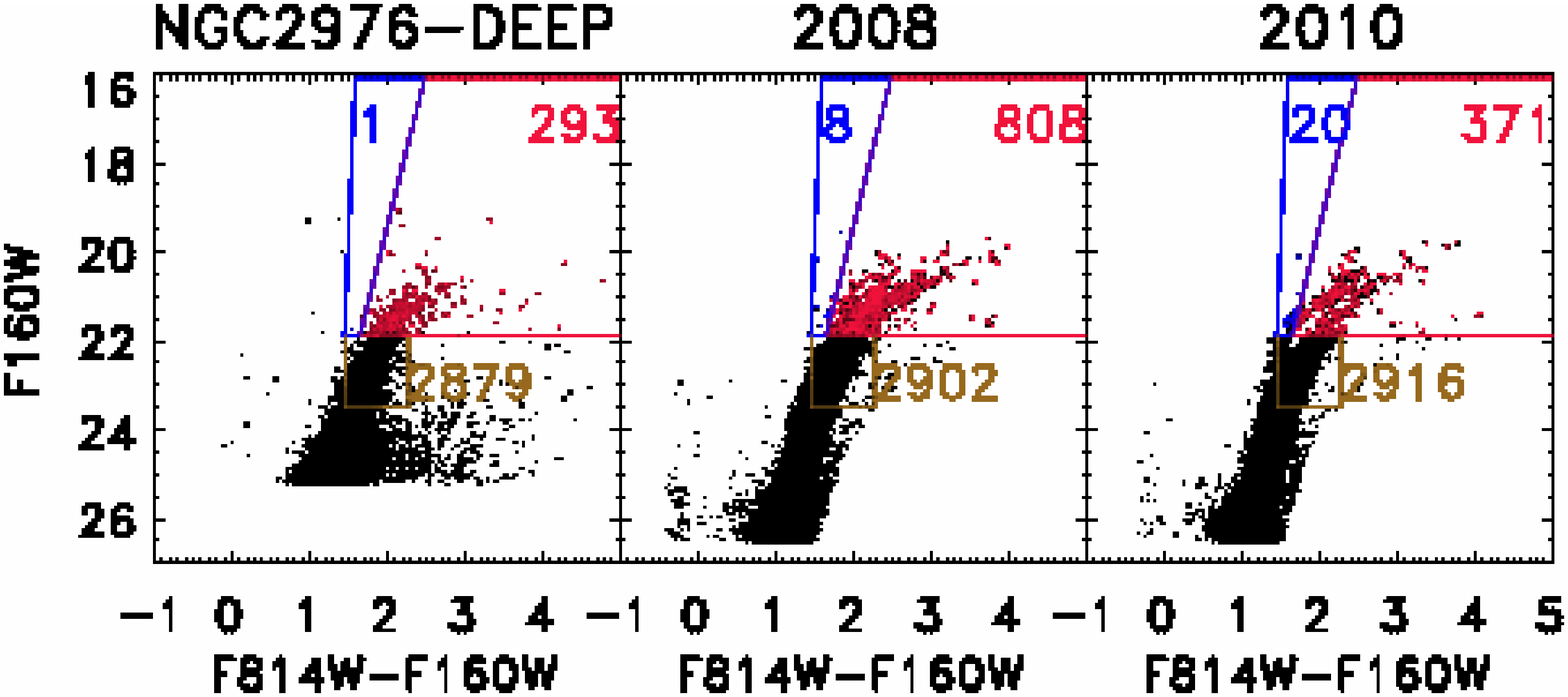}
\caption{continued  }
\end{figure*}

Next, we visually identify a roughly linear shift between the two coordinate systems and apply the transformation to our subset of matching stars.  This comparison acts both as a visual check that the star lists we are using are well constructed, and also as a first pass at determining the final transformation. 

The final transformation is determined iteratively with the routine \emph{MATCH} developed by Michael Richmond, based on the method of triangles described in \citet{Valdes95}.   First we find a linear fit between the two coordinate systems.  We then use that solution as a starting point for a quadratic solution.  We find that a cubic solution is generally unnecessary for the transformation between the distortion corrected WFC3 and ACS images.  

After determining the transformation between the two coordinate systems with our set of 150 matching stars, we apply the transformation to the entire NIR dataset, bringing it into the optical coordinate system.  The final step is to then use a separation criteria to determine if there is a good match.  We find that a separation of $0.07\arcsec$ works well across the entire field.  Typically 90\% of the stars in the NIR catalogue are well-matched to a star in the optical catalogue.  Of the remaining 10\%, the bulk are either located in the wings of saturated stars or in the chip gap in the ACS camera.   The $F160W$ vs. $(F814W - F160W)$ CMDs for all of the program galaxies are shown in Figure \ref{fig:CMD}.  

\section{The Contribution of TP-AGB and RHeB Stars to the NIR Flux of Galaxies}
The primary goal of this paper is to constrain the contribution of late stage stellar evolution to the  NIR fluxes of galaxies with different metallicities and star formation histories.  To do so, we must  (1) determine the total flux falling in the WFC3 F160W filter for each galaxy, and (2) determine the flux from the TP-AGB and RHeB stars in the same area.  

\begin{figure*}
\figurenum{2}
\centering
\includegraphics[scale=0.35]{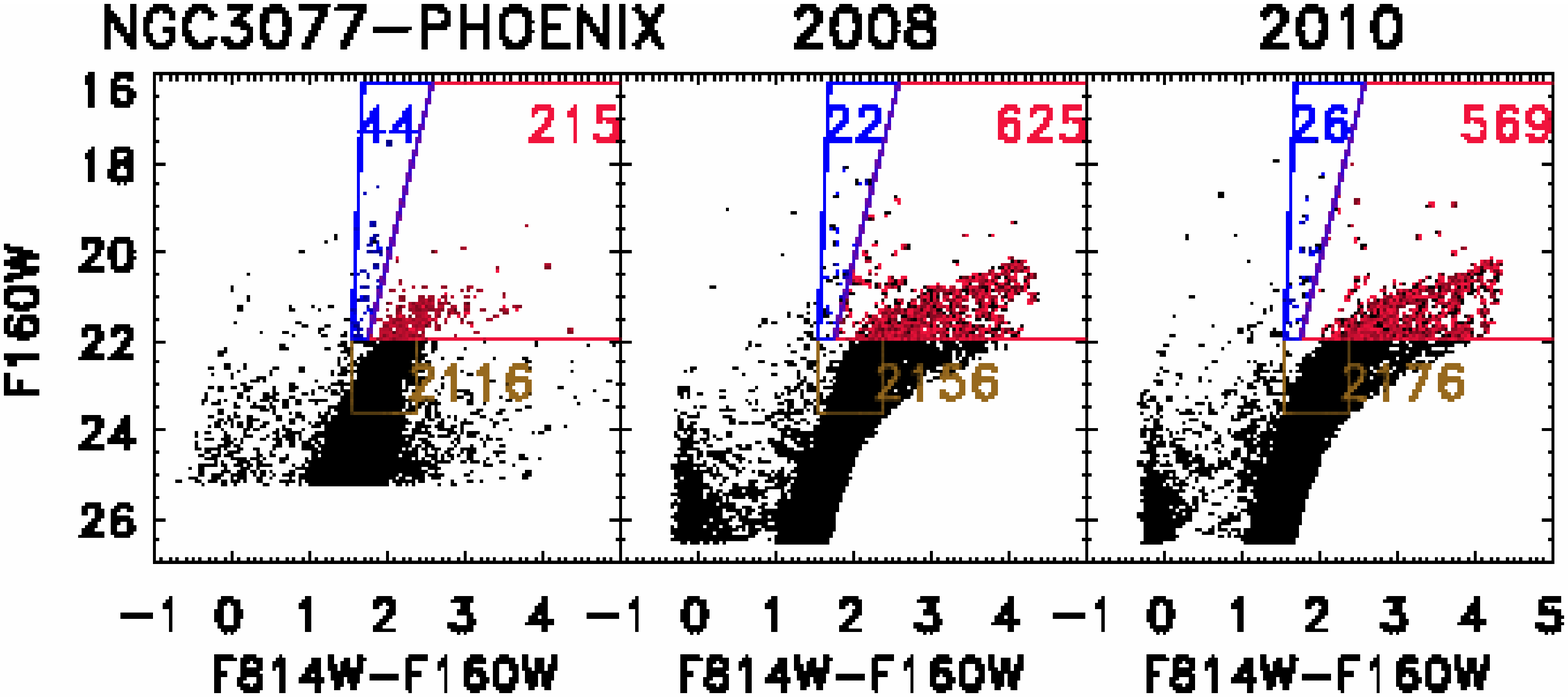}
\vspace{1mm}
\includegraphics[scale=0.35]{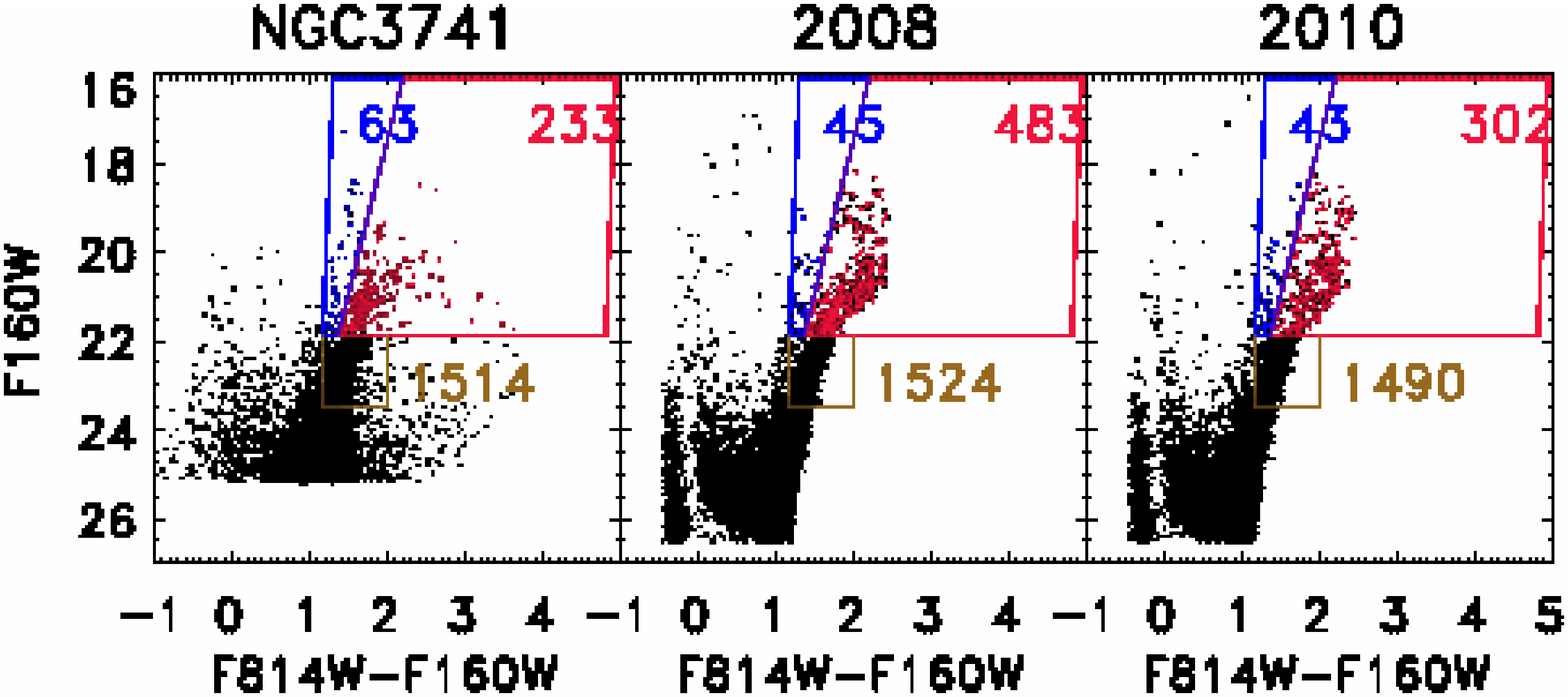}
\vspace{1mm}
\includegraphics[scale=0.35]{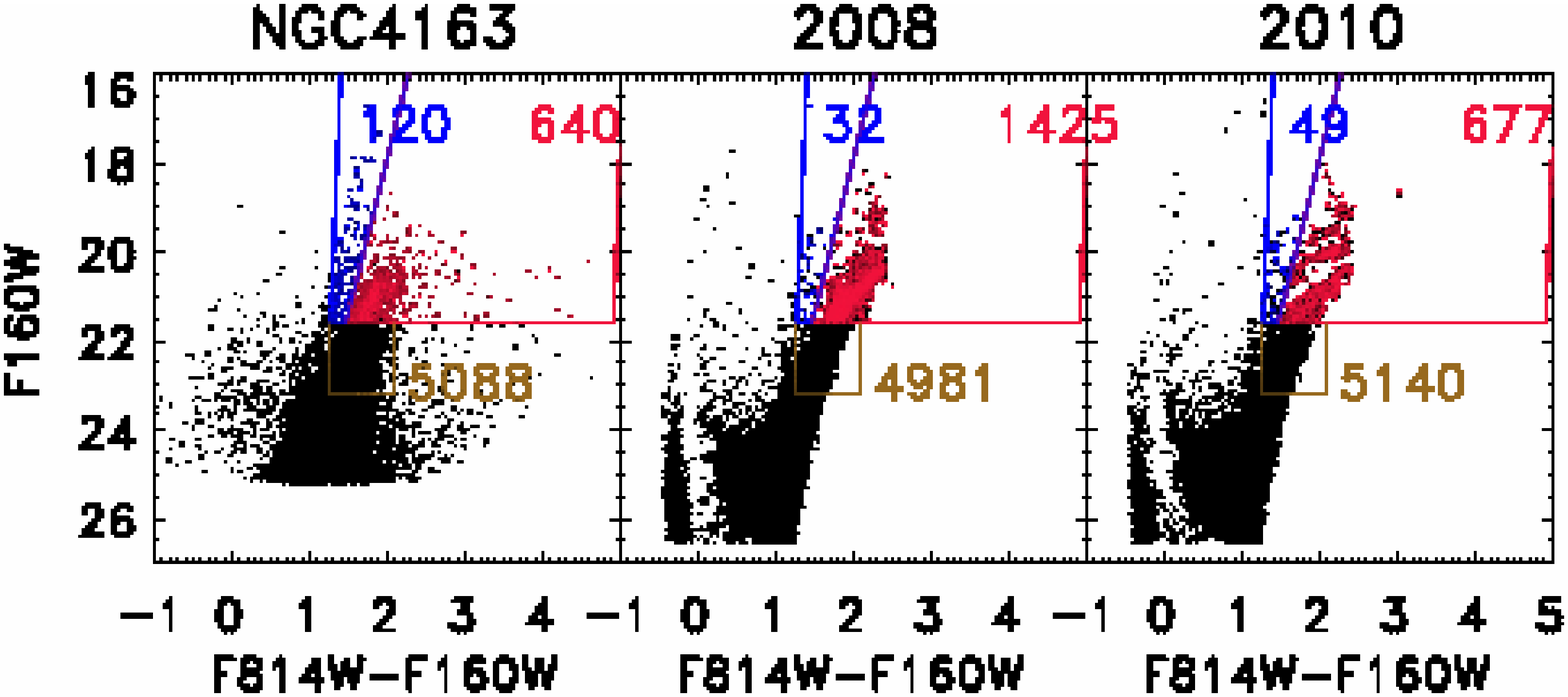}
\caption{continued  }
\end{figure*}

Determining the total fluxes of the sample fields is actually non-trivial, as many of the images do not contain a clean measure of the background sky.  Although we can easily determine the flux of the individual bright stars in each frame, we cannot directly photometer the stars that are too faint to detect.  Instead we chose to model the contribution from the faint end, by generating synthetic NIR CMDs based on the optically derived SFHs, as described below.  We will use these synthetic CMDs to both construct a total flux for each field, and to test model prescriptions for the TP-AGB and RHeB stars. 

\subsection{Synthetic NIR CMDs}
We create synthetic NIR CMDs for each galaxy based on the SFHs derived from very deep optical \HST\ imaging (see Section \ref{sec:sfh}).    We input the measured SFHs, reddening values, and distance moduli into CalcSFH to produce model Hess diagrams in the $F814W$ and $F160W$ filters.  We then sample these model Hess Diagrams with the routine NoisyCMD \citep{Dolphin02} to generate synthetic  photometry of each galaxy field down to K dwarfs.  

NoisyCMD uses the Padova isochrones \citep{Marigo08} with updated bolometric corrections and Teff-color relations \citep{Girardi08} to populate the model CMDs.  However, NoisyCMD does not include the effects of long lived thermal-pulses \citep[i.e. 10,000 years, see ][]{Wagenhuber98} which can scatter up to 20\% of the lower mass TP-AGB stars to lower luminosities. 

\begin{figure*}
\figurenum{2}
\centering
\includegraphics[scale=0.35]{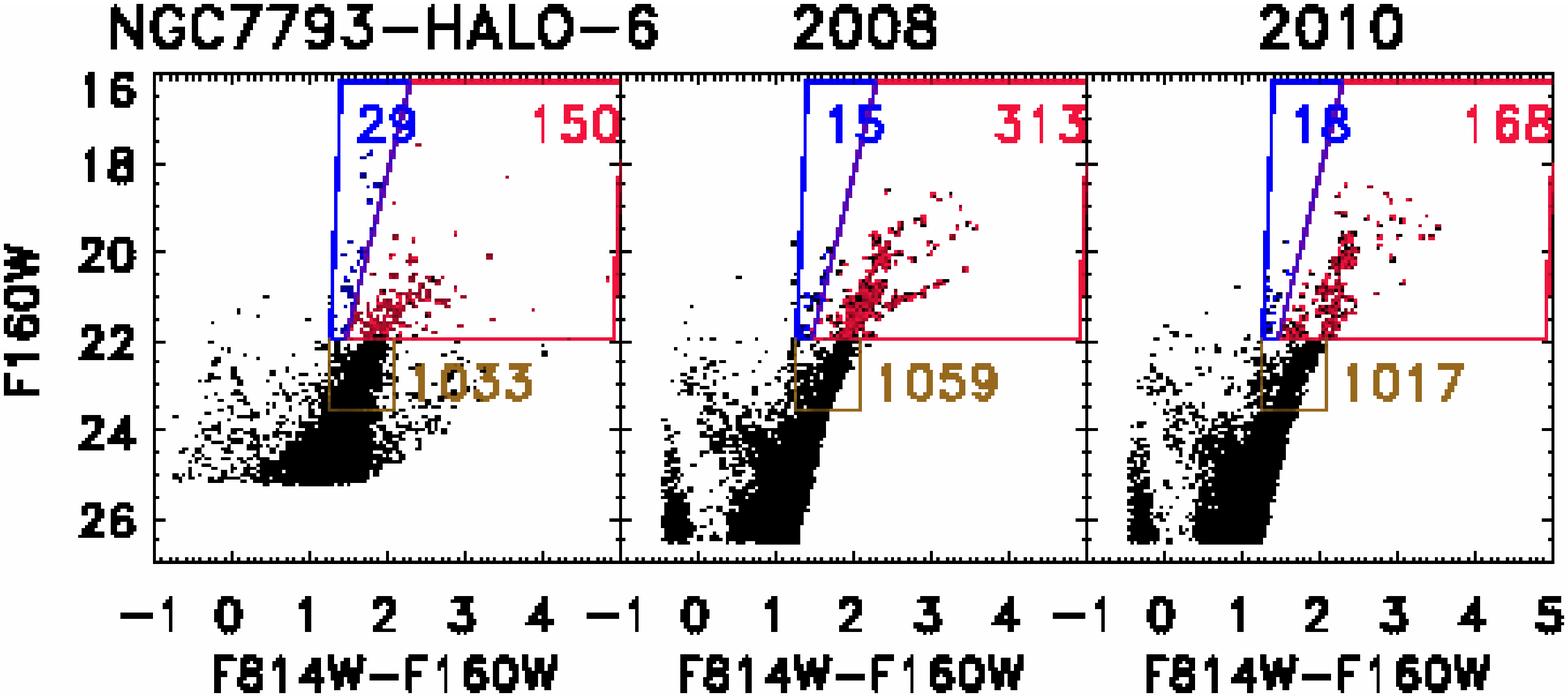}
\vspace{1mm}
\includegraphics[scale=0.35]{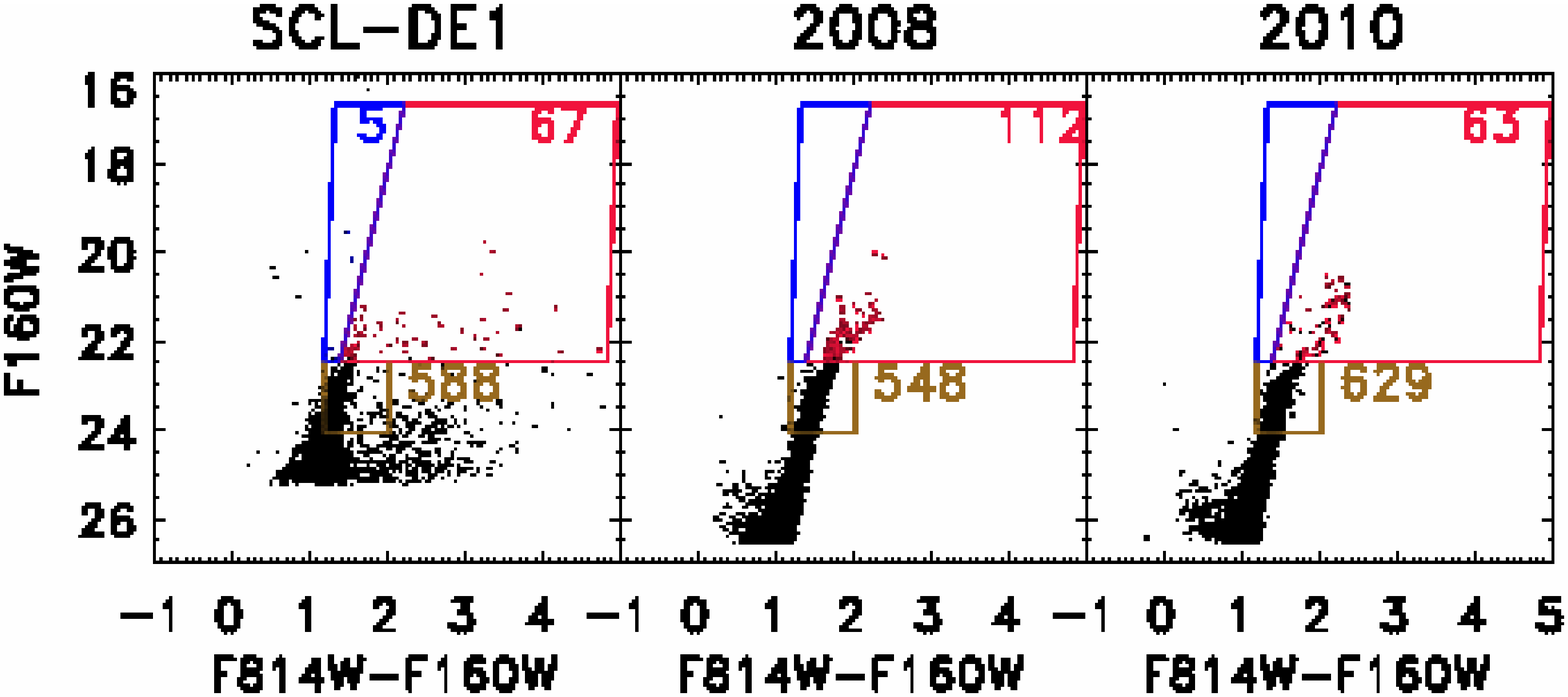}
\caption{continued }
\end{figure*}

\begin{deluxetable*}{lcccc}
\tabletypesize{\small}
\tablecaption{Total Flux and Numbers of RGB Stars in Each Galaxy \label{tab:rgb}}
\tablehead{\colhead{Galaxy} & \colhead{Total F160W Flux \tablenotemark{a}} & \colhead{\# RGB} & \colhead{\# RGB} & \colhead{\# RGB}\\
& \colhead{ergs cm$^{-2}$ s$^{-1}$}& \colhead{data} & \colhead{model 2008} & \colhead{model 2010}}
\startdata
          DDO71 &  7.550e-16 &   1958 $\pm$     44 &   1940 $\pm$     44 &   1906 $\pm$     43 \\
          DDO78 &  1.330e-15 &   3050 $\pm$     55 &   2913 $\pm$     53 &   3166 $\pm$     56 \\
          DDO82 &  4.570e-15 &   9605 $\pm$     98 &   9320 $\pm$     96 &   9581 $\pm$     97 \\
     ESO540-030 &  4.090e-16 &    741 $\pm$     27 &    756 $\pm$     27 &    728 $\pm$     26 \\
          HS117 &  3.980e-16 &    844 $\pm$     29 &    805 $\pm$     28 &    899 $\pm$     29 \\
     IC2574-SGS &  4.570e-15 &   7187 $\pm$     84 &   7299 $\pm$     85 &   7018 $\pm$     83 \\
          KDG73 &  1.810e-16 &    449 $\pm$     21 &    452 $\pm$     21 &    370 $\pm$     19 \\
          KKH37 &  6.830e-16 &   1214 $\pm$     34 &   1176 $\pm$     34 &   1293 $\pm$     35 \\
       M81-DEEP &  9.590e-16 &   1032 $\pm$     32 &   1024 $\pm$     32 &   1076 $\pm$     32 \\
  NGC0300-WIDE1 &  6.390e-15 &   2516 $\pm$     50 &   2409 $\pm$     49 &   2516 $\pm$     50 \\
 NGC2403-HALO-6 &  8.190e-16 &    695 $\pm$     26 &    716 $\pm$     26 &    679 $\pm$     26 \\
   NGC2976-DEEP &  1.450e-15 &   2879 $\pm$     53 &   2902 $\pm$     53 &   2916 $\pm$     54 \\
NGC3077-PHOENIX &  1.270e-15 &   2116 $\pm$     46 &   2156 $\pm$     46 &   2176 $\pm$     46 \\
        NGC3741 &  1.100e-15 &   1514 $\pm$     38 &   1524 $\pm$     39 &   1490 $\pm$     38 \\
        NGC4163 &  3.780e-15 &   5088 $\pm$     71 &   4981 $\pm$     70 &   5140 $\pm$     71 \\
 NGC7793-HALO-6 &  6.920e-16 &   1033 $\pm$     32 &   1059 $\pm$     32 &   1017 $\pm$     31 \\
        SCL-DE1 &  3.000e-16 &    588 $\pm$     24 &    548 $\pm$     23 &    629 $\pm$     25 \\
      UGC4305-1 &  3.410e-15 &   3664 $\pm$     60 &   3588 $\pm$     59 &   3725 $\pm$     61 \\
      UGC4305-2 &  3.250e-15 &   4087 $\pm$     63 &   4072 $\pm$     63 &   4064 $\pm$     63 \\
        UGC4459 &  1.250e-15 &   1815 $\pm$     42 &   1771 $\pm$     42 &   1787 $\pm$     42 \\
        UGC5139 &  1.500e-15 &   2783 $\pm$     52 &   2821 $\pm$     53 &   2647 $\pm$     51 \\
        UGC8508 &  1.960e-15 &   2094 $\pm$     45 &   2053 $\pm$     45 &   2097 $\pm$     45 \\
        UGCA292 &  2.470e-16 &    354 $\pm$     18 &    324 $\pm$     18 &    341 $\pm$     18 \\
\enddata
\tablenotetext{a}{Data brighter than F160W$=23$ mag plus model fainter than F160W$=23$.}
\end{deluxetable*}

The Padova isochrones have been discussed in detail previously \citep{Girardi00, Marigo07, Marigo08}; here we include a brief description. The primary distinction of the Padova isochrones over previous efforts \citep[e.g.,][]{BC03, Raimondo05, Vazquez05} is the detailed characterization of several key aspects of the TP-AGB phase, including: hot bottom burning, third dredge up, and variable atmospheric opacities. These effects are crucial for tracking the evolution of TP-AGB stars, especially across the transition from oxygen rich to carbon rich phases, and the production of circumstellar dust. In addition, the Padova isochrones incorporate mass-loss from dust driven winds \citep{Winters00, Winters03}, and follow the TP-AGB evolution through the loss of the outer gaseous envelope. Circumstellar dust brings its own complications, and different prescriptions can lead to different outcomes as explained in  \citet{Marigo08}. In the present work, we adopt the isochrones without circumstellar dust, however, briefly discussing  the effects dust may cause in the star counts and integrated fluxes.The bolometric corrections are described in detail in \citet{Girardi02, Girardi08}. They were generated from the spectrophotometric standards assembled in \citet{Bohlin07}, and a large library of spectral fluxes assembled in \citet{Girardi02} and \citet{Aringer08}.

%The Padova isochrones have been discussed in detail previously \citep{Girardi00, Marigo07, Marigo08}; here we include a brief description.  The primary distinction of the Padova isochrones over previous efforts \citep[e.g.,][]{BC03, Raimondo05, Vazquez05} is the detailed characterization of several key aspects of the TP-AGB phase, including: hot bottom burning, third dredge up, and variable atmospheric opacities.  These effects are crucial for tracking the evolution of TP-AGB stars, especially across the transition from oxygen rich to carbon rich phases, and the production of circumstellar dust. In addition, the Padova isochrones incorporate mass-loss from dust driven winds \citep{Winters00, Winters03}, and follow the TP-AGB evolution through the loss of the outer gaseous envelope. However, they do not currently include prescriptions for self-reddening from circumstellar dust. The bolometric corrections are described in detail in \citet{Girardi02, Girardi08}.  They were generated from the spectrophotometric standards assembled in \citet{Bohlin07}, and a large library of spectral fluxes assembled in \citet{Girardi02}.     

 %Note that deeper model CMDs do not result in appreciably larger total flux.

In an effort to account for any systematic offset between the models and the data, we produce two iterations of the model photometry.  In the first iteration, we generate model photometry of the more luminous stars.  We compare the numbers of luminous RGB stars in the model to the number of RGB stars in the data (see brown box in Figure \ref{fig:CMD}), and calculate a scaling between the two.  We then re-run the models to very faint levels, applying the scaling to the SFHs.  This assures a good match between the models and the data on the RGB.  The scalings we calculate are typically less than 10\% and give some indication of the uncertainty on the total luminosity we are measuring for each galaxy.  %Changing the SFH's by their estimated uncertainties gives another measure on the uncertainty of the total fluxes, which result in typical differences smaller than 5\%. 

 We create two different synthetic CMDs.  The first (middle panel of Figure \ref{fig:CMD}) is based on the 2008 Padova isochrones \citep{Marigo08}, which form the basis for several commonly adopted stellar population synthesis codes in use today. The second includes the updated Padova TP-AGB models  of \citet{Girardi10}, which has a shorter lifetime for low mass, low metallicity TP-AGB stars.  This latter model effectively lowers the total number of TP-AGB stars in the model CMDs (right panel of Figure \ref{fig:CMD}).

The model CMDs are shown in Figure \ref{fig:CMD} and (by design) are well-matched to the observed CMDs at the RGB (see Table 2).  

\subsection{Total 1.6 \um\ Fluxes}
While the model CMDs do a good job reproducing the well understood main sequence and red giant branch stars, they do not necessarily reproduce the most luminous stars which are in phases of late stage stellar evolution (i.e.  the TP-AGB and He burning phases that we are investigating).  We therefore adopt a total flux for each galaxy based on a hybrid of model + data fluxes, such that:
\begin{eqnarray}
\textrm{Total Stellar Flux} &=& \textrm{Observed Flux of Luminous Stars} + \nonumber \\ 
&&\textrm{Model Flux of Faint Stars}
\end{eqnarray}
We chose a splice point between the data and the model at $F160W=23$ mags, typically over a magnitude below the TRGB in $F160W$.  At this flux level, the HST magnitudes are well measured, and are good to within 0.05 mags.  This splice point is also much brighter than the typical 50\% completeness limit of $F160W \approx 25$ mags.

When calculating the observed flux, we exclude the small number (if any) of  extremely IR bright stars, $6-8$ magnitudes brighter than the TRGB (set by eye for this paper).  These stars are likely to be foreground stars (see Section 3.6 below).  Their flux is also excluded from the calculated contribution from the TP-AGB and RHeB phases.  In general the foreground numbers are expected to be small in the TP-AGB and RHeB regions of the CMD for fields of this small angular size \citep[e.g.][]{Girardi10}.  

\begin{figure*}
\figurenum{2}
\centering
\includegraphics[scale=0.35]{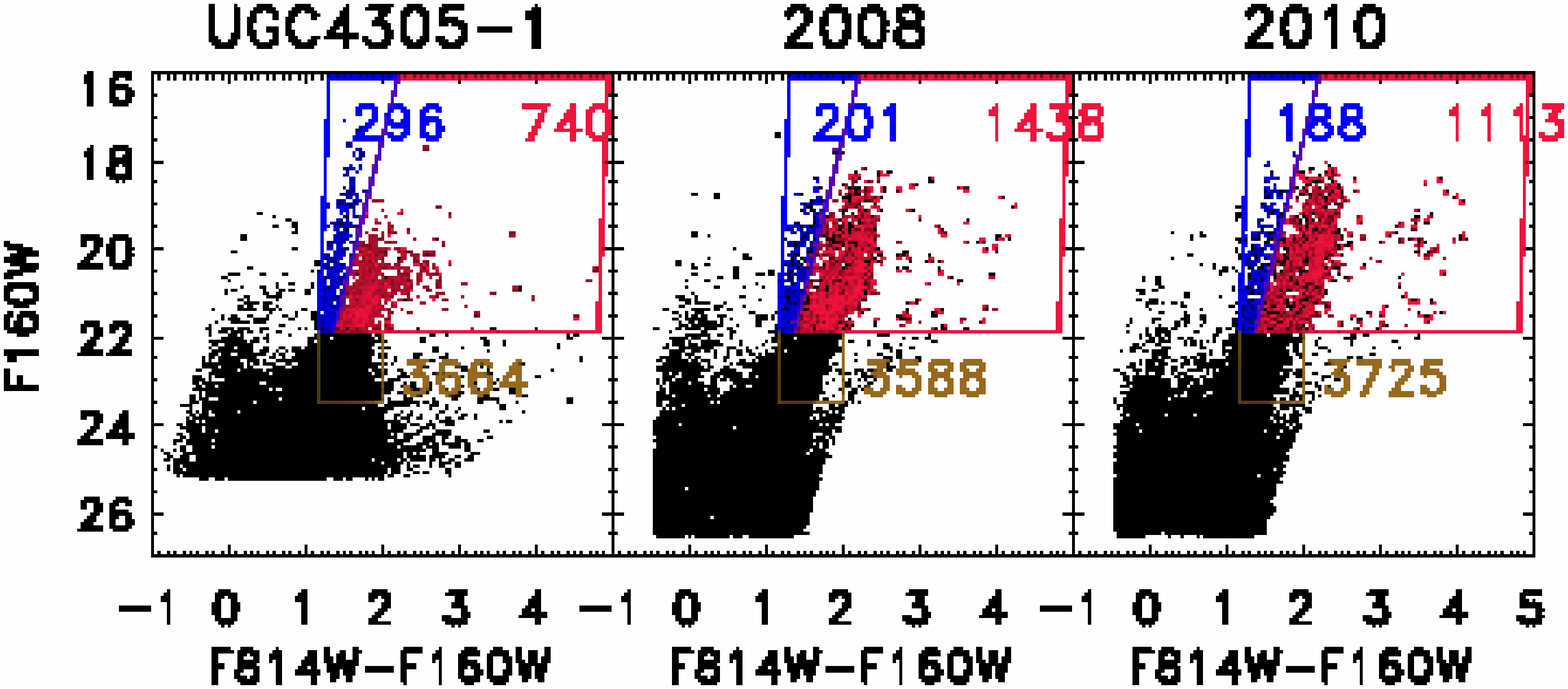}
\vspace{1mm}
\includegraphics[scale=0.35]{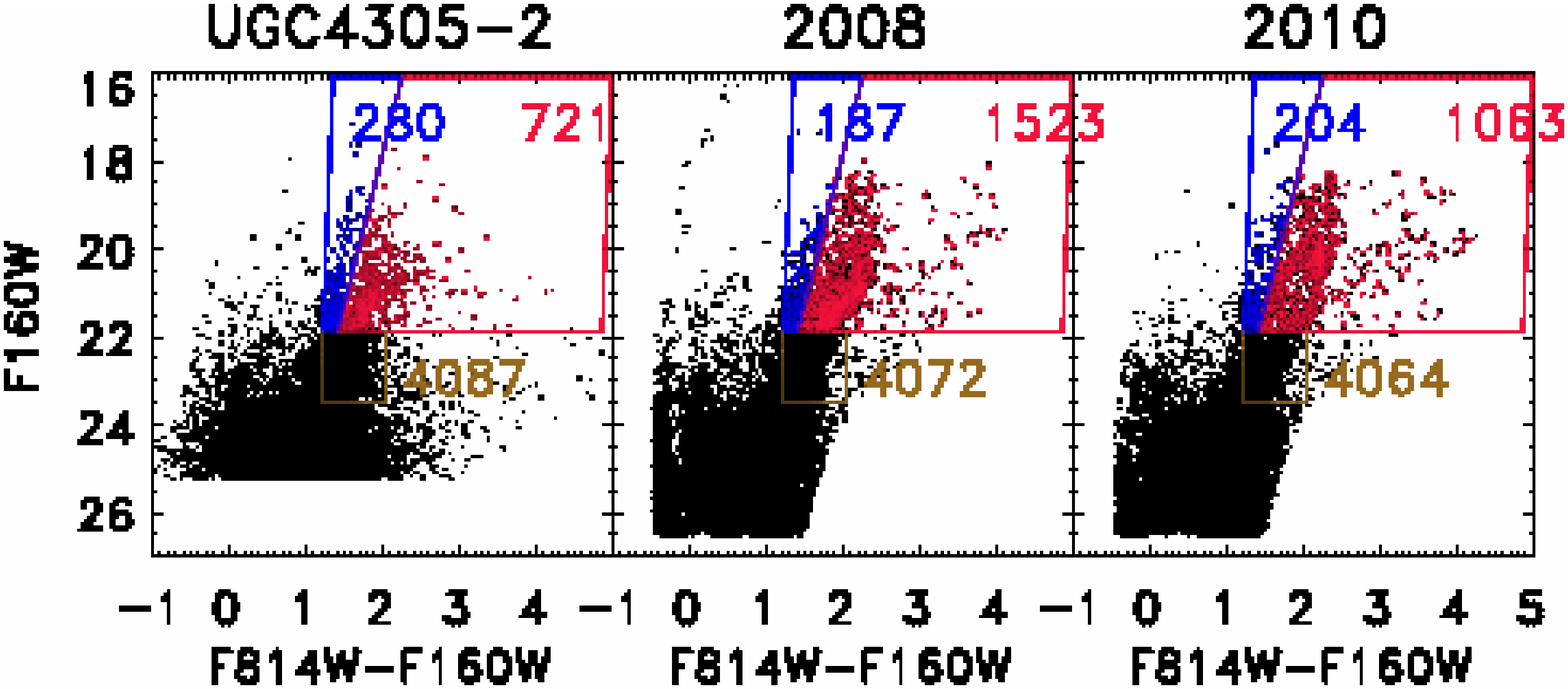}
\vspace{1mm}
\includegraphics[scale=0.35]{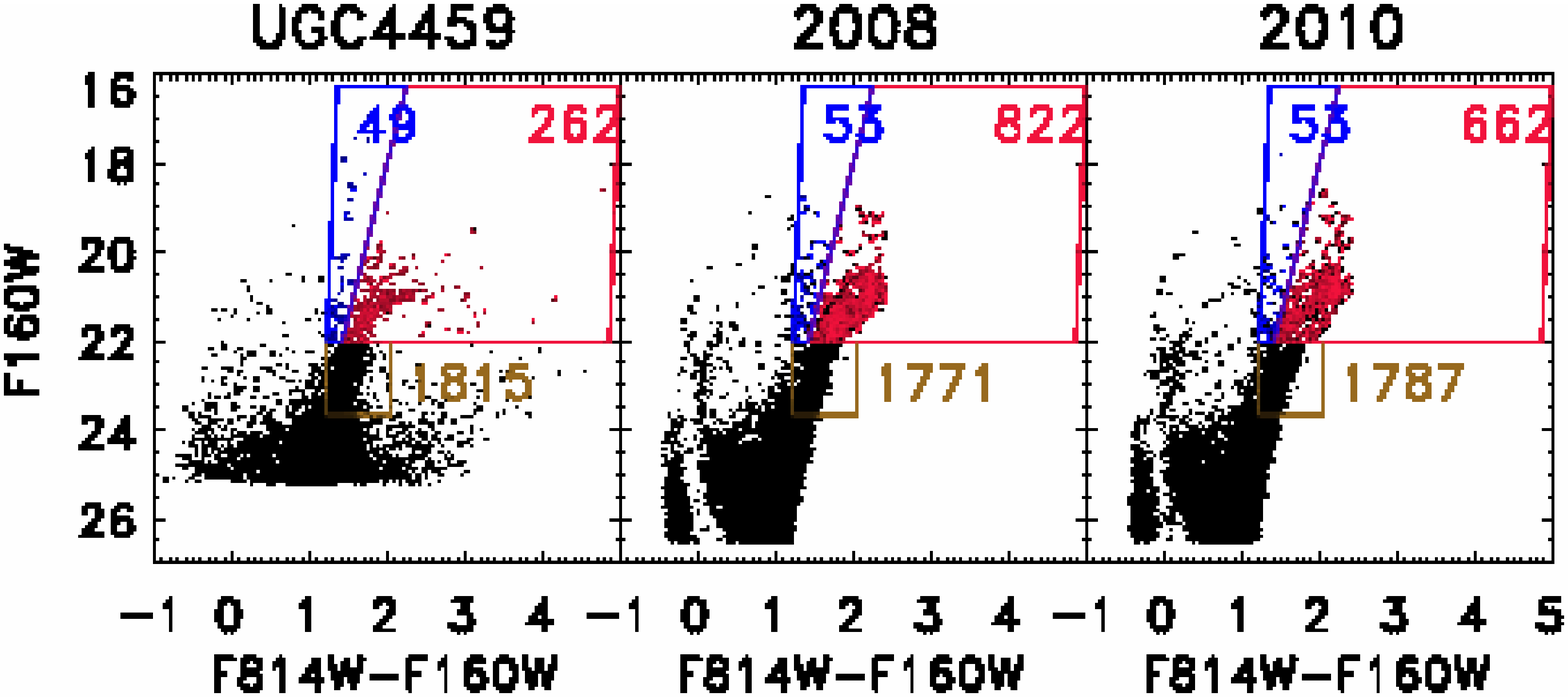}
\caption{continued }
\end{figure*}

To give an example of the type of calculation we will be making in the following sections, Figure \ref{fig:cum} shows the stellar (model + data) cumulative F160W flux fraction as a function of stellar magnitude for program galaxy UGC4305-1.   This plot is divided into three regions.  The left-most region shows the contribution of stars brighter than the TRGB.  The right-most region gives the contribution from faint MS stars.  The central region is dominated by RGB stars, but also contains some MS, fainter AGB, and fainter core helium burning stars.  This plot demonstrates one of the reasons why galaxy modelers have preferred to use NIR fluxes to constrain stellar masses.  The bulk of the light is from well-modeled RGB and MS stars.  However, even in this galaxy, which has a well-developed RGB, stars brighter than the TRGB contribute more than 30\% of the light.  At high-$z-$ where the RGB has had little time to develop, we expect that the NIR luminous RHeB and TP-AGB stars will contribute significantly larger fractions to the total.   

The total $F160W$ fluxes for each galaxy are given in Table \ref{tab:rgb}.  %In addition to determining total fluxes of the galaxies, we use the model CMDs to test stellar evolution models at the RHeB and TP-AGB phases.

\begin{figure*}
\figurenum{2}
\centering
\includegraphics[scale=0.35]{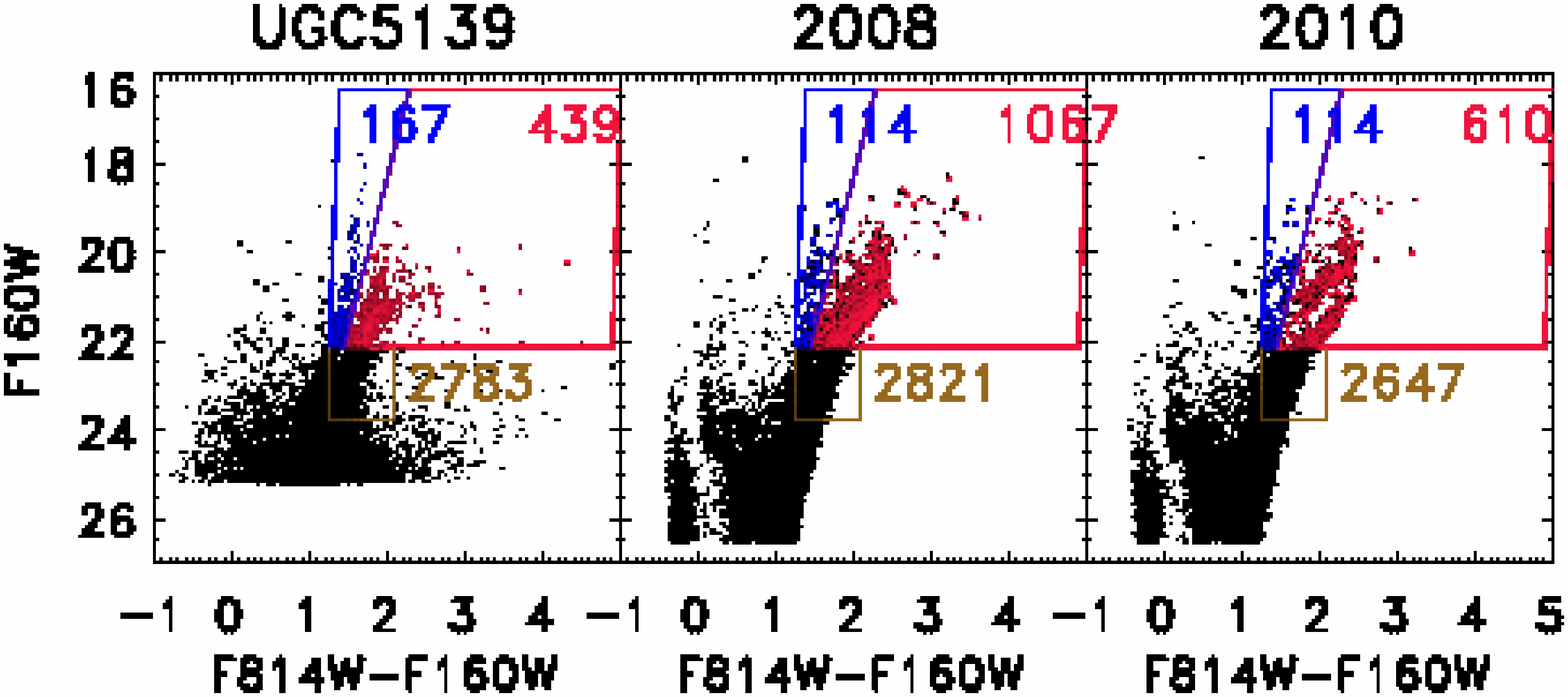}
\vspace{1mm}
\includegraphics[scale=0.35]{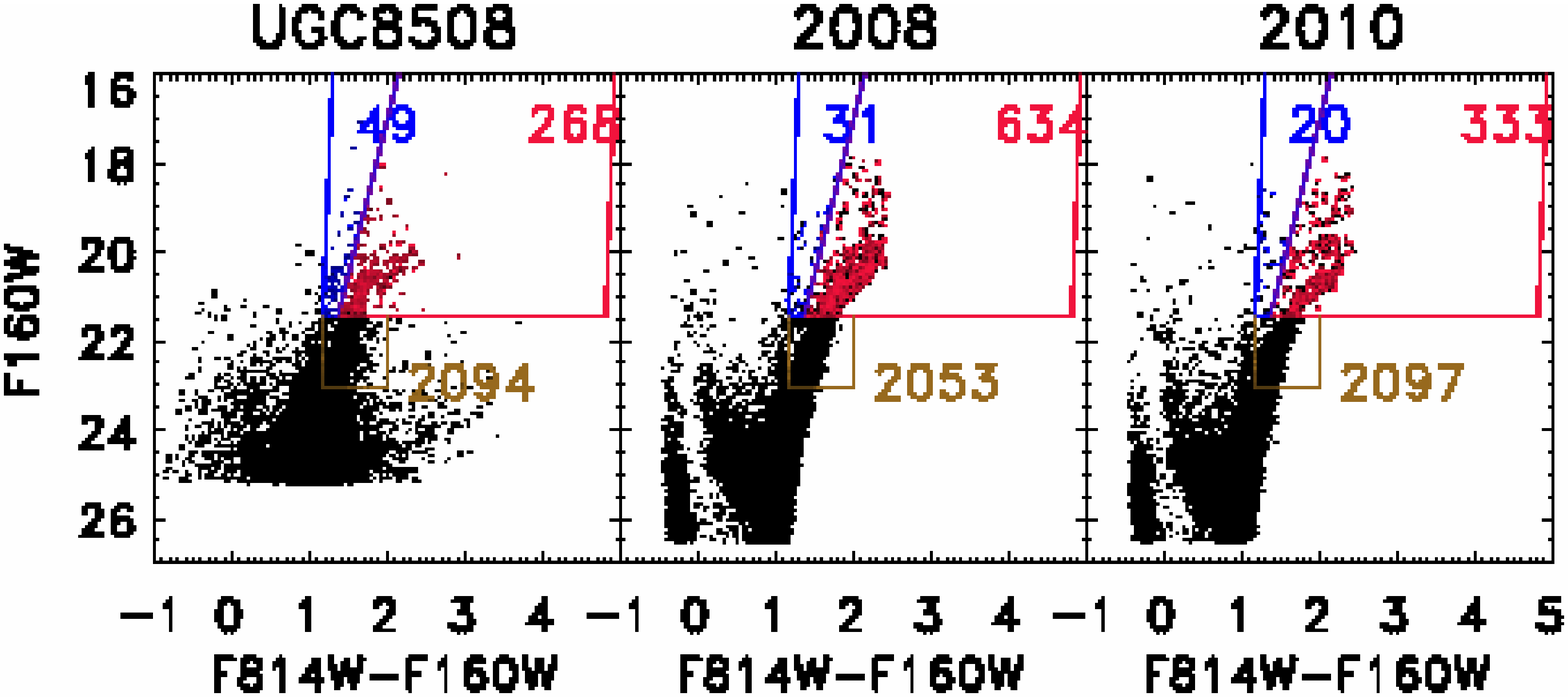}
\vspace{1mm}
\includegraphics[scale=0.35]{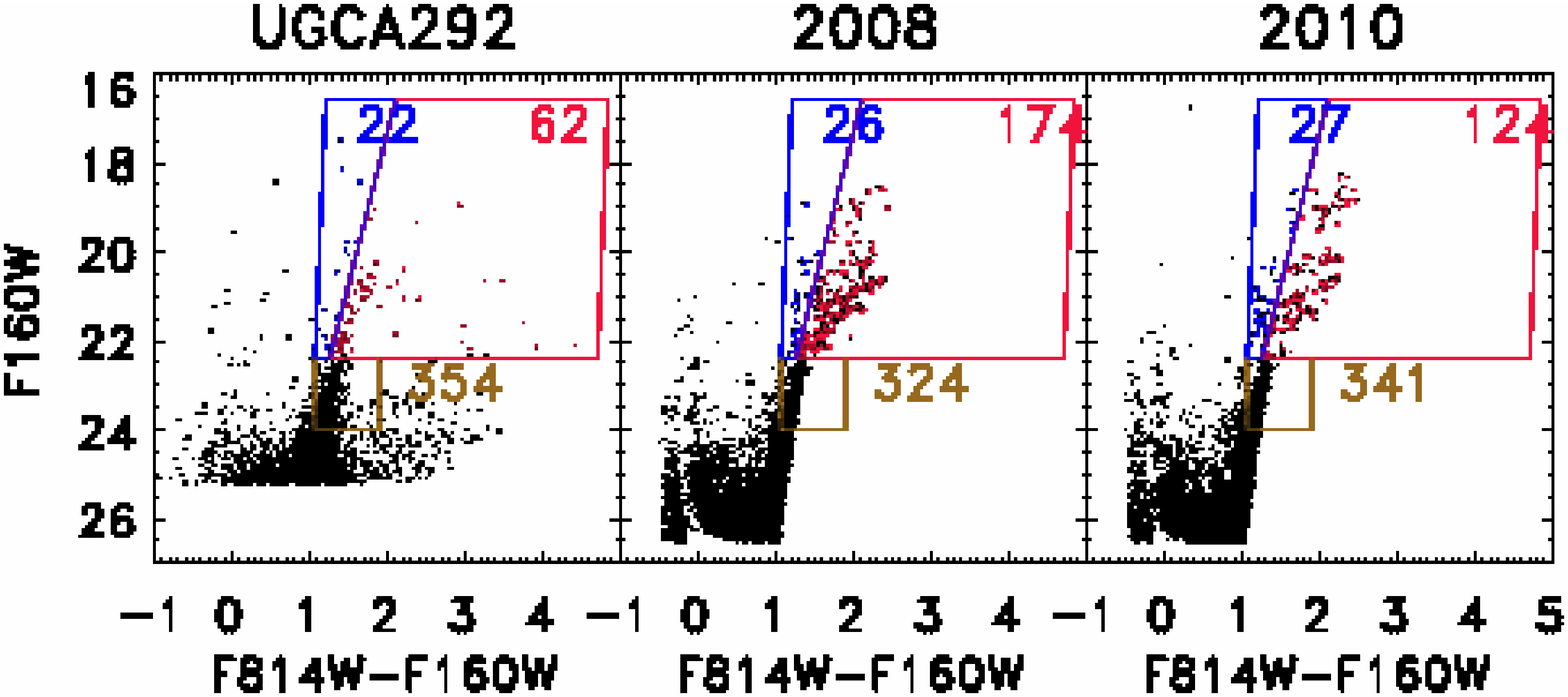}

\caption{continued  }
\end{figure*}

\subsection{Identifying Late Stage Stellar Evolution Sequences}

We identify stars on the TP-AGB and RHeB sequences by selecting them in color-magnitude space.  For the purposes of this paper, we are interested in those stars that might affect stellar mass estimates of high redshift galaxies, and thus focus on stars more luminous than the TRGB.    AGB stars fainter than these limits exist, but will not contribute significantly to the total light because their numbers will be dwarfed by the more numerous longer-lived RGB stars.   

Ideally we would use the IR observations to select a complete sample of the NIR luminous TP-AGB stars in each system.  The TP-AGB stars are more luminous in the IR bands compared with the optical data, and their colors are less affected by dust reddening than in the optical.  Thus a more complete census should be possible in the NIR compared to the optical, which may actually miss large numbers of TP-AGB stars \citep{Boyer09}.  Note: even the NIR can miss the most dust obscured AGB stars \citep{Boyer09}, but these stars will not affect the NIR luminosities of our galaxies because they are NIR faint.

Unfortunately, there is a problem with using the NIR CMDs to cleanly identify TP-AGB stars; the TP-AGB sequence above the TRGB has a similar IR color as the younger RHeB sequence thus making these two stellar classes hard to distinguish (Figure \ref{fig:AllCMD}).  
To cleanly identify a complete set of luminous TP-AGB stars, we therefore select them in the $(F814W - F160W)$ CMDs (Figure \ref{fig:CMD}), which have a much larger color separation between the TP-AGB and RHeB sequences.  

To select TP-AGB and RHeB stars, we create regions in the CMD space that isolate these sequences (Figure \ref{fig:CMD}).  To define the regions, we use a galaxy which has well populated TP-AGB and RHeB sequences, UGC~4305 (Figure \ref{fig:AllCMD}).  As discussed above, the faint-end limit is set by the TRGB, and the bright limit is set to exclude luminous foreground stars.  We shift these regions for each subsequent galaxy, applying a vertical shift to account for differences in distance and TRGB flux, and a horizontal shift to account for reddening variations. In addition to the RHeB and TP-AGB sequences, we also include a box that contains a large fraction of the upper RGB sequence. Tables \ref{tab:rgb} - \ref{tab:rheb} give the numbers of luminous RGB, AGB, and RHeB stars in each galaxy. 

We find that the $(F814W - F160W)$ CMDs contain roughly the same number of luminous stars (brighter than the TRGB) as the NIR only CMDs.  Thus, we are unlikely to be missing large numbers of TP-AGB stars in the final analysis, although rare highly reddened TP-AGB stars could be absent.

\subsection{Fraction of NIR Light Produced by TP-AGB Stars} 
Figure \ref{fig:agbfrac} shows the fraction of the $F160W$ light produced by the TP-AGB as a function of the fraction of young  stars in each galaxy (as estimated from the SFH routine CalcSFH).  The fractional flux contribution from TP-AGB stars in this sample varies from $\sim1$\% to $17$\%, with a trend of increasing contribution by the TP-AGB with an increasing fraction of young stars.  Uncertainties on the flux fractions are derived from the Poisson uncertainties of the numbers of TP-AGB stars and their typical fluxes. 

\begin{deluxetable*}{l|ccc|ccc}
\tabletypesize{\small}
\tablecaption{AGB Star Properties of Each Galaxy \label{tab:agb}}
\tablehead{\colhead{Galaxy} & \colhead{\# AGB} & \colhead{$\frac{\# model}{\# data}$} & \colhead{$\frac{\# model}{\# data}$} & \colhead{$f_{AGB}/f_{tot}$} & \colhead{$\frac{f_{model}}{f_{data}}$} & \colhead{$\frac{f_{model}}{f_{data}}$}\\
& \colhead{data} & \colhead{model 2008} & \colhead{model 2010}& \colhead{data} & \colhead{model 2008} & \colhead{model 2010}}
\startdata
          DDO71 &    146 $\pm$     12 &  3.62 $\pm$  1.94 &  1.53 $\pm$  0.82 &  0.08 $\pm$  0.01 &  4.48 $\pm$  2.99 &  1.96 $\pm$  1.31 \\
          DDO78 &    273 $\pm$     16 &  3.14 $\pm$  1.51 &  2.64 $\pm$  1.27 &  0.08 $\pm$  0.01 &  2.99 $\pm$  1.68 &  2.81 $\pm$  1.58 \\
          DDO82 &   1046 $\pm$     32 &  2.44 $\pm$  1.33 &  1.07 $\pm$  0.59 &  0.10 $\pm$  0.00 &  2.96 $\pm$  1.35 &  2.22 $\pm$  1.02 \\
     ESO540-030 &     69 $\pm$      8 &  5.86 $\pm$  2.12 &  4.14 $\pm$  1.50 &  0.07 $\pm$  0.01 &  8.51 $\pm$  4.33 &  6.70 $\pm$  3.41 \\
          HS117 &     63 $\pm$      7 &  2.57 $\pm$  1.09 &  2.05 $\pm$  0.87 &  0.06 $\pm$  0.01 &  3.20 $\pm$  1.44 &  2.85 $\pm$  1.28 \\
     IC2574-SGS &   1504 $\pm$     38 &  1.72 $\pm$  0.63 &  1.03 $\pm$  0.38 &  0.17 $\pm$  0.00 &  2.22 $\pm$  0.59 &  1.97 $\pm$  0.52 \\
          KDG73 &     60 $\pm$      7 &  1.82 $\pm$  0.69 &  1.25 $\pm$  0.47 &  0.13 $\pm$  0.02 &  2.13 $\pm$  1.21 &  1.49 $\pm$  0.85 \\
          KKH37 &    109 $\pm$     10 &  2.62 $\pm$  1.14 &  1.34 $\pm$  0.58 &  0.10 $\pm$  0.01 &  2.32 $\pm$  1.04 &  1.54 $\pm$  0.69 \\
       M81-DEEP &    157 $\pm$     12 &  4.79 $\pm$  0.73 &  4.94 $\pm$  0.75 &  0.08 $\pm$  0.01 &  5.89 $\pm$  0.93 &  7.43 $\pm$  1.17 \\
  NGC0300-WIDE1 &    465 $\pm$     21 &  1.99 $\pm$  0.51 &  1.80 $\pm$  0.46 &  0.15 $\pm$  0.00 &  2.61 $\pm$  0.58 &  2.75 $\pm$  0.61 \\
 NGC2403-HALO-6 &    182 $\pm$     13 &  1.69 $\pm$  0.71 &  1.37 $\pm$  0.58 &  0.17 $\pm$  0.01 &  1.74 $\pm$  0.70 &  1.80 $\pm$  0.72 \\
   NGC2976-DEEP &    293 $\pm$     17 &  2.76 $\pm$  2.01 &  1.27 $\pm$  0.92 &  0.10 $\pm$  0.01 &  3.00 $\pm$  2.49 &  1.43 $\pm$  1.19 \\
NGC3077-PHOENIX &    215 $\pm$     14 &  2.91 $\pm$  1.20 &  2.65 $\pm$  1.09 &  0.09 $\pm$  0.01 &  2.74 $\pm$  1.15 &  2.65 $\pm$  1.11 \\
        NGC3741 &    233 $\pm$     15 &  2.07 $\pm$  0.54 &  1.30 $\pm$  0.34 &  0.15 $\pm$  0.01 &  2.60 $\pm$  0.82 &  2.04 $\pm$  0.64 \\
        NGC4163 &    640 $\pm$     25 &  2.23 $\pm$  1.26 &  1.06 $\pm$  0.60 &  0.12 $\pm$  0.00 &  2.48 $\pm$  1.42 &  1.67 $\pm$  0.96 \\
 NGC7793-HALO-6 &    150 $\pm$     12 &  2.09 $\pm$  0.87 &  1.12 $\pm$  0.47 &  0.15 $\pm$  0.01 &  2.38 $\pm$  0.90 &  1.75 $\pm$  0.66 \\
        SCL-DE1 &     67 $\pm$      8 &  1.67 $\pm$  1.14 &  0.94 $\pm$  0.64 &  0.07 $\pm$  0.01 &  1.53 $\pm$  1.19 &  1.11 $\pm$  0.87 \\
      UGC4305-1 &    740 $\pm$     27 &  1.94 $\pm$  0.44 &  1.50 $\pm$  0.34 &  0.15 $\pm$  0.00 &  2.81 $\pm$  0.66 &  3.17 $\pm$  0.75 \\
      UGC4305-2 &    721 $\pm$     26 &  2.11 $\pm$  0.58 &  1.47 $\pm$  0.40 &  0.17 $\pm$  0.00 &  2.75 $\pm$  0.69 &  2.64 $\pm$  0.66 \\
        UGC4459 &    262 $\pm$     16 &  3.14 $\pm$  0.77 &  2.53 $\pm$  0.62 &  0.13 $\pm$  0.01 &  2.82 $\pm$  0.90 &  2.82 $\pm$  0.90 \\
        UGC5139 &    439 $\pm$     20 &  2.43 $\pm$  1.05 &  1.39 $\pm$  0.60 &  0.14 $\pm$  0.01 &  3.21 $\pm$  1.20 &  2.38 $\pm$  0.89 \\
        UGC8508 &    268 $\pm$     16 &  2.37 $\pm$  0.91 &  1.24 $\pm$  0.48 &  0.14 $\pm$  0.00 &  2.70 $\pm$  1.21 &  1.67 $\pm$  0.75 \\
        UGCA292 &     62 $\pm$      7 &  2.81 $\pm$  0.77 &  2.00 $\pm$  0.55 &  0.17 $\pm$  0.02 &  3.69 $\pm$  1.06 &  4.39 $\pm$  1.26 \\
\enddata
\end{deluxetable*}

Figure \ref{fig:agbfrac} considers recent star formation on two timescales  --- 2 Gyr (left), and 0.3 Gyr (right).  The shorter timescale tracks the lifetime of the most massive TP-AGB stars (e.g. $M > 3$ \msun), while the longer timescale tracks the more common but less massive (e.g. $M= 2-3$ \msun) TP-AGB stars. Interestingly the scatter in the plot is reduced for the shorter star formation timescale. The reduced scatter in this second version of the plot may be indicating that the trend is driven by the most massive TP-AGB stars.  Alternatively it may just be showing that the SFHs are better constrained for the youngest ages.  Larger samples with better constrained SFHs at older ages could be useful for explaining this interesting result.
  
 \begin{figure}
\centering
\includegraphics[scale=0.55]{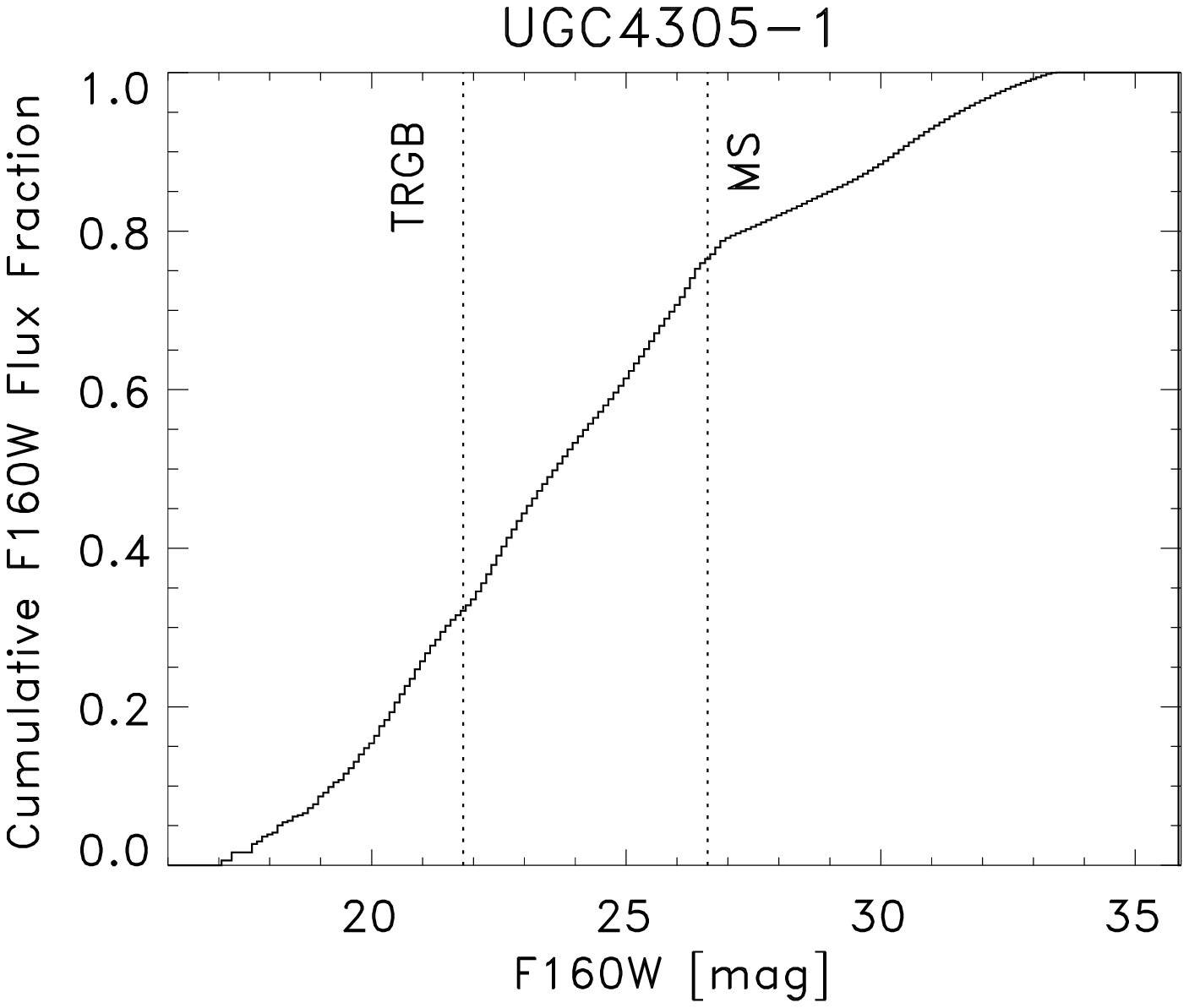}
\caption{\label{fig:cum} The cumulative F160W flux fraction as a function of stellar magnitude for program galaxy UGC4305-1.  The contribution from stars brighter than the TRGB is given to the left of the first vertical line and includes light from TP-AGB and RHeB stars.  The contribution from the faint MS is given to the right of the second vertical line.  The central region is dominated by RGB stars but also contains contributions from brighter MS, fainter AGB and fainter core helium burning stars. The bulk of the light is from well modeled MS and RGB stars, however, even in this galaxy more than 30\% of the F160W light is from stars brighter than the TRGB, namely TP-AGB and RHeB stars.
}
\end{figure}

While there is a strong trend with fractional age, there does not appear to be an equivalent trend with metallicity. The metallicity shown in this plot is the mean expected metallicity for stars that are 1 Gyr old (as estimated by CalcSFH).   Low and high metallicity systems both appear to be following the same general trends of increasing TP-AGB contribution with increasing fraction of young stars.   
  
There is evidence that SFH uncertainties are contributing to the scatter in the left hand version of this plot.  Two galaxies, ESO540-030 and HS117, in particular appear discrepant in the left-hand panel of Figure \ref{fig:agbfrac}, showing less TP-AGB light than their SFH might imply.  However, the SFH is highly uncertain for ESO540-030.  For instance the best fit SFH predicts a moderately high metallicity ([m/H] $> -0.8$) for the youngest populations of this low-mass galaxy.  This is likely an over-estimate, as the metallicity is significantly lower for the bulk of cosmic time.  While other studies have also suggested higher metallicities for the most recent stars \citep{Jerjen01}, the values they derive are still [m/H] $< -1$. In the case of HS117 a handful of extremely bright RHeB stars may also be lowering the AGB contribution to the total.  As we discuss in Section 3.6, these are likely foreground stars, further complicating the interpretation of these results. However, when we replot this Figure, now against the fraction of stars younger than 0.3 Gyrs (right hand panel of Figure \ref{fig:agbfrac}), ESO540-030 and HS117 are no longer significantly deviant.   

\begin{figure*}
\centering
\includegraphics[scale=0.6]{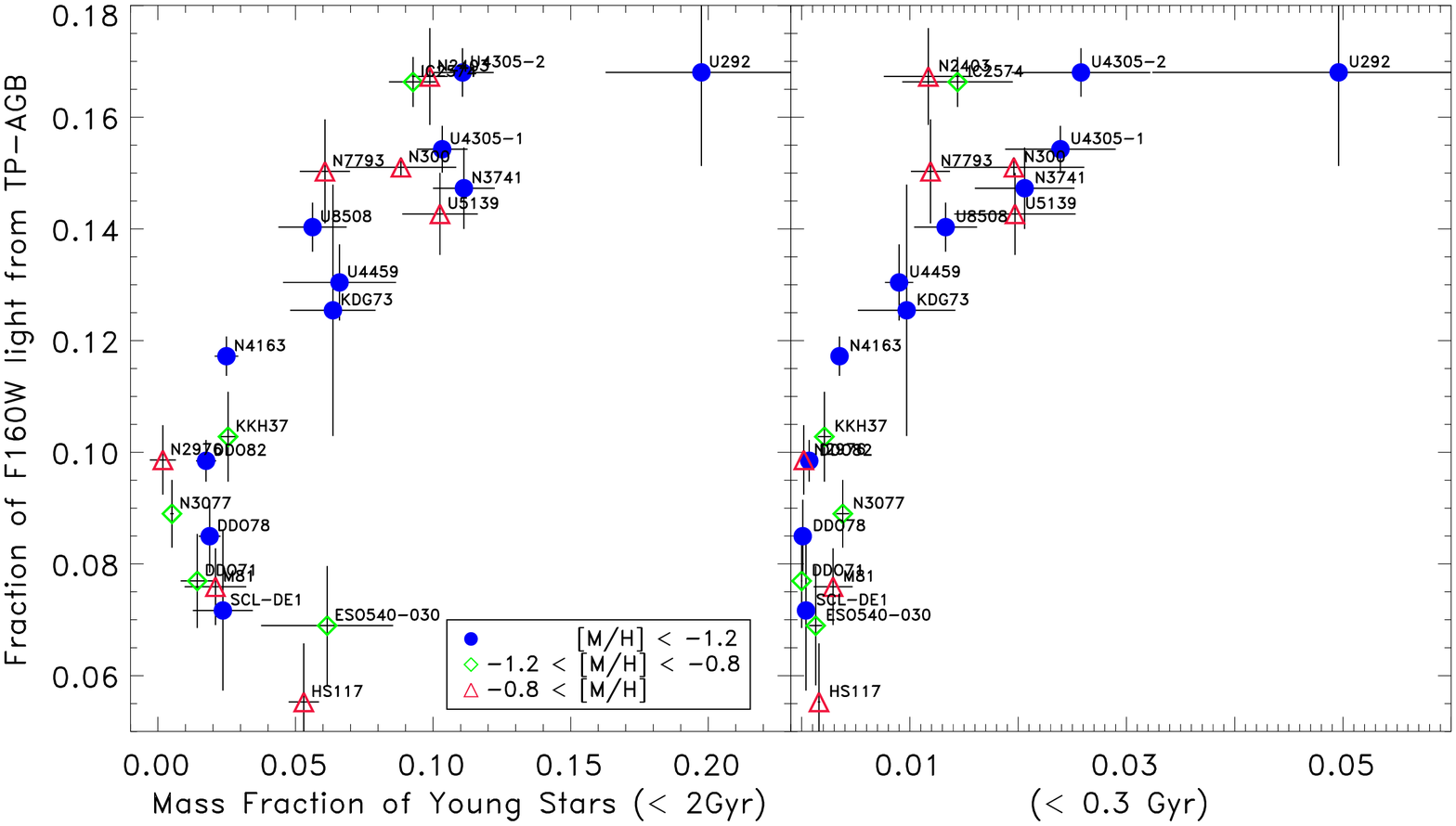}
\caption{\label{fig:agbfrac} The fraction of $F160W$ flux produced by TP-AGB stars as a function of the young stellar populations in each galaxy.  Mass fractions of stars less than 2 Gyrs (typical age range for TP-AGB brighter than TRGB) is shown on the left, while mass fractions of stars less than 0.3 Gyrs (age range for the most massive and luminous TP-AGB stars) is shown on the right. Galaxies with a higher fraction of young stars tend to show a larger contribution from TP-AGB stars then more evolved galaxies.  The TP-AGB contribution reaches as high as  17\% even in this sample which contain a well developed red giant branch.  In high-$z$ galaxies where the RGB has not developed, the TP-AGB contribution is likely to be significantly higher. We examine the influence of metallicity on the trend, using the expected metallicity of the 1 Gyr old population from the measured SFHs, and find no obvious trend with metallicity.  Galaxies ESO540-030 and HS117 are somewhat deviant in the first version of this plot but not in the second, demonstrating the difficulties in interpreting the effects of rare but luminous stars on a stellar population. }
\end{figure*}

Another galaxy, UGCA292, is also somewhat anomalous in Figure \ref{fig:agbfrac}.  UGCA929 is predicted to have the largest fraction of young stars, but its TP-AGB population does not account for a correspondingly large flux fraction compared with the other galaxies. However, UGCA292 is the least-populated galaxy in the sample (see Fig. 2), which causes not only the increased uncertainties depicted in the figure, but also potential difficulty in the derivation of the SFH.%{\bf NOTE: THESE GALAXIES INDICATE THE NEED OF EVALUATING ERRORS DUE TO THE UNCERTAIN SFH. THERE MIGHT BE NO HUGE DISCREPANCY AFTER ALL.}

The observed TP-AGB flux fractions are summarized in Table \ref{tab:agb}.  Poisson uncertainties are also quoted.

\subsection{ Fraction of NIR Light Produced by RHeB Stars}
Figure~\ref{fig:RHeBfrac} shows the contribution of RHeB stars to the 1.6 \um\ fluxes of galaxies as a function of the fraction of young stars.  Here we only plot the smaller age range, using the mass fraction of stars younger than 0.3 Gyrs.  This represents the timescale (or progenitor mass range $> 3.5$ \msun) over which RHeBs contribute significant fractions of the IR luminosity of galaxies. While less massive stars will go through a core helium burning phase, they will not reach luminosities brighter than the TRGB, and will instead populate the horizontal branch or red clump of the CMD. 

Again we see a similar trend where galaxies with a higher fraction of ongoing or recent star formation tend to show a larger contribution from the RHeB phases of stellar evolution.  Interestingly, the contribution of RHeB stars can match or even exceed the contribution from the TP-AGB phase of stellar evolution, reaching  as high as 21\% of the total in NGC~2403.  While there is not a strong trend in the RHeB flux fraction with metallicity, there may be some favoring of lower flux fractions for galaxies with low metallicity.  Such a trend could indicate that RHeBs are rarer in low metallicity systems. Comparing the red to blue helium burning fractions as a function of galaxy metallicity could shed more light on this issue \citep{McQuinn11}.  However, significantly more galaxies, especially galaxies with larger fractions of young stellar populations, should be used to determine if this trend indeed exists.   

As with the TP-AGB stars HS117 is somewhat deviant in this plot (although now in the opposite direction as in the TP-AGB plot.  In this case, a handful (4 stars) of extremely bright stars are pushing the RHeB flux fraction higher than 10\% even though the SFH suggests little star formation at these young ages.   With these small numbers it is hard to draw significant conclusions, as foreground stars could be important.  In the following section we will attempt to statistically account for any foreground stars.

The other galaxy that might be considered somewhat deviant in this plot is UGC~292.  However, once again, the large uncertainty in the SFH at young ages mean that this galaxy could actually belong closer to the main locus of points.

These results are also summarized in Table \ref{tab:rheb}.

\subsection{Foreground Stars}
With any CMD studies, foreground (or background) contamination can pose a problem.  There may be Milky Way stars that have similar colors and luminosities as the stars in the program galaxies.  This effect could tend to artificially increase the numbers of stars on the RHeBs or TP-AGB sequences.  We actually benefit from the fact that our fields are spatially very small, only 4.7 square arcmin. This area is  much smaller than is needed to image a typical Local Group dwarf galaxy ($> 200$ square arcmin) and thus has much lower field contamination.  However, contamination may still be important for the RHeB region, where the observed numbers can be small. %The most luminous RHeBs in the sample have NIR colors similar to the TRGB.  Thus the two types of contaminants that we would be most susceptible  to are  foreground TRGB stars or main sequence M dwarfs.  For a Milky Way TRGB star ($M_H \sim -2$ to $-4$) to be as faint as the brightest RHeB in our sample it would need to be at a distance $>100$ Kpc.  Likewise for a M dwarf ($M_H \sim 5.5$ to 7.5) to be as bright as a RHeB in our sample, it would need to be within 300 pc of the earth.  These are fairly tight bounds and suggest that contamination is likely to be small for our galaxies.  

\begin{figure*}
\includegraphics[scale=0.6]{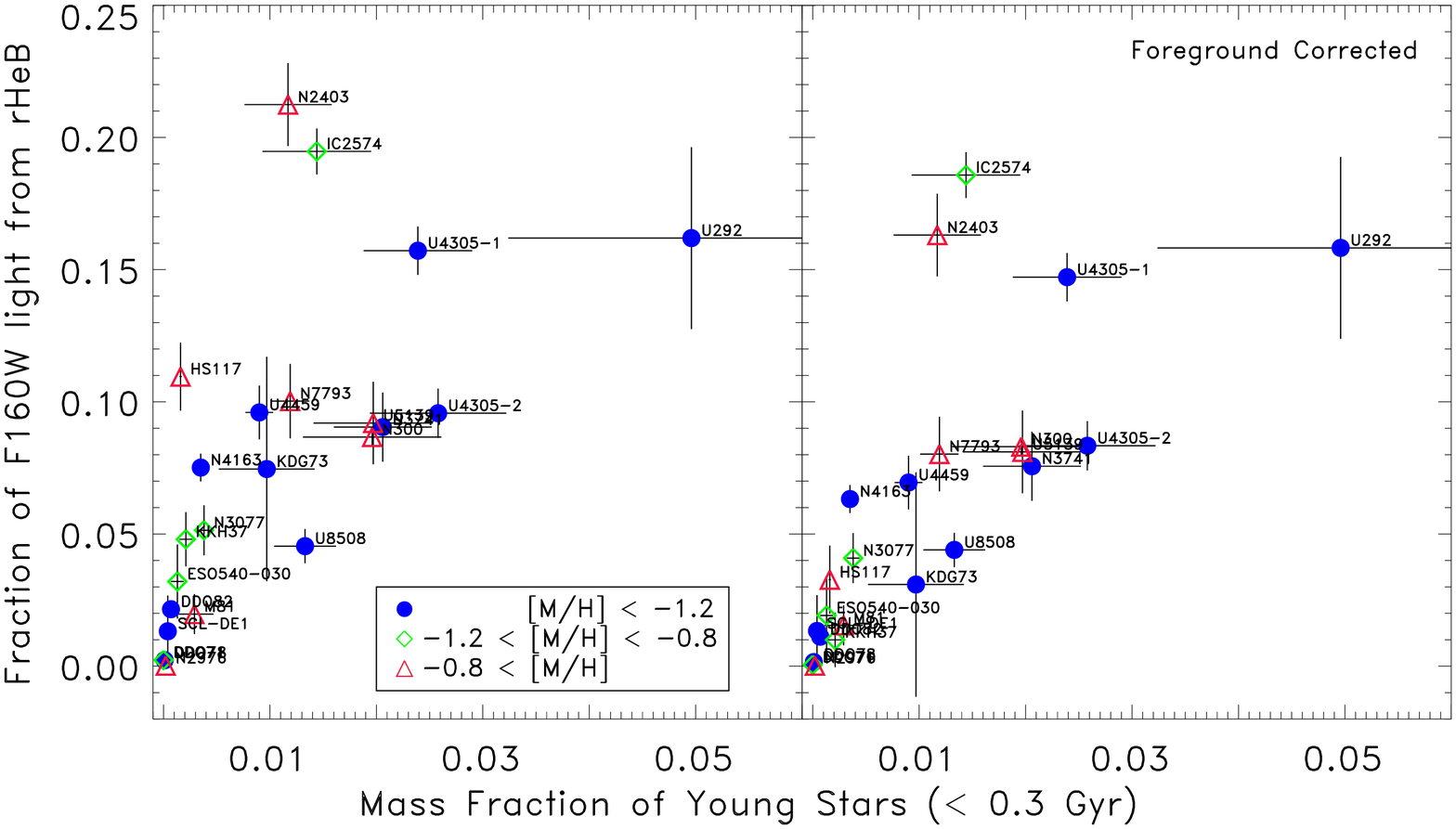}
\caption{\label{fig:RHeBfrac} Same as Figure \ref{fig:agbfrac} only now for the RHeB stars. Here we only plot against the shorter timescale of 0.3 Gyrs, as this is the timescale for RHeBs. The left panel shows the raw results, while the right applies a statistical correction for foreground contamination as estimated by TRILEGAL (see Table \ref{tab:foreground}).   Again there is a correlation, RHeBs contribute a larger fraction of the $F160W$ flux in galaxies with young stellar populations.  The flux contributions from RHeBs can be as large as those from TP-AGB stars and reach as high as 21\% of the total, or 18\% after a foreground correction.  As with the TP-AGB there is no strong trend in RHeB flux fractions with metallicity although there may be a slight favoring of larger RHeB flux fractions for more metal rich systems. }
\end{figure*}

\begin{deluxetable*}{l|ccc|ccc}[ht]
\tabletypesize{\small}
\tablecaption{RHeB Star Properties of Each Galaxy \label{tab:rheb}}
\tablehead{\colhead{Galaxy} & \colhead{\# RHeB} & \colhead{$\frac{\# model}{\# data}$} & \colhead{$\frac{\# model}{\# data}$} & \colhead{$f_{RHeB}/f_{tot}$} & \colhead{$\frac{f_{model}}{f_{data}}$} & \colhead{$\frac{f_{model}}{f_{data}}$}\\
& \colhead{data} & \colhead{model 2008} & \colhead{model 2010}& \colhead{data} & \colhead{model 2008} & \colhead{model 2010}}
\startdata
          DDO71 &      2 $\pm$      1 &  0.50 $\pm$  1.87 &  0.50 $\pm$  1.87 &  0.00 $\pm$  0.00 &  0.00 $\pm$  0.00 &  0.43 $\pm$  0.86 \\
          DDO78 &      5 $\pm$      2 &  0.20 $\pm$  0.81 &  0.20 $\pm$  0.81 &  0.00 $\pm$  0.00 &  0.00 $\pm$  0.00 &  0.44 $\pm$  0.88 \\
          DDO82 &     39 $\pm$      6 &  1.51 $\pm$  2.74 &  0.49 $\pm$  0.88 &  0.02 $\pm$  0.00 &  0.22 $\pm$  0.33 &  0.23 $\pm$  0.19 \\
     ESO540-030 &     10 $\pm$      3 &  0.10 $\pm$  0.09 &  0.10 $\pm$  0.09 &  0.03 $\pm$  0.01 &  0.00 $\pm$  0.00 &  0.06 $\pm$  0.12 \\
          HS117 &      8 $\pm$      2 &  0.12 $\pm$  0.12 &  0.25 $\pm$  0.23 &  0.11 $\pm$  0.01 &  0.01 $\pm$  0.00 &  0.02 $\pm$  0.03 \\
     IC2574-SGS &    479 $\pm$     21 &  0.46 $\pm$  0.13 &  0.43 $\pm$  0.13 &  0.19 $\pm$  0.01 &  0.30 $\pm$  0.08 &  0.23 $\pm$  0.08 \\
          KDG73 &     18 $\pm$      4 &  0.17 $\pm$  0.07 &  0.44 $\pm$  0.19 &  0.07 $\pm$  0.04 &  0.13 $\pm$  0.14 &  0.36 $\pm$  0.29 \\
          KKH37 &     15 $\pm$      3 &  0.40 $\pm$  0.24 &  0.27 $\pm$  0.16 &  0.05 $\pm$  0.01 &  0.14 $\pm$  0.21 &  0.12 $\pm$  0.10 \\
       M81-DEEP &     16 $\pm$      4 &  0.38 $\pm$  0.90 &  0.06 $\pm$  0.15 &  0.02 $\pm$  0.01 &  0.33 $\pm$  0.66 &  0.30 $\pm$  0.30 \\
  NGC0300-WIDE1 &    117 $\pm$     10 &  0.44 $\pm$  0.13 &  0.38 $\pm$  0.12 &  0.09 $\pm$  0.00 &  0.33 $\pm$  0.24 &  0.43 $\pm$  0.25 \\
 NGC2403-HALO-6 &     50 $\pm$      7 &  0.34 $\pm$  0.17 &  0.42 $\pm$  0.21 &  0.21 $\pm$  0.02 &  0.24 $\pm$  0.15 &  0.20 $\pm$  0.12 \\
   NGC2976-DEEP &      1 $\pm$      1 &  8.00 $\pm$  4.79 & 20.00 $\pm$ 11.96 &  0.00 $\pm$  0.00 & 21.88 $\pm$ 11.54 & 19.80 $\pm$ 14.85 \\
NGC3077-PHOENIX &     44 $\pm$      6 &  0.50 $\pm$  0.17 &  0.59 $\pm$  0.20 &  0.05 $\pm$  0.01 &  0.65 $\pm$  0.26 &  0.72 $\pm$  0.29 \\
        NGC3741 &     63 $\pm$      7 &  0.71 $\pm$  0.18 &  0.68 $\pm$  0.18 &  0.09 $\pm$  0.01 &  0.46 $\pm$  0.14 &  0.41 $\pm$  0.15 \\
        NGC4163 &    120 $\pm$     10 &  0.27 $\pm$  0.12 &  0.41 $\pm$  0.18 &  0.08 $\pm$  0.01 &  0.22 $\pm$  0.12 &  0.23 $\pm$  0.09 \\
 NGC7793-HALO-6 &     29 $\pm$      5 &  0.52 $\pm$  0.17 &  0.62 $\pm$  0.21 &  0.10 $\pm$  0.01 &  0.14 $\pm$  0.19 &  0.10 $\pm$  0.20 \\
        SCL-DE1 &      5 $\pm$      2 &  0.20 $\pm$  0.19 &  0.20 $\pm$  0.19 &  0.01 $\pm$  0.01 &  0.00 $\pm$  0.00 &  0.08 $\pm$  0.08 \\
      UGC4305-1 &    296 $\pm$     17 &  0.68 $\pm$  0.15 &  0.64 $\pm$  0.14 &  0.16 $\pm$  0.01 &  0.45 $\pm$  0.10 &  0.41 $\pm$  0.12 \\
      UGC4305-2 &    280 $\pm$     16 &  0.67 $\pm$  0.14 &  0.73 $\pm$  0.16 &  0.10 $\pm$  0.01 &  0.66 $\pm$  0.15 &  0.56 $\pm$  0.17 \\
        UGC4459 &     49 $\pm$      7 &  1.08 $\pm$  0.33 &  1.08 $\pm$  0.33 &  0.10 $\pm$  0.01 &  0.42 $\pm$  0.18 &  0.48 $\pm$  0.21 \\
        UGC5139 &    167 $\pm$     12 &  0.68 $\pm$  0.21 &  0.68 $\pm$  0.21 &  0.09 $\pm$  0.02 &  0.71 $\pm$  0.32 &  0.82 $\pm$  0.28 \\
        UGC8508 &     49 $\pm$      7 &  0.63 $\pm$  0.37 &  0.41 $\pm$  0.24 &  0.05 $\pm$  0.01 &  0.59 $\pm$  0.40 &  0.75 $\pm$  0.26 \\
        UGCA292 &     22 $\pm$      4 &  1.18 $\pm$  0.40 &  1.23 $\pm$  0.41 &  0.16 $\pm$  0.03 &  0.47 $\pm$  0.17 &  0.59 $\pm$  0.22 \\
\enddata
\end{deluxetable*}

To statistically estimate the foreground contamination, we run TRILEGAL \citep{Girardi05} which models the Milky Way contamination for a given field size, in a given input direction.  We run TRILEGAL with the canonical settings including a thin disk component, a bulge component, and a halo component.   We only consider model foreground stars that are brighter than $F160W=23$ mag, roughly the magnitude cutoff for real data in our calculation of the total fluxes of the program galaxies.  For every star in the foreground model, we determine if there is a real star within 0.5 mags in CMD-space.  If there is, we flag the closest one as a potential foreground star. Thus we only account for plausible foreground stars. For instance a model foreground star that is 0.5 mags brighter than the brightest actual star will be assigned the flux of the actual star.  Likewise, model stars that are brighter still, will not be considered at all.

Table \ref{tab:foreground} presents statistical estimates for the numbers and fluxes of foreground stars in the direction of each galaxy in our sample.  The foreground stars have been classified by the region of the CMD in which they are found, i.e. RHeB and TP-AGB. An estimate of the total flux from all plausible foreground stars is also provided in the table, including stars that are not in the RHeB and TP-AGB regions. 

The TRILEGAL models typically predict fewer than 5 plausible foreground stars in the TP-AGB region of the CMD.  Although for KKH~37 there may be twice that number.  Foreground stars account for fewer than 5\% of the AGB flux in these galaxies, except for KKH~37 where the foreground may be contributing as much as  20\%.  Because the foreground contribution is smaller than the Poisson uncertainties on the numbers of TP-AGB stars, we will not make any special attempt to account for it in our additional analysis of the TP-AGB.

Typically there are on the order of 5 foreground stars in the RHeB region, with as many as 10 in NGC~2403.  As half of the program galaxies contain fewer than 30 stars in the RHeB region, foreground could potentially be responsible for a large fraction of the measured RHeB flux.  For instance, HS117, which is thought to have little recent star formation, is found to have 8 RHeB stars 4 of which are very luminous and make up the over 90\% of the flux.  However, the TRILEGAL model predicts that at least three of those luminous stars are foreground.  The one remaining RHeB star could also well be foreground given the small number statistics.  Removing the foreground stars from the RHeB sequences actually improves the correlation between the flux contribution of RHeBs and the star formation histories of the galaxies as shown in the right hand panel of Figure \ref{fig:RHeBfrac}.  After foreground correction HS117 is no longer a deviant point in the plot. Because of the uncertainties introduced by foreground stars on the RHeB sequences, we will only consider galaxies with more than 30 RHeBs for the remainder of the RHeB analysis.      

%Contamination in the AGB and RHeB regions of the CMDs for our sample is negligible.  For instance in the direction of UGC4305 no Milky Way foreground stars are predicted at the colors and luminosities of the regions of interest.  

%In addition to this statistical test, if significant flux were being contributed by foreground stars we would not expect to have a clear trend of increasing RHeB contribution with the measured SFH.  Rather we would expect the foreground stars to wash out any such trends.  As a result, it is likely that most of the observed RHeBs are real.  In addition, we only consider those galaxies with significant numbers of RHeBs ($>30$ stars) for our conclusions (although there is little difference in considering the complete sample), again reducing the influence of contamination on our results.

\subsection{Comparison with Models}
In addition to quantifying the flux contributions of TP-AGB and RHeB stars to the NIR luminosities of galaxies, we check if these results can be reproduced by SPS codes based on the 2008 and 2010 Padova isochrones.  For each galaxy, we use the best-fit optically derived SFH to model the stellar content in the NIR. We then compare the numbers and fluxes of the synthetic TP-AGB and RHeBs to the real data.  

\begin{figure}
\includegraphics[scale=0.5]{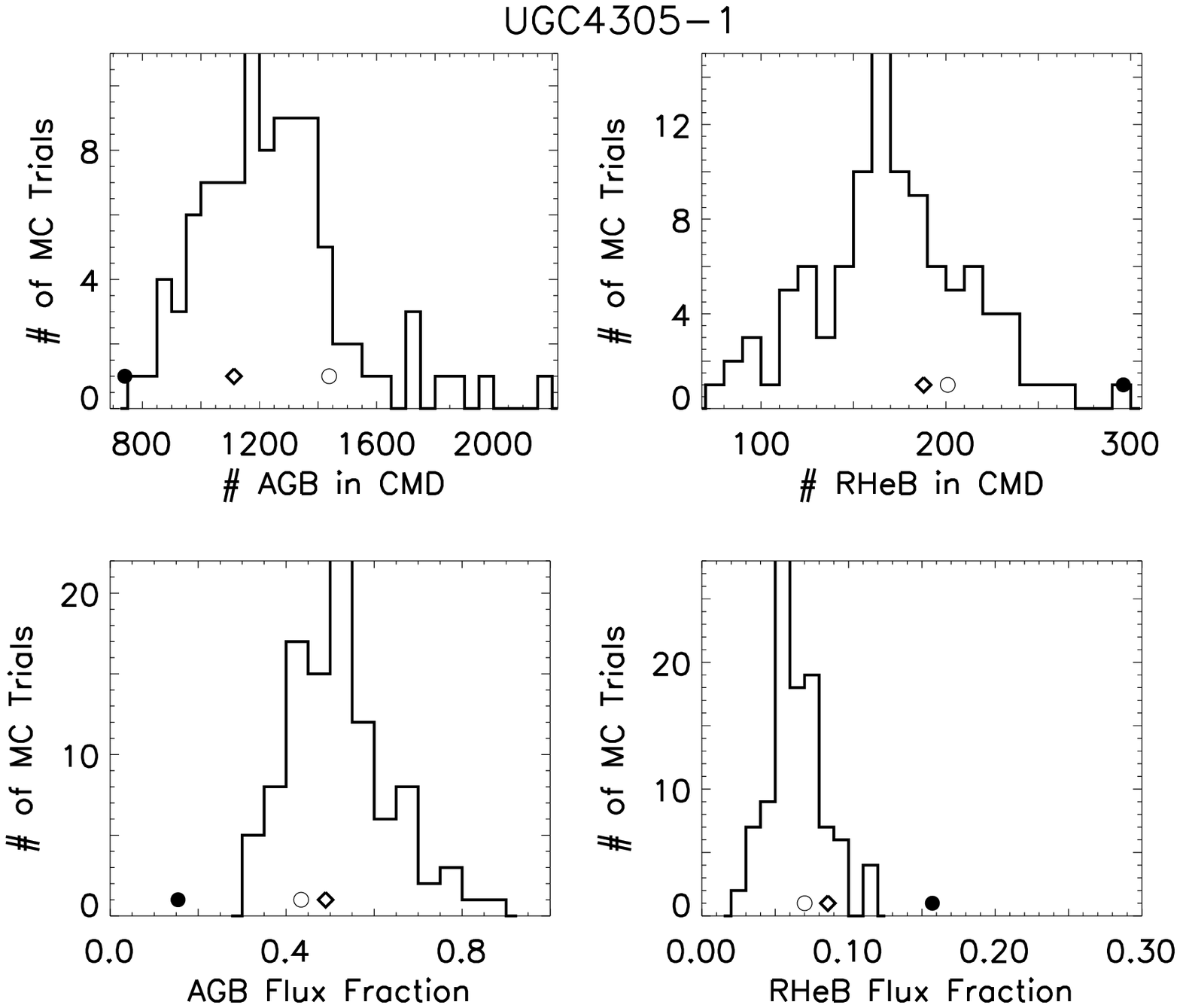}
\caption{\label{fig:modeldatacomp} A comparison of the observed numbers and fluxes of TP-AGB and RHeB stars to model predictions for galaxy UGC4305-1.  The observed data are shown as filled circles.  The SPS model results based on the 2008 (open circles) and 2010 (diamonds) Padova isochrones are also shown.  In addition, we plot the results from 100 MC simulations of the 2010 model (histograms) that span the range of acceptable SFH as determined by CALCSFH.  While the observations are generally offset from the simulation results, the scatter in the simulations is typically large.  We take the standard deviations  of the 100 MC simulations as a measure of the uncertainty in the model results.
}
\end{figure}

To determine the uncertainties on the numbers and fluxes of synthetic TP-AGB and RHeB stars, we create 100 additional model CMDs for each galaxy.  Each of these 100 models is created with a different SFH chosen to span the range of acceptable SFHs as determined by CalcSFH. The adopted uncertainty of the model measurement is then given by the standard deviation of the 100 Monte Carlo results.

Figure~\ref{fig:modeldatacomp} shows an example model/data comparison for UGC~4305-1.  The four panels plot the observed numbers (top) and flux contributions (bottom) of TP-AGB and RHeB stars for actual data (filled circles), and the best-fit SPS models from the 2008 (open circles) and 2010 (diamonds) Padova isochrones.  Also shown are the results for 100 MC simulations based on the 2010 models (histograms) that span the full range of acceptable SFHs.  For UGC4305-1 the SPS models tend to over-predict the observed numbers and fluxes of the TP-AGB stars and under-predict the observed numbers and flux contributions of RHeB stars, although the uncertainties on the model results (i.e. the widths of the histograms) are large.  %The full model/data comparisons are shown in Figures \ref{fig:agbfracfake} and \ref{fig:RHeBfracfake} and discussed below.
  
%\subsubsection{Numbers of TP-AGB Stars}

%{\bf THIS IS A NEW SECTION, ITS GOAL IS TO EVINCE THAT THE TP-AGB NUMBERS ARE BEING WELL PREDICTED BY 2010 MODELS:

\begin{figure*}
\centering
\includegraphics[scale=0.7]{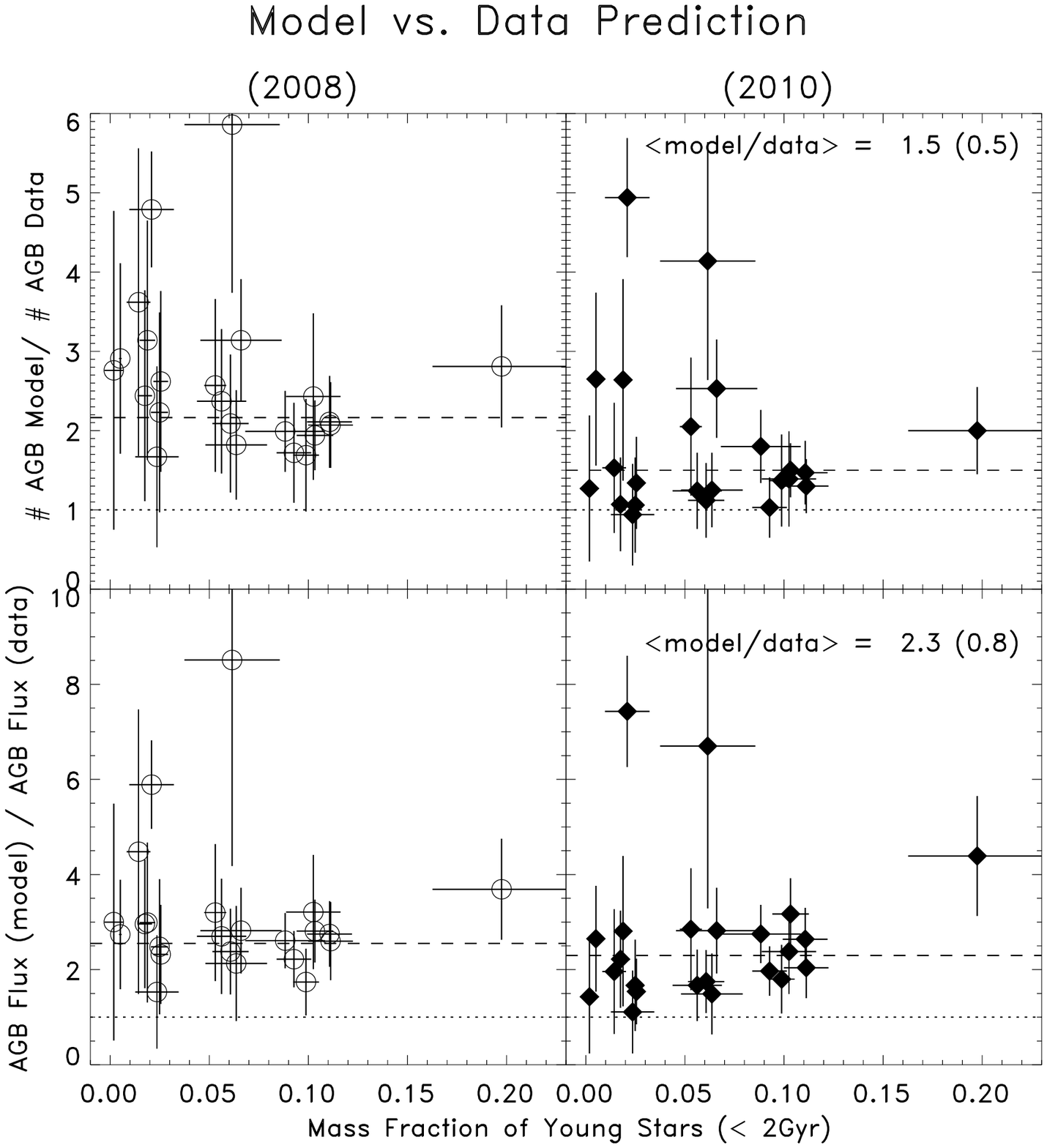}
\caption{\label{fig:agbfracfake} A model/data comparison of the numbers and $F160W$ fluxes of the TP-AGB stars in our sample galaxies.  SPS model results based on the optically derived best-fit SFH and the 2008 Padova isochrones are shown as circles (left).  SPS models based on the 2010 Padova isochrones are shown as diamonds (right).  Uncertainties on the model/data ratios based on 100 MC simulations that span the range of acceptable input SFHs are shown.  While the 2008 models tend to over-predict the numbers of TP-AGB stars, especially for galaxies with little on-going star formation, the 2010 version of the models reproduce the data reasonably well;  65\% of the models are equivalent to the data to within the measured uncertainties (see dotted 1/1 line).  The weighted mean model/data number ratio (dashed line) is 1.5 (for the 2010 model) with a standard deviation of 0.5.  While the updated best-fit SPS models do a reasonable job with the numbers of stars, they tend to over-predict the fluxes of the TP-AGB stars by a larger weighted mean factor of 2.3 with a standard deviation of 0.8, similar to the results from the 2008 model. }
\end{figure*}

Figure \ref{fig:agbfracfake} compares the predicted-to-observed numbers of TP-AGB stars for the full sample of galaxies.  Results from the best-fit SPS models based on both the 2008 (circles) and 2010 (diamonds) Padova isochrones are shown. It is clear that 2008 models tend to overestimate the numbers of TP-AGB stars, presenting excesses in the numbers of TP-AGB stars by factors of $1.7 - 6$. The situation is largely remedied in the 2010 models: indeed, for 65~\% of the galaxies the ratio between modeled and observed numbers is consistent with unity to within the measured uncertainties. However, there remains a small systematic bias to larger numbers in the 2010 models by a weighted mean factor of 1.5 with a standard deviation 0.5.  The weighted mean factor drops to $1.4\pm0.5$ if we remove the two most discrepant galaxies.  Considering the relatively large uncertainties in the SFH and metallicities of these galaxies, this  agreement can be considered quite good. 

%Moreover, we note that the most discrepant points, in the case of the 2010 models, correspond to the galaxies ESO540-030 with an excess factor of 4.14, and M81-DEEP with 4.94. These two galaxies are discrepant in most of the plots presented later in this paper. M81-DEEP is unusual in that it is the most metal-rich system in the sample (see Figure~\ref{fig:mherrors}). At the very high metallicity end, it is possible that the models tuned to the LMC could have problems.  The metallicity of ESO540-030 is not unusually high,  but its CMD is characterized by a large fraction of extremely red objects all along the RGB sequence, so the accuracy of its SFH may be called in question. If we neglect these two galaxies, we note the following trends: first, the over-prediction of the TP-AGB stars in 2008 models tend to decrease with the mass fraction of young stars. This trend is however completely absent in the 2010 models. This means that the overproduction problem of the 2008 models has been widely corrected in the 2010 models -- even though the correction only applies to the old metal poor AGB stars. 

%\subsubsection{Fluxes of TP-AGB Stars}
Figure \ref{fig:agbfracfake} also compares the observed vs.\ predicted flux contribution of TP-AGB stars. The best-fit SPS models show a larger offset in TP-AGB flux than in numbers, with models generally over-predicting the flux contribution.  The typical offset for the 2008 models is a factor three, but  as expected, the 2010 models do a better job of reproducing the data, especially for galaxies with little on-going star formation. However, for galaxies with recent star formation, the 2010 models appear to be just as discrepant as the 2008 models, with offsets reaching a factor of 2 - 3 or larger.  The weighted mean offset between the model/data flux ratio is 2.3 with a standard deviation of 0.8.  %In the discussion section, we will examine possible explanations for these results.

Two galaxies present very high model/data discrepancies for both number and flux contribution of AGB stars, ESO540-030 with a model/data flux ratio $= 6.70$ and M81-DEEP with model/data flux ratio $=7.43$.  These galaxies are discussed further in Section 4.1.  The third most discrepant galaxy, UGCA292 with model/data flux ratio $=4.39$,  is the least populated galaxy in our sample.  Therefore its large measurement uncertainties are driven by both small number statistics and large uncertainties in the SFH.  If we remove these galaxies from the weighted mean we derive an overall model/data flux ratio of $2.2 \pm 0.8$. 

\begin{deluxetable*}{lcccccc}
\tabletypesize{\small}
\tablecaption{Statistical Estimates of Foreground Star Contamination from TRILEGAL \label{tab:foreground}}
\tablehead{\colhead{Galaxy} & \multicolumn{2}{c}{TP-AGB Region} & \multicolumn{2}{c}{RHeB Region} & \multicolumn{2}{c}{All Foreground\tablenotemark{a}}\\
& \colhead{\#}& \colhead{flux\tablenotemark{b}} & \colhead{\#} & \colhead{flux\tablenotemark{b}}& \colhead{\#} & \colhead{flux\tablenotemark{b}}}
\startdata
               DDO71&  3&  1.56e-18&  1&  1.36e-18&  9&  5.01e-18\\
               DDO78&  3&  1.96e-18&  1&  9.44e-19&  6&  3.62e-18\\
               DDO82&  2&  1.47e-18&  8&  4.81e-17& 14&  5.04e-17\\
          ESO540-030&  1&  6.91e-19&  3&  5.37e-18&  7&  6.52e-18\\
               HS117&  1&  1.98e-19&  4&  3.21e-17&  7&  4.66e-17\\
          IC2574-SGS&  4&  6.59e-18&  6&  5.35e-17& 15&  6.78e-17\\
               KDG73&  3&  1.12e-18&  6&  8.33e-18& 12&  1.37e-17\\
               KKH37& 11&  1.21e-17&  5&  2.64e-17& 25&  3.99e-17\\
            M81-DEEP&  3&  1.23e-18&  3&  4.17e-18&  9&  5.87e-18\\
       NGC0300-WIDE1&  2&  4.85e-18&  3&  2.64e-17& 10&  3.80e-17\\
      NGC2403-HALO-6&  0&  0.00e+00& 10&  4.83e-17& 10&  4.83e-17\\
        NGC2976-DEEP&  1&  3.39e-18&  0&  0.00e+00&  7&  7.33e-18\\
     NGC3077-PHOENIX&  0&  0.00e+00&  5&  1.40e-17&  7&  1.54e-17\\
             NGC3741&  1&  2.54e-18&  2&  1.79e-17&  5&  2.13e-17\\
             NGC4163&  1&  2.66e-18&  5&  4.83e-17& 12&  5.21e-17\\
      NGC7793-HALO-6&  1&  1.64e-18&  8&  1.53e-17& 13&  1.79e-17\\
             SCL-DE1&  4&  1.15e-18&  0&  0.00e+00&  8&  3.10e-18\\
           UGC4305-1&  6&  5.38e-18&  5&  4.17e-17& 18&  5.11e-17\\
           UGC4305-2&  6&  5.46e-18&  5&  4.53e-17& 19&  6.39e-17\\
             UGC4459&  5&  5.91e-18&  5&  3.63e-17& 13&  4.52e-17\\
             UGC5139&  4&  2.26e-18&  5&  1.81e-17& 13&  2.10e-17\\
             UGC8508&  4&  3.17e-18&  2&  3.08e-18&  9&  6.82e-18\\
             UGCA292&  6&  2.63e-18&  1&  1.80e-18& 10&  5.61e-18\\
\enddata
\tablenotetext{a}{Foreground Brighter than $F160W =23$ [mag]}
\tablenotetext{b}{ergs cm$^{-2}$ s$^{-1}$}

\end{deluxetable*}

\begin{figure*}
\centering
\includegraphics[scale=0.7]{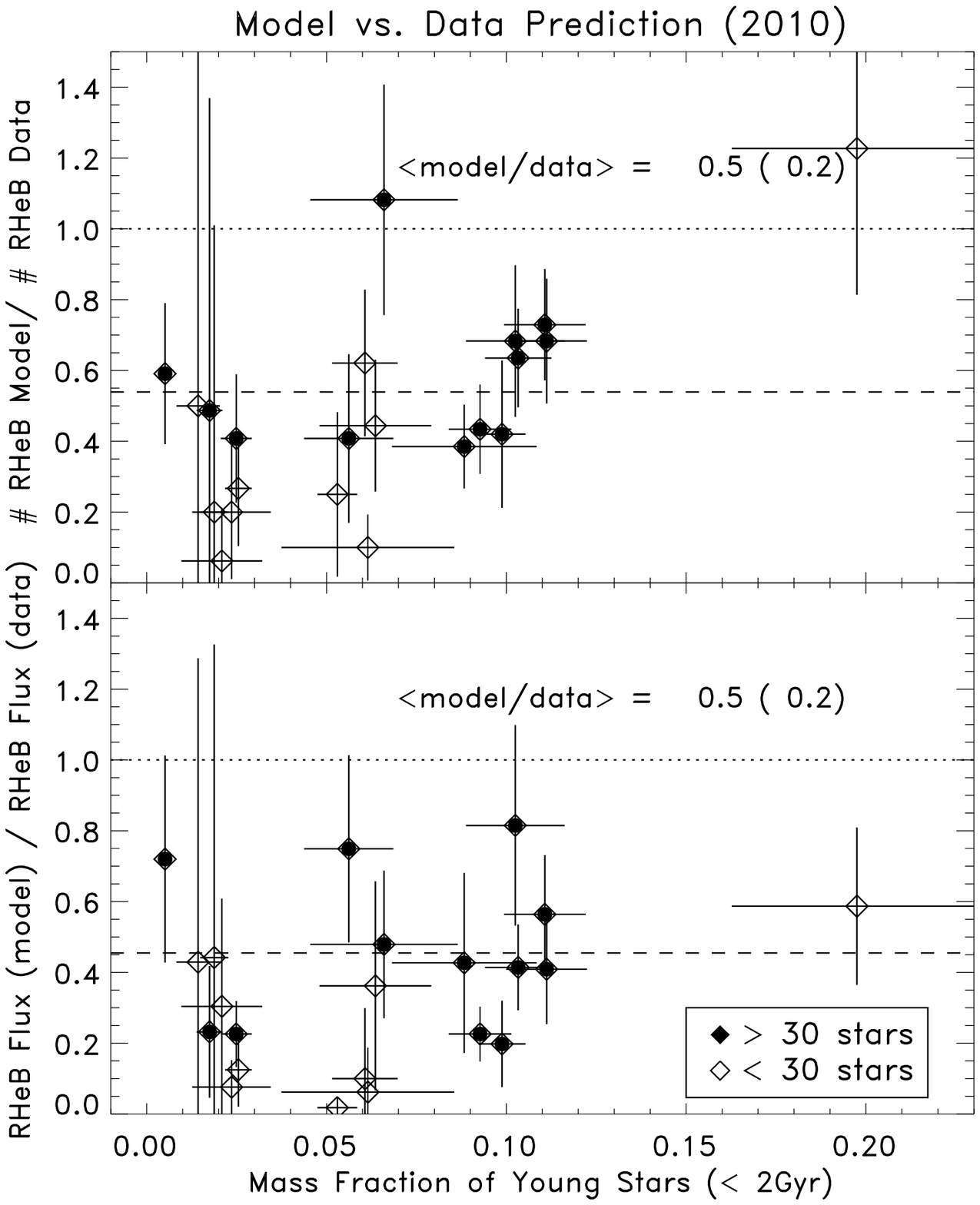}
\caption{\label{fig:RHeBfracfake} Same as Figure \ref{fig:agbfracfake}, only now for the RHeB model/data comparison.  Because there is no difference in the RHeB populations of the 2008 and 2010 models only the 2010 version is shown here (diamonds).  The best-fit SPS models tend to under-predict the numbers and fluxes of the RHeB stars.  The weighted mean fractional model/data differences for galaxies with significant numbers of RHeBs ($>30$ stars, filled diamonds) is $0.5 \pm 0.2$, for both the numbers and fluxes of  RHeBs. }
\end{figure*}

%\subsubsection{Numbers and Fluxes of RHeB Stars}
Figure \ref{fig:RHeBfracfake} compares the data to the SPS model predictions for RHeB stars.   In this case, the best-fit SPS models tend to under-predict the numbers and flux contribution of stars; both are under-estimated by a mean factor of $2.0\pm0.6$ (for galaxies with larger than 30 RHeB stars, shown as filled diamonds in the plot).  There is no appreciable difference between the SPS models based on the 2008 and 2010 Padova isochrones, so only the 2010 models are shown in this figure.  Likewise, correcting for foreground contamination does not alter these results as they are based on the galaxies with the largest number of RHeB stars.  Explanations for the model/data differences will be explored in the discussion section.

%{\bf TO BE CHECKED: An important point is that most of the simulations present HeB stars much more spread in color, and generally redder, than the observations. See for instance NGC3077-PHOENIX in Fig.2: it has about 10 model stars which are likely RHeB and are counted as TP-AGB. UGC4305-1 is another case: it has a very narrow observed red sequence, and a very broad modeled one. 

%What is causing this discrepance? The errors in the SFHs, for sure. In order to have a well-defined red HeB sequence, you need also a small spread in [Fe/H] at young ages. As shown in Table~1 and Figure~\ref{fig:mherrors}, the [Fe/H] dispersion at young ages, as derived from the best-fitting MATCH solutions, are as a rule of the order of $\sigma\sim0.25$~dex SFHs. These imply too large [Fe/H] intervals for the He-burning models, and hence too broad a red sequence in the models. This has to be revised, since an error in the SFH is being interpreted in terms of a discrepancy in the model lifetimes. Running MATCH with the Zinc option on could have prevented this broad red sequences from happening.

%Another important point: without error bars, these comparisons may be misleading. The typical numbers of RHeB stars are small enough to ensure large error bars in all galaxies.}

\section{Discussion}
Our results show that both the TP-AGB and RHeB sequences can contribute significantly to the NIR flux of a galaxy.  Even for galaxies with well-developed red giant branch populations,  the combination of these two late stages of stellar evolution can make up almost 40\% of the NIR light, while comprising negligible stellar mass.  As a result, these phases must be well-calibrated to accurately estimate stellar masses of galaxies.  Making matters worse, the parent populations for RHeBs and the most luminous TP-AGB stars are comprised of massive (e.g. M $> 3.5$ \msun) young stars with short lifetimes.  We therefore expect significant variations in the NIR M/L ratio of galaxies on very short timescales (e.g. $<300$ Myrs), especially in the early universe, where these stars will dominate the light.  

In addition to affecting M/L ratios, these rare but luminous populations may also be responsible for some of the scatter in key NIR galaxy scaling relations such as NIR metallicity-luminosity relations \citep[e.g.][]{Salzer05}, and NIR versions of the Tully-Fisher relation \citep[e.g.][]{Conselice05}.  Typically the scatter in these NIR relations is smaller by $\sim30\%$ compared to the optical, from avoiding the large scatter induced by dust obscuration and very young blue stars.  However, significant scatter remains --- e.g. more than the formal uncertainty in metallicity \citep{Salzer05}.  \citet{Salzer05} demonstrate that the scatter in the NIR does not correlate with the instantaneous star formation rate (as measured by Balmer line strengths).  However, some of the remaining scatter could well be attributed to differences in the ratios of evolved luminous stars.  For instance we show here that galaxies with recent star formation and large RHeB populations could be as much as 20\% brighter at 1.6 \um\ compared with similar mass galaxies with little recent star formation.  Likewise the younger TP-AGB stars also can significantly impact the NIR flux of a galaxy on longer timescales.  As the scatter in the \citet{Salzer05} metallicity-luminosity relation is $\sim0.2$ dex, luminous evolved stars could well be playing a significant role.  

\begin{figure}
\centering
\includegraphics[scale=0.5]{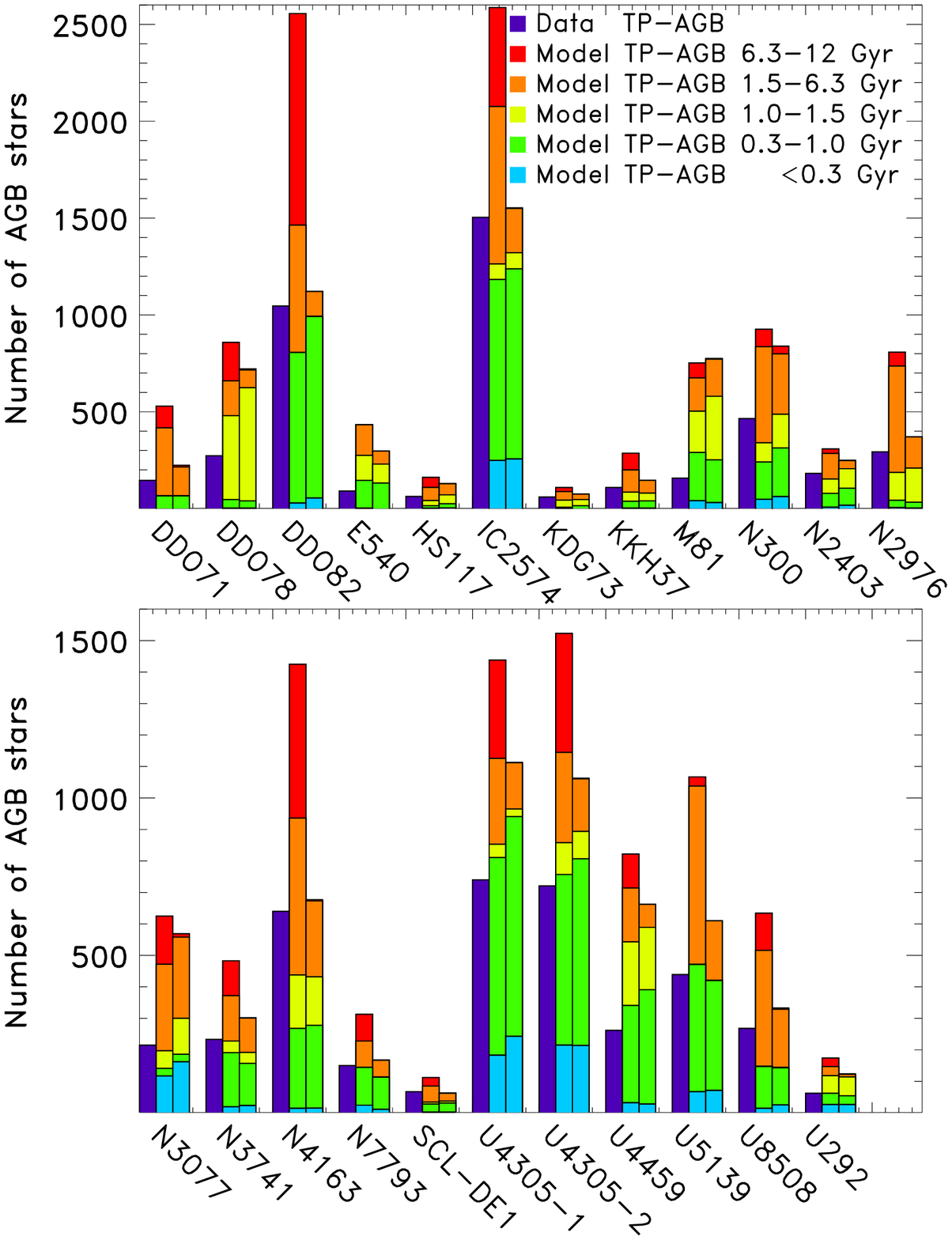}
\caption{\label{fig:AGBage} The numbers of TP-AGB stars in each galaxy (purple) compared to the 2008 (middle of each group of three) and 2010 (right of each group of three) Padova model predictions.  The total number of TP-AGB stars for each model are divided into different age bins, as indicated in the legend.  In all cases the models over-predict the numbers of TP-AGB stars.  The main difference between the 2010 and 2008 model is that the 2010 model has corrected the lifetimes of low-mass (old) TP-AGB stars.  These histograms show the effect of this correction, there are very few low-mass (old) luminous TP-AGB stars in the 2010 models.  With this correction, about 65\% of the 2010 models are in rough agreement with the actual TP-AGB star counts in the observed data. }
\end{figure}

%\subsection{Understanding Model and Data Discrepancies}

Unfortunately, we have shown that the best-fit SPS models of our sample galaxies --- based on the optically derived SFHs --- currently have difficulties recovering the NIR flux contributions from RHeB and TP-AGB stars. The best-fit SPS models tend to over-predict the NIR fluxes from TP-AGB stars and under-predict the NIR fluxes from RHeB stars. This latter discrepancy is particularly worrisome as the short-lived luminous RHeB stars are susceptible to rapid variations in the SFR.  For both the RHeB and TP-AGB, several factors may be contributing to the model/data discrepancies. Some may be driven by difficulties in measuring accurate SFHs for these galaxies, especially on the short timescales for which these stars live. Difficulties with SPS modeling (converting SFH into a CMD), including oversimplifications in the description of the $Teff$--luminosity variations during thermal pulse cycles and in the description of extinction by circumstellar dust, may also play a role. Additionally, the stellar evolution codes on which the SPS models are based may require further updates for these difficult to model stages of evolution.  We now examine the model/data discrepancies in more detail and explore the implications for studies of galaxies in the local and high redshift universe. 

\subsection{TP-AGB Model and Data Discrepancies} 
\citet{Girardi10} showed that SPS models, based on the 2008 Padova isochrones, over-predicted the numbers of TP-AGB stars in optical CMDs of ten old, metal poor low-mass galaxies.  They found that the predicted numbers of TP-AGB stars could be reconciled with the data if the lifetimes of the low-mass TP-AGB stars were significantly reduced from roughly 4 Gyrs to roughly 1 Gyr, i.e. reduced to lifetimes similar to those of more massive AGBs.  This conclusion formed the basis of the newly released Padova 2010 isochrones. 

Now we have expanded this investigation to 23 galaxies with a wide range of SFHs, including dwarfs and spirals, and have examined the behavior of the TP-AGB in the NIR.  We confirm the \citet{Girardi10} result that SPS models based on the 2008 Padova isochrones over-predict the numbers of TP-AGB stars.  We also find that SPS models based on the 2010 version of the Padova isochrones have largely eliminated the over-prediction of TP-AGB numbers.  While the new SPS models still show a small systematic bias to larger numbers than the data, $\sim65\%$ of the sample galaxies have model number predictions that overlap the data to within the uncertainties (Figure \ref{fig:agbfracfake}).  

Figure \ref{fig:AGBage} shows this comparison in more detail.  For each galaxy, we plot histograms of the numbers of TP-AGB stars in the data and compare to the results from the 2008 and 2010 models, only now the model histograms are sub-divided by the age (mass) of the synthetic star.  The 2010 model effectively eliminates the oldest (lowest mass) TP-AGB stars bringing the predicted numbers of TP-AGB stars more in line with the data.  We stress that it may be possible to reduce the 2010 model numbers further while using the same 2010 Padova isochrones, by means of reasonable changes to the SPS code: e.g., by adding temperature--luminosity variations driven by the thermal pulse cycles, and/or dust obscuration variations. These effects alone could be able to reduce the model numbers by a good $\sim20$\% percent, and will be further explored by Rosenfield et al. (in prep.).      

We note that the most discrepant points, in the case of the 2010 models, correspond to the galaxies ESO540-030 with an excess factor of 4.14, and M81-DEEP with 4.94. These two galaxies are discrepant in most of the plots presented in this paper. M81-DEEP is unusual in that it is the most metal-rich system in the sample. While the mean metallicity at 1 Gyr is listed at $-0.4$ in Table 1, this estimate is misleading.  It hides the fact that most (75\%) of the TP-AGB stars (which span a range of ages) have metallicities near solar (i.e. $[M/H] > -0.1$), producing the very red TP-AGB branch in the modeled CMDs.  At the very high metallicity end, it is possible that the models tuned to the LMC could have problems. 

Issues at high metallicity may also be affecting the modeled CMDs of several other galaxies in this sample.  For instance, galaxies DDO82, IC2574-SGS, and NGC~300 all show plumes of very red TP-AGB stars in their modeled CMDs. These plumes are not nearly as well-populated or obvious in the observed CMDs of these galaxies. While the mean metallicities for the modeled TP-AGB populations in these galaxies are metal poor, these red plumes are metal rich (roughly solar). This result again suggests issues at the high metallicity end, either with the measured SFHs or the stellar evolution codes.   

The metallicity of ESO540-030 is not unusually high,  but its modeled CMD is characterized by a large fraction of extremely red objects all along the RGB sequence, so the accuracy of its SFH may be called in question. If we neglect ESO540-030 and M81-DEEP, the overproduction problem of the 2008 models has been largely corrected in the 2010 models -- even though the correction only applies to the old metal poor AGB stars. 

While we have largely accounted for the differences in TP-AGB number, there continues to be discrepancies between the predicted and observed TP-AGB fluxes.  How can we understand that the best-fit SPS models reasonably reproduce the TP-AGB numbers, with typically less than $\sim50$\% excess, while the excess in F160W flux is much larger ($>200$\%)? The answer is that the excess occurs mainly in the form of relatively few but very luminous TP-AGB stars that are predicted to exist in the model but that are not observed in the data. This phenomenon can be seen in the CMDs of Figure 2.  

\begin{figure*}
\centering
\includegraphics[scale=0.7]{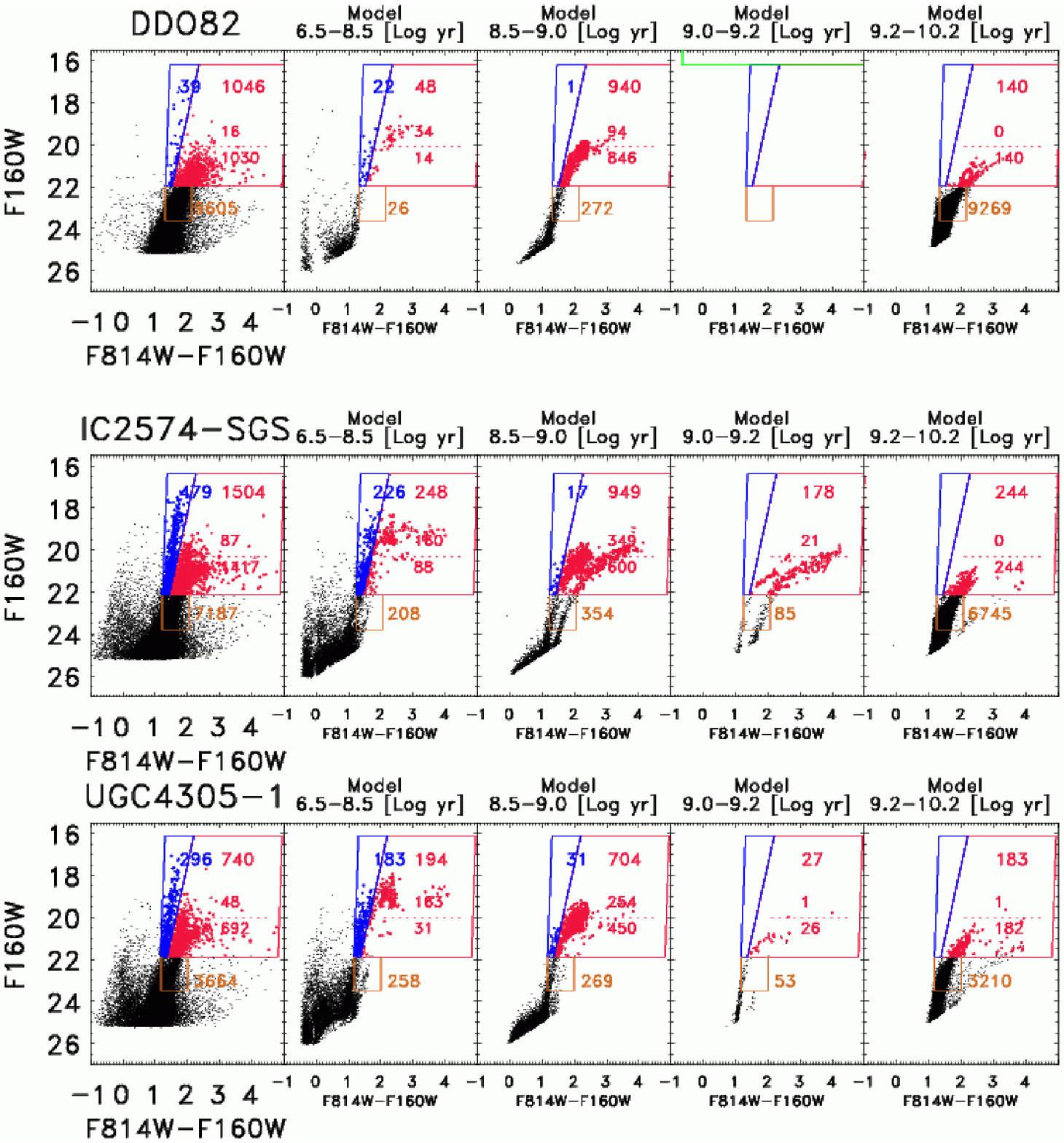}
\caption{\label{fig:RHeBdetail} CMDs of three sample galaxies, selected to have had significant recent star formation, compared to SPS models (based on the 2010 Padova isochrones), with the models divided into different time bins.  Colored regions are same as Figure \ref{fig:CMD}, only now we add an additional luminosity cut on the TP-AGB bin to compare the numbers of the most luminous TP-AGB stars with the model predictions. These high luminosity TP-AGB stars are primarily at young ages ($<1$ Gyr).  Typically the models contain more high luminosity TP-AGB stars than the data.  
}
\end{figure*}

To better understand these predictions, Figure \ref{fig:RHeBdetail} re-plots the 2010 models for 3 of the sample galaxies, now broken up into different time bins.  The most luminous TP-AGB stars are found at the youngest ages, in the $ < 0.3$ Gyr age bins, which correspond to the highest progenitor masses (M $>3.5$ \msun).  Additional luminous TP-AGB stars are found in the $0.3 -1$ Gyr age bins.  Very few of the oldest ($>1$ Gyr), or lowest mass, TP-AGB stars reach the brightness levels of their more massive counterparts. The models appear to over-predict the numbers of the brightest TP-AGB stars (above the dashed line) by large factors, primarily at the youngest ages ($< 1$ Gyr).      

Unfortunately, due to a paucity of stars, constraints derived from TP-AGB stars in star clusters of the Magellanic Clouds are relatively poor for the age interval between 0.1 and $\sim0.5$ Gyr  \citep{Girardi07}.  Therefore, Figures 2 and \ref{fig:RHeBdetail}  could be providing precious information at where specifically the TP-AGB models need further improvement, and where additional observational constraints are the most urgent.  

Other issues may also contribute significantly to the model/data mismatch.  (1) Lack of a description for the pulse cycle luminosity variations (e.g. low-luminosity dip and flash luminosity) in the NoisyCMD code leads to a fraction ($\simeq 20 - 30 \%$) of the simulated AGB stars brighter than they should be. While the TP-AGB tracks in Marigo \& Girardi (2007)  do follow in detail the flash-driven luminosity variations, these features are not included in the stellar isochrones or in NoisyCMD, where the whole TP-AGB evolution is assigned the pre-flash maximum quiescent luminosity predicted by the core mass - luminosity relation. (2) Dust obscuration from circumstellar TP-AGB envelopes could affect both the observational selection criteria and/or the model predictions. In fact, the observational data could be "missing" TP-AGB stars, with self-extinction hiding NIR luminous TP-AGB stars in the optical data (F814W) data we use to select them. Additionally, model prescriptions for the dustiest phases of the TP-AGB are not included in NoisyCMD, thus they may be predicted to be more luminous at NIR wavelengths, than if proper dust modeling were used.  (3) The SFHs we have derived for our galaxies may be wrong and thus predict incorrect number of AGB stars. Accurate constraints on the SFHs on the short timescales of the most massive stars are difficult to obtain even from very deep CMDs.

We now explore the last two possible limitations, and leave further examination of the stellar evolution codes to future papers in this series.

\subsubsection{Are the $F814W - F160W$ CMDs Missing NIR Luminous AGB Stars?}

\citet{Boyer09} demonstrated that optical searches will miss a large fraction of the most dust obscured AGB stars.  As part of our search criteria, we have used some optical data (\HST\ $F814W$), and therefore are likely to have missed the most dust obscured sources.  However, for our purposes we are only concerned about those missing TP-AGB stars that are actually luminous at 1.6 \um\ (i.e.~brighter than the TRGB). To test if there are large numbers of NIR luminous AGB stars missing from our samples, we return to the NIR only CMDs ($F110W - F160W$).  We find that there are typically several red TP-AGBs with luminosities brighter than the TRGB that were missed in the optical-IR CMD search.  However the total numbers of missing stars are typically fewer than the Poisson uncertainties of the original count.  In addition, because these are among the most dust obscured, they tend to not be among the most luminous TP-AGBs at 1.6 \um. Therefore missing AGB stars in the data cannot account for the model/data differences in the NIR flux.

What about the role of dust obscuration on the model itself?  Real AGB stars can experience significant self-induced dust obscuration and drop out of both optical and NIR CMDs \citep{Boyer09}.  In fact, some model prescriptions predict that for progenitor ages of $100 -200$ Myrs, TP-AGB stars can spend nearly half of their lives as highly dust obscured objects \citep{Marigo08}.  However, these dust prescriptions have not been included in the SPS codes we are using to model the CMD.  Therefore we expect that some of the model TP-AGB stars would be much fainter at NIR wavelengths if the proper dust prescriptions were included.  

\begin{figure*}
\centering
\includegraphics[scale=0.75]{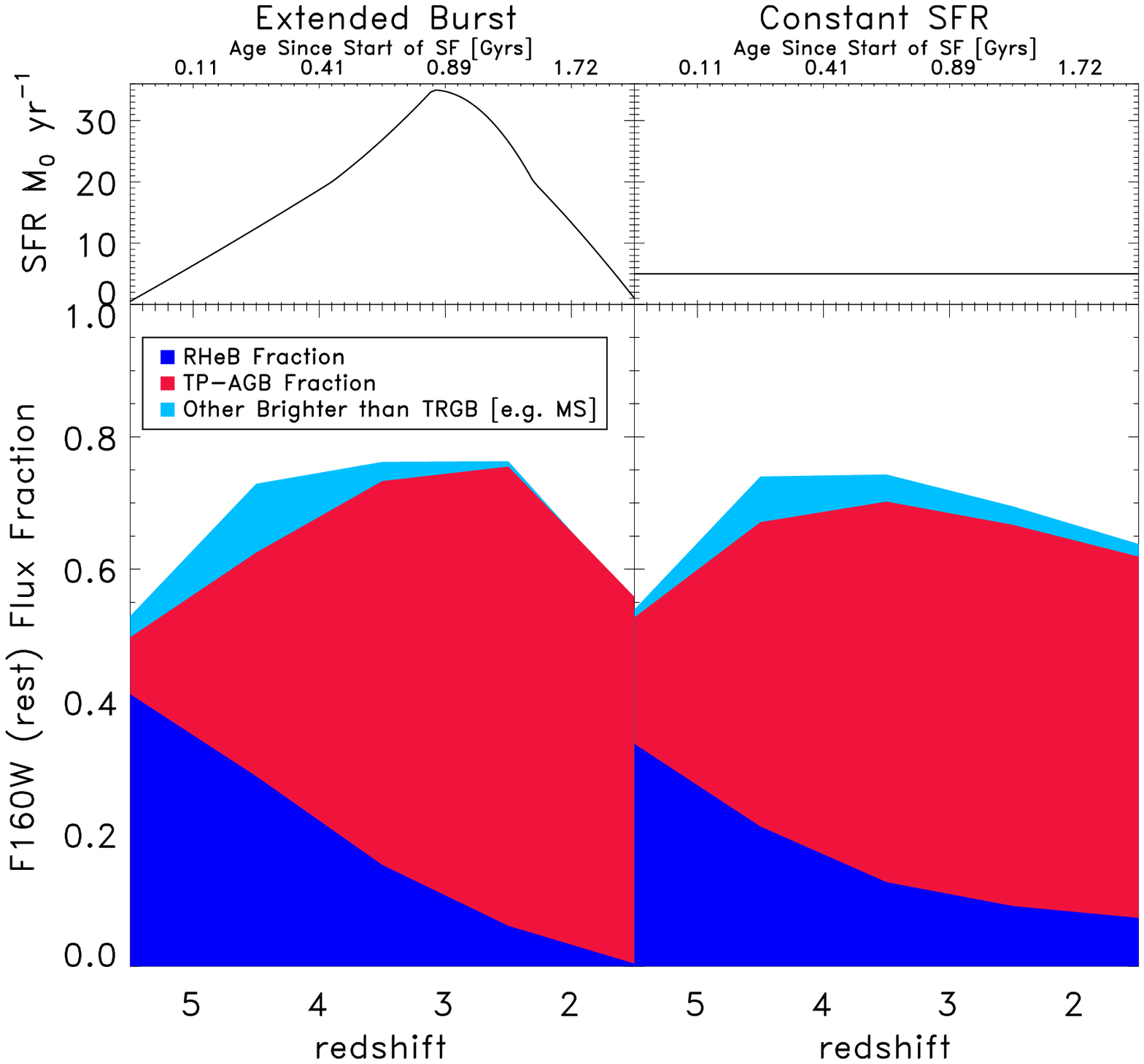}
\caption{\label{fig:highz} The fraction of NIR light arising from IR luminous stars as a function of redshift as predicted by the 2008 (and 2010) Padova isochrones.   We generate model populations at high redshift with two different star formation histories.  Star formation begins at $z=6$ for both models, however one undergoes an extended burst that peaks at $z=3$ with a SFR of 30 M$\_{\odot}$ yr$^-1$ (left), and the other has a constant SFR of 5 $M\_{\odot}$ yr$^-1$ (right).  Top panels show the SFHs, while bottom panels show the predicted flux contributions from, RHeBs (blue), TP-AGB stars (red) and additional MS stars brighter than the TRGB (stars). While RHeBs dominate at early times (producing 40\% of the light at $z=5.5$) , TP-AGB stars dominate by $z=4$, reaching a peak of almost 70\% of the $F160W$ flux in the extended burst model and over 50\% for the constant SFR.  These models are representative of the types of stellar population synthesis currently used when fitting high-$z$ CMDs.  However, they may need to be corrected to account for the observational results presented in this paper. }
\end{figure*}

One way to explore this issue further is to track the carbon-to-oxygen (C/O) ratio in the model AGB populations.  Over the course of an AGB star's lifetime, carbon is dredged up from the interior to the surface, changing the overall C/O ratio.  When this ratio exceeds unity the star is termed a carbon star.  The additional carbon makes it much easier to form dust in the stellar atmosphere.  Therefore the reddest TP-AGB stars tend to be associated with carbon-rich populations \citep[e.g.][]{Nikolaev00, Cioni06, Bonanos09, Boyer11}.  

We use the FAKE routine  \citep[which is part of the MATCH package][]{Dolphin02} to track the C/O ratio, metallicity, age, and mass of the artificial TP-AGB stars.  What we find is that the most luminous model TP-AGB stars, which are typically quite blue, are metal-poor and carbon-rich.  Compared to the metal-rich AGB, metal-poor stars typically need less carbon dredge up to reach a C/O ratio greater than unity.  However, because the SPS models we are using do not include prescriptions for circumstellar dust, the colors of these artificial carbon stars appear to be driven by metallicity, rather than their C/O ratio.  Being metal poor, these TP-AGB stars are blue, even though they should have significant amounts of circumstellar dust.  In contrast, the reddest TP-AGB model stars, especially in galaxies DDO82, IC2574-SGS, and M81, are oxygen-rich stars with high (solar) metallicity.  High metallicity means that more carbon needs to be dredged up to to become a carbon star, so these very red TP-AGB stars are actually modeled as oxygen-rich.  Again, their red color is driven by their metallicity rather than their C/O ratio. 

These results are counter to the observationally-driven expectations that the reddest stars will be carbon-stars and the bluer ones will be oxygen-rich.  A more complete treatment of self-obscuration by dust will be explored further in the next paper in this series, and may be key for accurate modeling of the TP-AGB, even in the NIR.

\subsubsection{Are the Measured SFHs of the Sample Accurate?}

If the input SFHs are incorrect, the SPS models are unlikely to match the data. We now test if the observed discrepancies between the model and real AGB stars can be explained by uncertainties in the SFHs of our galaxies.  To do so, we systematically lower the SFRs for intermediate aged populations ($<2$ Gyrs) to bring them in line with the predicted numbers and fluxes of TP-AGB stars and then recalculate the model CMDs.  We find that by systematically lowering the SFRs by roughly the uncertainties in a given age metallicity bin, we can reduce the model predicted AGB contributions by roughly a factor of 2.  While this solution appears to fix much of the data/model discrepancies on the AGB, it actually creates a larger problem in another region of the CMD, namely the main sequence turn-off (MSTO) for intermediate aged populations.  The systematically lower SFRs now under-predict the numbers of stars on the MSTO by a factor of two. The number of TP-AGB stars and intermediate aged MS stars are therefore in tension.  However, the MSTO is much better understood than the AGB, and is better populated making it more robust to stochastic fluctuations. It is therefore more likely that the original SFHs were reasonable.  
%As the models tend to over-predict the AGB,

In addition to forcing a SFH with a systematically lower SFR at young ages, we also have run MC simulations that span the full range of input SFHs deemed acceptable by the SPS modeling routine CalcSFH.  As described previously, we use the MC simulations to derive the true uncertainties on the numbers and fluxes from the TP-AGB (see Figure \ref{fig:agbfrac}).  While the uncertainties on any one galaxy are large (e.g. 20-70\%), they are typically not large enough to account for the offset in flux from the data, especially for galaxies with significant recent star formation.  In all cases, the best-fit SFH systematically over-predicts the TP-AGB flux.  The weighted average of the model over-prediction in flux is $230\%$ with a standard deviation of 80\% for the ensemble of galaxies.

 \begin{figure*}
%\centering
\includegraphics[trim=0 20 0 40,scale=0.75]{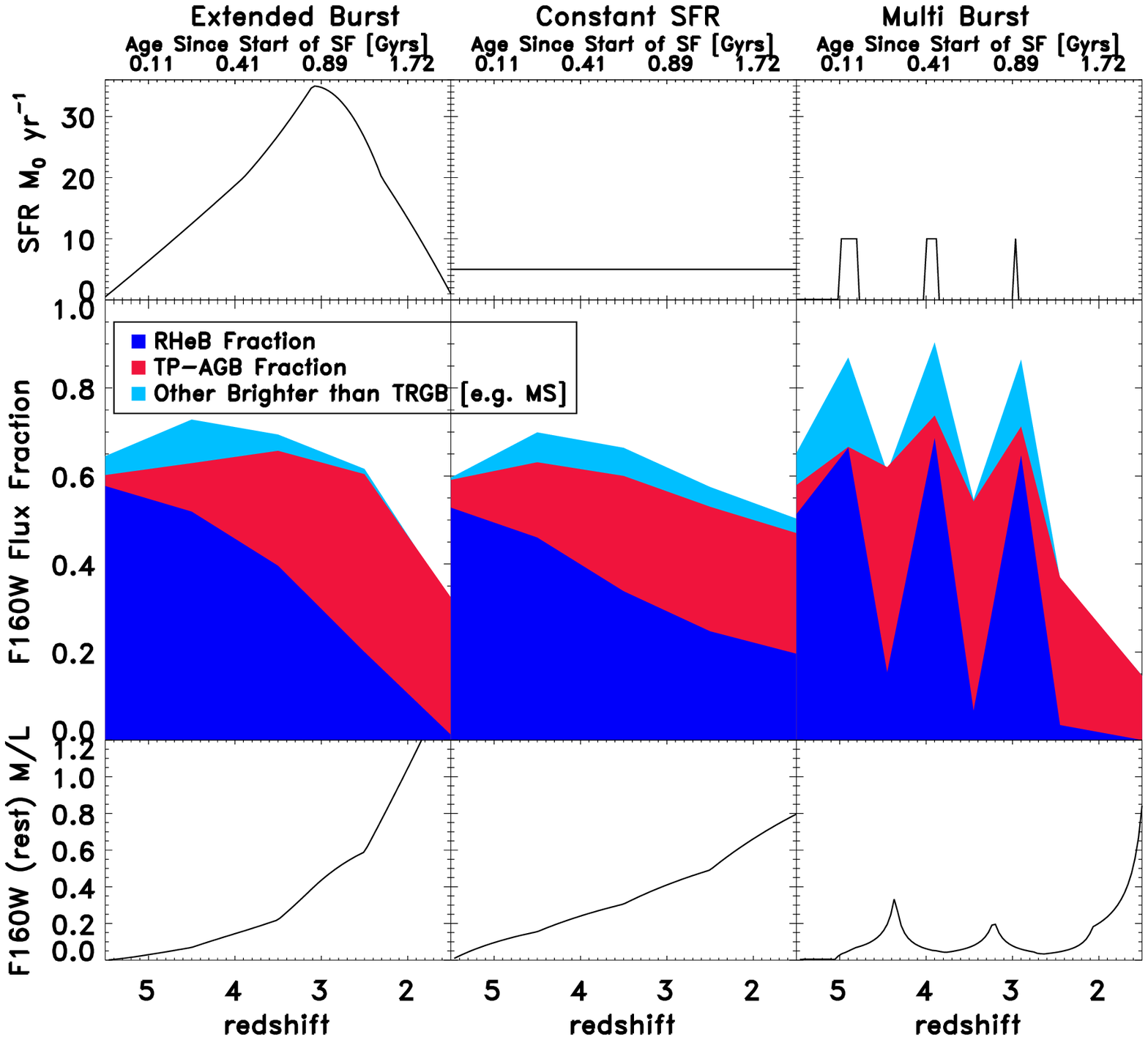}
\caption{\label{fig:highz2} Same as Figure \ref{fig:highz}, only now we include corrections to the model predictions for TP-AGB and RHeB fluxes.  RHeB flux contributions are increased by a factor of two, while the TP-AGB contributions are reduced by a similar factor compared to the 2010 Padova models.  The over-all contribution of massive stars to the $F160W$ flux is only reduced slightly from a peak of  80\% in Figure \ref{fig:highz} to a peak of 70\% here.  However the makeup of the population has changed dramatically.  Now the RHeB contribution dominates or is comparable to the TP-AGB contribution for much of the observed time-period, except for late times in the extended burst model.  This has significant implications for SEDs of galaxies at high redshift.  While TP-AGB stars preferentially increase the NIR-to-optical flux ratio of a population, RHeBs are significantly bluer.  In addition, RHeB are often accompanied by similar numbers of blue HeBs \citep{McQuinn11}.  Therefore RHeBs and their blue counterparts are likely to increase both the optical and NIR luminosities of galaxies, and lower their M/L ratios across the full SED.   Also shown are the NIR M/L ratios of these model galaxies (bottom). The M/L ratio varies rapidly by over a factor of 10 between $z=5$ and 2 for the extended burst model (and by roughly a factor of $\sim7$ for a constant SFR).  To emphasize how these populations can rapidly change the M/L ratio of a population we now also show a multi-burst model (right).  In the multi-burst model the M/L ratio oscillates by a factor of two as the bursts turn on and off.  The amplitude of this oscillation would be even larger for larger burst strengths, or longer quiescent periods between bursts. Note: the widths of the spikes in the sand diagram for the multi-burst model (right, middle), are artificially wide because of the course redshift sampling. 
}
\end{figure*}

\subsection{RHeB Model and Data Discrepancies} 
Our analysis in \S 3.4 shows that RHeB stars are also a significant contributor to the NIR flux in galaxies with ongoing star formation.  However, the best-fit SPS models do not capture the properties of RHeB populations in multiple ways.  First, the SPS models tend to under-predict the numbers of RHeB stars (see Table 4).  Second, they tend  to under-predict the luminosities of individual RHeB stars. This result can be easily seen in Figure \ref{fig:RHeBdetail}, where the most luminous RHeB stars in IC2574, for instance, are almost a magnitude brighter than the most luminous model RHeB stars.  Third, the model RHeBs appear to be redder than the data in these NIR filters.  This result can also be seen in Figure \ref{fig:RHeBdetail}.  While the observed RHeBs in IC2574 are clearly separated from the red edge of the RHeB selection region, the model RHeB stars tend to hug the red edge of the of this region, with a color offset of $\sim0.2$ mags. Finally the RHeB branches in the observational data tend to be tight sequences and thus are easily captured by the narrow boxes in CMD space.  The model RHeB sequences, on the other hand, tend to be spread out in color, forming much looser sequences, which may contribute to stars falling out of the same boxes in the CMD.

 Figure \ref{fig:RHeBdetail} also reveals several additional issues, including the problematic effects of photometric uncertainties when selecting subregions of the CMD.  In galaxies IC2754-SGS and UGC4305-1, a significant number of stars in the $0.3 - 1$ Gyr time bin scatter into the RHeB selection region, even though they are likely to be AGB stars.   This effect also drives the large model/data number ratio for RHeBs in galaxy NGC~2976, where there is only one real RHeB star but a larger number of predicted stars. However, removing these spurious stars from the RHeB classification would only tend to strengthen our conclusion that the models tend to under-predict the RHeB flux in these galaxy.  Of course, there may be a similar contribution of TP-AGB to the RHeB bin in the actual data, in which case no correction would be necessary.  %The other two examples in Figure \ref{fig:RHeBdetail} show less contamination from old stars entering the RHeB selection region.  

An even more important issue with the RHeB models is that some model stars may be sufficiently red to leave the RHeB selection region altogether and instead fall into the TP-AGB selection region.  For instance,  the combined numbers of model RHeB and TP-AGB stars in the earliest time bin of Figure \ref{fig:RHeBdetail} is very similar to the total number of RHeBs in the data, in all three cases shown here.  If these young red stars actually belong on the RHeB sequence, but are being added to the TP-AGB sequence, then the effect will be to over-predict the TP-AGB flux, and under-predict the RHeB flux, a trend that exists in  the data.   %The problems with model RHeB stars will be discussed further in an upcoming paper (Rosenfield et al. 2011, in preparation).  Regardless of the physical explanation, our results require that the model RHeB flux contributions must be increased by roughly a factor of two to match the data. 

As with TP-AGB results, these data/model differences could be the result of a wide range of issues:  (1) there could be a population of foreground stars contributing to the observed data, that do not exist in the model;  (2) there could be problems with measuring accurate SFHs or correctly populating CMDs over the short timescales of RHeBs; or  (3) there could be lingering issues with the stellar evolution codes, for example, the lifetimes of the most massive and luminous RHeBs could be under-predicted.  The first issue was examined in detail in Section 3.6, where we found that foreground contamination could likely bias the observations when only a few RHeBs are observed.  However, foregound contamination was found to be small compared to the observations of well-populated RHeB sequences. We examine potential issues with modeling young SFHs below but leave any updates to the stellar evolution code for the next paper in this sequence Rosenfield et al. (in preparation).

%\subsubsection{Are Foreground Stars Contaminating the Observed RHeB Sequences?}

\subsubsection{Are the SFHs and SPS Codes Accurate for the Youngest Populations?}
Two of the most obvious issues with the RHeB branches are that the best-fit SPS models are too diffuse in color space and do not contain stars bright enough to match the data.  Both of these differences could potentially be caused by inaccurate SFHs at these young ages.  For instance, our best-fit SFHs have a relatively broad metallicity spread (typically 0.25 dex) even at young ages.  This metallicity spread will cause the modeled RHeB branch to be broader than if the metallicity spread were narrower.  We estimate that dropping the metallicity spread to 0 dex could increase the numbers of synthetic RHeB stars by as much as 25\%.  While not enough to account for a factor of 2 discrepancy between the models and the data, this change would tend to improve the match between the models and data.  To test this possibility more rigorously, we recalculate the optically  derived best-fit SFHs with CalcSFH only now we force the narrowest possible spread on the input metallicities at young ages (using the -zinc flag, giving an effective spread in metallicity of 0.1 dex).  The SPS models based on these new SFHs still under-predict the RHeB flux by roughly the same factor as before.

Another issue is that the SPS models do not contain enough very luminous RHeB stars.  The stellar evolution models can make stars that are as bright as the data, but only at the youngest ages, e.g. $< 50$ Myrs.  It may therefore be that the best-fit SFHs are systematically under-predicting the SFRs of our galaxies at the youngest ages.  This bias would tend to decrease the numbers of the most luminous RHeBs and significantly decrease the predicted RHeB flux contribution.  

To test if the SFHs are wrong at very young ages, we rerun the SFHs only now increasing the SFRs at the youngest ages.  Doubling the SFRs at these ages is enough to roughly match the actual RHeB flux observed in the real galaxies.  Unfortunately, doing so increases the numbers of very young main sequence and main sequence turn-off stars as well.  Compared to the data, these new models have roughly five times the numbers of stars in these additional regions of the CMD.  Thus it is unlikely that SFH problems alone can account for the model/data discrepancies.

\subsection{Implications for Observations at High Redshift}
Our results demonstrate that short lived TP-AGB and RHeB stars can contribute significant fractions of the NIR light of local galaxies, while contributing negligible amounts of stellar mass.  We expect that the flux contributions of these stars will be much greater at high-redshift where star formation rates are high, and the RGB is less developed.  In this section we attempt to put some constraint on the contribution of TP-AGB and RHeB stars to the rest-frame NIR fluxes of galaxies at high redshift.  We again focus on the rest-frame NIR fluxes where these stars are most luminous and where rest-frame 1.6 \um\ fluxes of high-$z$ galaxies are well constrained from deep \Spitzer\ imaging in the $3-8$ \um\ observed-frame bands.  This wavelength regime is also where the future James Webb Space Telescope will be operating. 

First we show the predictions from the 2008/2010 Padova isochrones, as these, and similar isochrones, are the foundation for interpreting colors and magnitudes at high-$z$.  Then, by reducing the contribution of TP-AGB flux and increasing the contribution of RHeB flux, we produce predictions that could account for the discrepancies between the SPS models and our observational datasets. Finally, we discuss these predictions in the context of SED fitting at high redshift.  

We caution that our observational constraints at low redshift are primarily tracing low metallicity systems, typically 1/10 -- 1/5 solar.  Massive high redshift systems such as sub-mm galaxies and BzK galaxies are likely to be at higher mean metallicity \citep[e.g.,][]{Swinbank04, Onodera10}, despite the overall decline in gas phase metallicity with redshift \citep[e.g.,][]{Erb06a, Moustakas11}.   However, our results are directly applicable to lower mass systems at high redshift such as $z=3$ Lyman break galaxies \citep{Erb06a, Mannucci09, Sommariva11} and gamma ray burst host systems \citep{Laskar11} which have metallicities 1/10 -- 1/2 solar.  Our results will also be applicable to the even lower mass systems that will be observed with the James Webb Space Telescope and the Thirty Meter Telescope.

%To do so, we make use of the same stellar population synthesis models we used to study the local galaxies, only now we apply two additional assumptions that are motivated by our earlier analysis.  First, based on our analysis of Figure \ref{fig:AGBage}, we assume that the 2010 input models are are over predicting the flux of high-mass (young) AGB stars by roughly a factor of three.  Second, we assume that the models under-produce the RHeB fluxes by roughly factor of two for all systems with ongoing star formation.  %This possibility was demonstrated in Figure \ref{fig:RHeBfracfake}, and although Figure \ref{fig:RHeBdetail} suggests a more complicated relationship, we adopt the factor of two as a benchmark.    

We use CalcSFH and NoisyCMD to generate model stellar populations at high redshift with two different star formation histories, an extended burst and a constant star formation history.  We assume that star formation begins at $z=6$ for both models, and we track the model galaxies through $z=1$, a time-frame of roughly 5 Gyrs.  The extended burst reaches a maximum star formation rate of 30 $M_{\odot}$ yr$^{-1}$ at $z=3$, and then declines to a $SFR =0\; M_{\odot}$ yr$^{-1}$ by $z=1$.  The constant star formation model has $SFR= 5\; M_{\odot}$ yr$^{-1}$ over the full time period.  These star formation rates are $2-4$ orders of magnitude higher than those measured in the dwarf galaxies in our study, but they are representative of the SFRs found in the high-z galaxies studied with current instrumentation.   In each case we assume that metallicity increases from [M/H] $=-0.7$ to 0 by $z=1$.  We create 5 burst and 5 constant SFR models at each redshift, and average the results for each set.

Figure \ref{fig:highz} shows the predicted fraction of the NIR luminosity arising from NIR luminous stars for high-$z$ galaxies modeled with the Padova 2008 isochrones (We use the Padova 2008 models here as they are most similar to the AGB-heavy models that have been used to date in interpreting high-z data.  However, the 2010 isochrones will give roughly equivalent results because there has not been enough time to create significant numbers of low-mass TP-AGB stars).  When star formation commences, RHeB stars rapidly become the dominant IR source in our model galaxies, providing $\sim40$\% of the $F160W$ light at $z=5.5$.  At that time, TP-AGB stars and massive main-sequence stars provide another small fraction ($\sim10$\%).  The remainder, $\sim50$\%,  comes from less massive main sequence stars.  By $z=3.5$, the contribution from RHeBs declines due to increasing numbers of luminous intermediate-aged TP-AGB stars.  At this redshift, TP-AGB stars are predicted to be the dominant NIR source, producing 50\% or more of the total. 

Figure \ref{fig:highz} gives an overview of the current predictions for massive-star flux contributions to high-$z$ galaxies.  These fractions are baked into current high-$z$ SED fitting and mass estimates.  However, our data suggest that there may be problems with these predictions. In particular, the RHeBs contribution may be under-predicted by as much as a factor of two, and the TP-AGB fraction may being over-predicted by a similar factor.  Taking these approximate corrections into account, a different picture emerges for the luminosities of high-$z$ galaxies,  as shown in Figure \ref{fig:highz2}.  Now the RHeB phase dominates or is comparable to the TP-AGB phase, except for late times in the extended burst model.  

Coincidentally, there is only a small change in the overall flux of NIR light from massive stars between Figures \ref{fig:highz} and \ref{fig:highz2}.   The decline in TP-AGB flux is effectively offset by the increase in RHeB flux, at least for these relatively short SFHs. The NIR M/L ratios are therefore not radically different between the 2008 models and the approximately corrected version.     However, the different make-up of the population has significant ramifications for SED fitting.   TP-AGB stars will tend to increase the NIR-to-optical flux ratio, because TP-AGB stars are very red (more luminous in the NIR than the optical).  In contrast, RHeB stars are significantly less red than TP-AGB stars.  In addition, the existence of RHeB stars usually implies the existence of Blue Helium Burning (BHeB) stars, which are very luminous at bluer wavelengths \citep[e.g.][]{Dohm-Palmer02}.  Massive core helium burning stars tend to evolve through blue and red loops before ending their fusion lifetimes, and the numbers of BHeB are typically comparable to the numbers of RHeBs at a given luminosity \citep{McQuinn11}. Thus together the red and blue core helium burning populations will tend to simultaneously increase both the optical and the NIR luminosity rather than affecting the NIR alone as expected for TP-AGB stars.  %Figure \ref{fig:MtoL} shows how rapidly the rest-frame F160W M/L ratio changes at high redshift for the two simple SFHs we have modeled. 

These results may help explain several puzzling outcomes for SED fitting at intermediate redshift.   For instance \citet{Kriek10} found that SEDs of $z\sim0.7$ post-starburst galaxies cannot be simultaneously fit in the optical and NIR with \citet{Maraston05} models, which have a large TP-AGB component. The \citet{Maraston05} models tend to favor larger NIR fluxes for a given optical flux, compared with actual observations.   However, both the blue and red sides of these SEDs can be fit with \citet{BC03} models, which have a smaller TP-AGB contribution.  Likewise,  \citet{Muzzin09} found that SEDs of massive compact galaxies at $z\sim2$ are better fit with \citet{BC03} models than updated versions that include a larger TP-AGB component.  In both cases the AGB-lite models tend to predict more stellar mass than the AGB-heavy models. 

Our results also suggest that AGB-lite models may be a better match to the data from local galaxies.  This alone may be enough to explain the \citet{Kriek10} and \citet{Muzzin09} results.  However, their conclusion, that these post-starburst galaxies contain higher stellar mass than the AGB heavy models would predict, may be incorrect.  If their sample galaxies contain some stellar populations younger than $\sim300$ Myrs, then red and blue HeB stars may also be contributing flux.  By increasing the flux at both the optical and NIR, these helium burning stars necessitate a lower M/L ratio across all wavelengths. The current AGB-lite models do not necessarily adequately capture the HeBs, and our data suggest that they must be considered when there has been recent star formation. 

Figure \ref{fig:highz2} also shows the evolution of the NIR M/L ratio for our example high-$z$ SFHs  (the extended burst model and the constant SFR model) after applying approximate corrections that increase the RHeB flux and decrease the TP-AGB flux.  For both SFHs, the M/L ratio changes rapidly, and by large factors --- over a factor of 10 between $z=5$ to 2 for the extended burst and a factor of 7 for the constant SFR.  Capturing this rapid evolution is essential for estimating accurate stellar masses and proper SED fitting at high-$z$.  For example, Figure \ref{fig:highz2} shows a multiple-burst SFH, where the M/L ratio oscillates by a factor of roughly 2 as the bursts turn on and off.  These oscillations amplitudes will be even stronger if the burst strengths are larger or the quiescent intervals are longer.

Again, we note that Figure \ref{fig:highz2} is intended to demonstrate the consequences of the largest reasonable changes to the stellar evolution codes, based on our observational data.  If some of the model/data differences described in the previous sections arise from poor SFH measurements, or issues with the SPS modeling, then the changes we are describing here will be smaller than we are reporting.  A final assessment of the model/data comparison, and model updates will be provided in future papers in this series.

\section{Conclusions}
We report on the contribution of evolved stars to the NIR fluxes of 23 nearby galaxies observed with \HST\ WFC3 (NIR) and ACS (optical).  The \HST\ color-magnitude diagrams separate different phases of stellar evolution, including the red giant branch, the asymptotic giant branch and luminous red helium burning stars.  The CMDs also provide markers of the star formation histories of each galaxy.  We use the optically derived SFHs to model the NIR CMD to the main sequence K dwarfs.  We  use these models to produce an estimate of the total 1.6 \um\ flux for  each galaxy by combining the observed NIR fluxes for luminous stars with the model fluxes of fainter stars.   We then calculate the contribution of TP-AGB and RHeB stars to this total flux.  

While the RHeB and TP-AGB sequences represent negligible stellar mass, they can account for as much as 21\% (or 18\% after foreground correction) and 17\% (respectively) of the 1.6 \um\ fluxes for  these local galaxies.  At higher redshift, when galaxies do not have well developed red giant branches, RHeBs and TP-AGB stars are expected to produce much higher fractions of the NIR light.  At high-$z$, the summation of all IR luminous sources (the rare stars brighter than the typical TRGB at that time) can be expected to be as high as 70\%, most of which comes from TP-AGB and RHeBs.  These massive stars can therefore produce short-lived dramatic changes in the NIR M/L ratio.  Stellar masses based on NIR fluxes could therefore be fraught with large systematic uncertainties.  Likewise, these evolved populations may contribute to the scatter of NIR galaxy scaling relations such as the NIR metallicity-luminosity and Tully-Fisher relations.

We compared our observational results to predictions from optically derived, best-fit SPS models \citep[based on the isochrones from Padova,][]{Marigo08,Girardi10}.  We confirmed the \citet{Girardi10} result that SPS models based on the 2008 Padova isochrones over-produce the old (low-mass) TP-AGB populations.  We also showed that SPS models based on the 2010 version of the Padova models have largely corrected the TP-AGB numbers problem; the SPS models of 65\% of our sample match the observed numbers within the measured uncertainties.  While a small systematic in the numbers remains ($N_{model}/N_{data} = 1.5 \pm0.5$, or $1.4\pm0.5$ after removing outliers) a larger bias exists in the predicted  $F160W$ fluxes from TP-AGB populations, with  $L_{model}/L_{data} \sim 2.3\pm0.8$ (or $2.2 \pm 0.8$ after removing outliers). 

This model/data discrepancy in TP-AGB flux is primarily driven by predictions of very luminous (and young) TP-AGB stars that are largely absent from the NIR data.  We show that uncertainties in the measured SFHs are unlikely to account for the bulk of this over-prediction. Part of the discrepancy is likely due to the fact that galaxy simulations in this paper do not take into account the TP-AGB pulse cycle luminosity variations and, more importantly, obscuration of the optical and NIR due to dusty circumstellar envelopes. Since both aspects are included in the original datasets of evolutionary tracks and isochrones, their effect will be explored in the next paper on this series.

This work also suggests that the typical uncertainties of $<50$~\% in the lifetimes of TP-AGB models, which seem to have been reached by present-day calibrations based on Magellanic Cloud data, may still not be good enough for the accurate modeling of distance galaxies and the derivation of their stellar masses, because of the possible large contribution of a few bright TP-AGB stars to the integrated light. If our next goal is to reduce typical errors in the lifetimes and fluxes of TP-AGB populations by a factor of at least two, for a wide enough range of stellar masses and metallicities, it is clear that such a goal cannot be reached using Magellanic Cloud data only. It requires large samples of nearby galaxies with well-constrained SFHs as those presented in, e.g., Gullieuszik et al. (2008b), Melbourne et al. (2010), Girardi et al. (2010), and in this paper.

We also examine the predictions of the best-fit SPS models for RHeB stars.  In general they predict fewer, fainter, and  redder RHeB stars than are  observed in our galaxies.  As a result the models typically under-predict the flux contribution of RHeBs by a factor of $2.0 \pm 0.6$  at $F160W$, for galaxies with significant RHeB populations.  As with the TP-AGB stars, issues with the measured SFHs are unlikely to account for all of the data/model discrepancies on the RHeB branch. However, lingering issues with how the SPS models populate the CMDs or with the stellar evolution isochrones themselves could resolve these differences.   

If we take our model/data differences at face value, we can make approximate predictions for how the future models might behave when used to study more distant galaxies.  For instance current AGB-heavy models predict higher rest-frame NIR-to-optical flux ratios than our data would support, a discrepancy has been seen in several high-$z$ SED fitting programs \citep[e.g.][]{Muzzin09,Kriek10}. These results will be explored further in Rosenfield et al. (in preparation), and, if needed, any updates to the Padova isochrones will be included at that time.           

\acknowledgments
Authors BFW, JD, PR, \& AD were  partially supported by HST NASA grant GO-11719 from the Space Telescope Science Institute, which is operated by the Association of Universities for Research in Astronomy, Incorporated, under NASA contract NAS5-26555. L.G. and P.M. thanks the support from contract ASI-INAF I/009/10/0

\bibliographystyle{apj}
\bibliography{/Users/jmel/bib/bigbib2}

\begin{thebibliography}{78}
\expandafter\ifx\csname natexlab\endcsname\relax\def\natexlab#1{#1}\fi

\bibitem[{REV(????)}]{REVTEX41Control}
 ????

\bibitem[{08(1)}]{apsrev41Control}
08. 1

\bibitem[{{Aringer} {et~al.}(2008){Aringer}, {Nowotny}, \&
  {H{\"o}fner}}]{Aringer08}
{Aringer}, B., {Nowotny}, W., \& {H{\"o}fner}, S. 2008, in EAS Publications
  Series, Vol.~28, EAS Publications Series, ed. {S.~Wolf, F.~Allard, \&
  P.~Stee}, 67--74

\bibitem[{{Blum} {et~al.}(2006){Blum}, {Mould}, {Olsen}, {Frogel}, {Werner},
  {Meixner}, {Markwick-Kemper}, {Indebetouw}, {Whitney}, {Meade}, {Babler},
  {Churchwell}, {Gordon}, {Engelbracht}, {For}, {Misselt}, {Vijh}, {Leitherer},
  {Volk}, {Points}, {Reach}, {Hora}, {Bernard}, {Boulanger}, {Bracker},
  {Cohen}, {Fukui}, {Gallagher}, {Gorjian}, {Harris}, {Kelly}, {Kawamura},
  {Latter}, {Madden}, {Mizuno}, {Mizuno}, {Nota}, {Oey}, {Onishi}, {Paladini},
  {Panagia}, {Perez-Gonzalez}, {Shibai}, {Sato}, {Smith}, {Staveley-Smith},
  {Tielens}, {Ueta}, {Van Dyk}, \& {Zaritsky}}]{Blum06}
{Blum}, R.~D., {Mould}, J.~R., {Olsen}, K.~A., {Frogel}, J.~A., {Werner}, M.,
  {Meixner}, M., {Markwick-Kemper}, F., {Indebetouw}, R., {Whitney}, B.,
  {Meade}, M., {Babler}, B., {Churchwell}, E.~B., {Gordon}, K., {Engelbracht},
  C., {For}, B., {Misselt}, K., {Vijh}, U., {Leitherer}, C., {Volk}, K.,
  {Points}, S., {Reach}, W., {Hora}, J.~L., {Bernard}, J., {Boulanger}, F.,
  {Bracker}, S., {Cohen}, M., {Fukui}, Y., {Gallagher}, J., {Gorjian}, V.,
  {Harris}, J., {Kelly}, D., {Kawamura}, A., {Latter}, W.~B., {Madden}, S.,
  {Mizuno}, A., {Mizuno}, N., {Nota}, A., {Oey}, M.~S., {Onishi}, T.,
  {Paladini}, R., {Panagia}, N., {Perez-Gonzalez}, P., {Shibai}, H., {Sato},
  S., {Smith}, L., {Staveley-Smith}, L., {Tielens}, A.~G.~G.~M., {Ueta}, T.,
  {Van Dyk}, S., \& {Zaritsky}, D. 2006, \aj, 132, 2034

\bibitem[{{Bohlin}(2007)}]{Bohlin07}
{Bohlin}, R.~C. 2007, in Astronomical Society of the Pacific Conference Series,
  Vol. 364, The Future of Photometric, Spectrophotometric and Polarimetric
  Standardization, ed. {C.~Sterken}, 315

\bibitem[{{Bonanos} {et~al.}(2009){Bonanos}, {Massa}, {Sewilo}, {Lennon},
  {Panagia}, {Smith}, {Meixner}, {Babler}, {Bracker}, {Meade}, {Gordon},
  {Hora}, {Indebetouw}, \& {Whitney}}]{Bonanos09}
{Bonanos}, A.~Z., {Massa}, D.~L., {Sewilo}, M., {Lennon}, D.~J., {Panagia}, N.,
  {Smith}, L.~J., {Meixner}, M., {Babler}, B.~L., {Bracker}, S., {Meade},
  M.~R., {Gordon}, K.~D., {Hora}, J.~L., {Indebetouw}, R., \& {Whitney}, B.~A.
  2009, \aj, 138, 1003

\bibitem[{{Boyer} {et~al.}(2009){Boyer}, {Skillman}, {van Loon}, {Gehrz}, \&
  {Woodward}}]{Boyer09}
{Boyer}, M.~L., {Skillman}, E.~D., {van Loon}, J.~{\relax Th}., {Gehrz}, R.~D.,
  \& {Woodward}, C.~E. 2009, \apj, 697, 1993

\bibitem[{{Boyer} {et~al.}(2011){Boyer}, {Srinivasan}, {van Loon}, {McDonald},
  {Meixner}, {Zaritsky}, {Gordon}, {Kemper}, {Babler}, {Block}, {Bracker},
  {Engelbracht}, {Hora}, {Indebetouw}, {Meade}, {Misselt}, {Robitaille},
  {Sewi{\l}o}, {Shiao}, \& {Whitney}}]{Boyer11}
{Boyer}, M.~L., {Srinivasan}, S., {van Loon}, J.~{\relax Th}., {McDonald}, I.,
  {Meixner}, M., {Zaritsky}, D., {Gordon}, K.~D., {Kemper}, F., {Babler}, B.,
  {Block}, M., {Bracker}, S., {Engelbracht}, C.~W., {Hora}, J., {Indebetouw},
  R., {Meade}, M., {Misselt}, K., {Robitaille}, T., {Sewi{\l}o}, M., {Shiao},
  B., \& {Whitney}, B. 2011, \aj, 142, 103

\bibitem[{{Bressan} {et~al.}(1994){Bressan}, {Chiosi}, \&
  {Fagotto}}]{Bressan94}
{Bressan}, A., {Chiosi}, C., \& {Fagotto}, F. 1994, \apjs, 94, 63

\bibitem[{{Bruzual} \& {Charlot}(2003)}]{BC03}
{Bruzual}, G., \& {Charlot}, S. 2003, \mnras, 344, 1000

\bibitem[{{Bruzual A}(2007)}]{Bruzual07}
{Bruzual A}, G. 2007, ArXiv Astrophysics e-prints

\bibitem[{{Bundy} {et~al.}(2005){Bundy}, {Ellis}, \& {Conselice}}]{Bundy05}
{Bundy}, K., {Ellis}, R.~S., \& {Conselice}, C.~J. 2005, \apj, 625, 621

\bibitem[{{Charlot} \& {Bruzual}(1991)}]{Charlot91}
{Charlot}, S., \& {Bruzual}, A.~G. 1991, \apj, 367, 126

\bibitem[{{Cioni} {et~al.}(2006){Cioni}, {Girardi}, {Marigo}, \&
  {Habing}}]{Cioni06}
{Cioni}, M., {Girardi}, L., {Marigo}, P., \& {Habing}, H.~J. 2006, \aap, 448,
  77

\bibitem[{{Cioni} {et~al.}(1999){Cioni}, {Habing}, {Loup}, {Epchtein}, \& {The
  Denis Consortium}}]{Cioni99}
{Cioni}, M.~R., {Habing}, H.~J., {Loup}, C., {Epchtein}, N., \& {The Denis
  Consortium}. 1999, in IAU Symposium, Vol. 190, New Views of the Magellanic
  Clouds, ed. Y.-H. {Chu}, N.~{Suntzeff}, J.~{Hesser}, \& D.~{Bohlender},
  385--+

\bibitem[{{Conroy} {et~al.}(2009){Conroy}, {Gunn}, \& {White}}]{Conroy09}
{Conroy}, C., {Gunn}, J.~E., \& {White}, M. 2009, \apj, 699, 486

\bibitem[{{Conselice} {et~al.}(2005){Conselice}, {Bundy}, {Ellis}, {Brichmann},
  {Vogt}, \& {Phillips}}]{Conselice05}
{Conselice}, C.~J., {Bundy}, K., {Ellis}, R.~S., {Brichmann}, J., {Vogt},
  N.~P., \& {Phillips}, A.~C. 2005, \apj, 628, 160

\bibitem[{{Dalcanton} {et~al.}(2012){Dalcanton}, {Williams}, {Melbourne},
  {Girardi}, {Dolphin}, {Rosenfield}, {Boyer}, {de Jong}, {Gilbert}, {Marigo},
  {Olsen}, {Seth}, \& {Skillman}}]{Dalcanton12}
{Dalcanton}, J.~J., {Williams}, B.~F., {Melbourne}, J.~L., {Girardi}, L.,
  {Dolphin}, A., {Rosenfield}, P.~A., {Boyer}, M.~L., {de Jong}, R.~S.,
  {Gilbert}, K., {Marigo}, P., {Olsen}, K., {Seth}, A.~C., \& {Skillman}, E.
  2012, \apjs, 198, 6

\bibitem[{{Dalcanton} {et~al.}(2009){Dalcanton}, {Williams}, {Seth}, {Dolphin},
  {Holtzman}, {Rosema}, {Skillman}, {Cole}, {Girardi}, {Gogarten},
  {Karachentsev}, {Olsen}, {Weisz}, {Christensen}, {Freeman}, {Gilbert},
  {Gallart}, {Harris}, {Hodge}, {de Jong}, {Karachentseva}, {Mateo}, {Stetson},
  {Tavarez}, {Zaritsky}, {Governato}, \& {Quinn}}]{Dalcanton09}
{Dalcanton}, J.~J., {Williams}, B.~F., {Seth}, A.~C., {Dolphin}, A.,
  {Holtzman}, J., {Rosema}, K., {Skillman}, E.~D., {Cole}, A., {Girardi}, L.,
  {Gogarten}, S.~M., {Karachentsev}, I.~D., {Olsen}, K., {Weisz}, D.,
  {Christensen}, C., {Freeman}, K., {Gilbert}, K., {Gallart}, C., {Harris}, J.,
  {Hodge}, P., {de Jong}, R.~S., {Karachentseva}, V., {Mateo}, M., {Stetson},
  P.~B., {Tavarez}, M., {Zaritsky}, D., {Governato}, F., \& {Quinn}, T. 2009,
  \apjs, 183, 67

\bibitem[{{Davidge}(2010)}]{Davidge10}
{Davidge}, T.~J. 2010, \apj, 718, 1428

\bibitem[{{Davis} {et~al.}(2007){Davis}, {Guhathakurta}, {Konidaris}, {Newman},
  {Ashby}, {Biggs}, {Barmby}, {Bundy}, {Chapman}, {Coil}, {Conselice},
  {Cooper}, {Croton}, {Eisenhardt}, {Ellis}, {Faber}, {Fang}, {Fazio},
  {Georgakakis}, {Gerke}, {Goss}, {Gwyn}, {Harker}, {Hopkins}, {Huang},
  {Ivison}, {Kassin}, {Kirby}, {Koekemoer}, {Koo}, {Laird}, {Le Floc'h}, {Lin},
  {Lotz}, {Marshall}, {Martin}, {Metevier}, {Moustakas}, {Nandra}, {Noeske},
  {Papovich}, {Phillips}, {Rich}, {Rieke}, {Rigopoulou}, {Salim},
  {Schiminovich}, {Simard}, {Smail}, {Small}, {Weiner}, {Willmer}, {Willner},
  {Wilson}, {Wright}, \& {Yan}}]{Davis07}
{Davis}, M., {Guhathakurta}, P., {Konidaris}, N.~P., {Newman}, J.~A., {Ashby},
  M.~L.~N., {Biggs}, A.~D., {Barmby}, P., {Bundy}, K., {Chapman}, S.~C.,
  {Coil}, A.~L., {Conselice}, C.~J., {Cooper}, M.~C., {Croton}, D.~J.,
  {Eisenhardt}, P.~R.~M., {Ellis}, R.~S., {Faber}, S.~M., {Fang}, T., {Fazio},
  G.~G., {Georgakakis}, A., {Gerke}, B.~F., {Goss}, W.~M., {Gwyn}, S.,
  {Harker}, J., {Hopkins}, A.~M., {Huang}, J., {Ivison}, R.~J., {Kassin},
  S.~A., {Kirby}, E.~N., {Koekemoer}, A.~M., {Koo}, D.~C., {Laird}, E.~S., {Le
  Floc'h}, E., {Lin}, L., {Lotz}, J.~M., {Marshall}, P.~J., {Martin}, D.~C.,
  {Metevier}, A.~J., {Moustakas}, L.~A., {Nandra}, K., {Noeske}, K.~G.,
  {Papovich}, C., {Phillips}, A.~C., {Rich}, R.~M., {Rieke}, G.~H.,
  {Rigopoulou}, D., {Salim}, S., {Schiminovich}, D., {Simard}, L., {Smail}, I.,
  {Small}, T.~A., {Weiner}, B.~J., {Willmer}, C.~N.~A., {Willner}, S.~P.,
  {Wilson}, G., {Wright}, E.~L., \& {Yan}, R. 2007, \apjl, 660, L1

\bibitem[{{Dohm-Palmer} \& {Skillman}(2002)}]{Dohm-Palmer02}
{Dohm-Palmer}, R.~C., \& {Skillman}, E.~D. 2002, \aj, 123, 1433

\bibitem[{{Dolphin}(2000)}]{Dolphin00}
{Dolphin}, A.~E. 2000, \pasp, 112, 1383

\bibitem[{{Dolphin}(2002)}]{Dolphin02}
---. 2002, \mnras, 332, 91

\bibitem[{{Dolphin} {et~al.}(2005){Dolphin}, {Weisz}, {Skillman}, \&
  {Holtzman}}]{Dolphin05}
{Dolphin}, A.~E., {Weisz}, D.~R., {Skillman}, E.~D., \& {Holtzman}, J.~A. 2005,
  ArXiv Astrophysics e-prints

\bibitem[{{Erb} {et~al.}(2006){Erb}, {Shapley}, {Pettini}, {Steidel}, {Reddy},
  \& {Adelberger}}]{Erb06a}
{Erb}, D.~K., {Shapley}, A.~E., {Pettini}, M., {Steidel}, C.~C., {Reddy},
  N.~A., \& {Adelberger}, K.~L. 2006, \apj, 644, 813

\bibitem[{{Fontana} {et~al.}(2006){Fontana}, {Salimbeni}, {Grazian},
  {Giallongo}, {Pentericci}, {Nonino}, {Fontanot}, {Menci}, {Monaco},
  {Cristiani}, {Vanzella}, {de Santis}, \& {Gallozzi}}]{Fontana06}
{Fontana}, A., {Salimbeni}, S., {Grazian}, A., {Giallongo}, E., {Pentericci},
  L., {Nonino}, M., {Fontanot}, F., {Menci}, N., {Monaco}, P., {Cristiani}, S.,
  {Vanzella}, E., {de Santis}, C., \& {Gallozzi}, S. 2006, \aap, 459, 745

\bibitem[{{Frogel} {et~al.}(1990){Frogel}, {Mould}, \& {Blanco}}]{Frogel90}
{Frogel}, J.~A., {Mould}, J., \& {Blanco}, V.~M. 1990, \apj, 352, 96

\bibitem[{{Gallart} {et~al.}(2008){Gallart}, {Stetson}, {Meschin}, {Pont}, \&
  {Hardy}}]{Gallart08}
{Gallart}, C., {Stetson}, P.~B., {Meschin}, I.~P., {Pont}, F., \& {Hardy}, E.
  2008, \apjl, 682, L89

\bibitem[{{Giavalisco} {et~al.}(2004){Giavalisco}, {Ferguson}, {Koekemoer},
  {Dickinson}, {Alexander}, {Bauer}, {Bergeron}, {Biagetti}, {Brandt},
  {Casertano}, {Cesarsky}, {Chatzichristou}, {Conselice}, {Cristiani}, {Da
  Costa}, {Dahlen}, {de Mello}, {Eisenhardt}, {Erben}, {Fall}, {Fassnacht},
  {Fosbury}, {Fruchter}, {Gardner}, {Grogin}, {Hook}, {Hornschemeier}, {Idzi},
  {Jogee}, {Kretchmer}, {Laidler}, {Lee}, {Livio}, {Lucas}, {Madau},
  {Mobasher}, {Moustakas}, {Nonino}, {Padovani}, {Papovich}, {Park},
  {Ravindranath}, {Renzini}, {Richardson}, {Riess}, {Rosati}, {Schirmer},
  {Schreier}, {Somerville}, {Spinrad}, {Stern}, {Stiavelli}, {Strolger},
  {Urry}, {Vandame}, {Williams}, \& {Wolf}}]{Giavalisco04}
{Giavalisco}, M., {Ferguson}, H.~C., {Koekemoer}, A.~M., {Dickinson}, M.,
  {Alexander}, D.~M., {Bauer}, F.~E., {Bergeron}, J., {Biagetti}, C., {Brandt},
  W.~N., {Casertano}, S., {Cesarsky}, C., {Chatzichristou}, E., {Conselice},
  C., {Cristiani}, S., {Da Costa}, L., {Dahlen}, T., {de Mello}, D.,
  {Eisenhardt}, P., {Erben}, T., {Fall}, S.~M., {Fassnacht}, C., {Fosbury}, R.,
  {Fruchter}, A., {Gardner}, J.~P., {Grogin}, N., {Hook}, R.~N.,
  {Hornschemeier}, A.~E., {Idzi}, R., {Jogee}, S., {Kretchmer}, C., {Laidler},
  V., {Lee}, K.~S., {Livio}, M., {Lucas}, R., {Madau}, P., {Mobasher}, B.,
  {Moustakas}, L.~A., {Nonino}, M., {Padovani}, P., {Papovich}, C., {Park}, Y.,
  {Ravindranath}, S., {Renzini}, A., {Richardson}, M., {Riess}, A., {Rosati},
  P., {Schirmer}, M., {Schreier}, E., {Somerville}, R.~S., {Spinrad}, H.,
  {Stern}, D., {Stiavelli}, M., {Strolger}, L., {Urry}, C.~M., {Vandame}, B.,
  {Williams}, R., \& {Wolf}, C. 2004, \apjl, 600, L93

\bibitem[{{Girardi} {et~al.}(2002){Girardi}, {Bertelli}, {Bressan}, {Chiosi},
  {Groenewegen}, {Marigo}, {Salasnich}, \& {Weiss}}]{Girardi02}
{Girardi}, L., {Bertelli}, G., {Bressan}, A., {Chiosi}, C., {Groenewegen},
  M.~A.~T., {Marigo}, P., {Salasnich}, B., \& {Weiss}, A. 2002, \aap, 391, 195

\bibitem[{{Girardi} {et~al.}(2000){Girardi}, {Bressan}, {Bertelli}, \&
  {Chiosi}}]{Girardi00}
{Girardi}, L., {Bressan}, A., {Bertelli}, G., \& {Chiosi}, C. 2000, \aaps, 141,
  371

\bibitem[{{Girardi} {et~al.}(2008){Girardi}, {Dalcanton}, {Williams}, {de
  Jong}, {Gallart}, {Monelli}, {Groenewegen}, {Holtzman}, {Olsen}, {Seth},
  {Weisz}, \& {the ANGST/ANGRRR Collaboration}}]{Girardi08}
{Girardi}, L., {Dalcanton}, J., {Williams}, B., {de Jong}, R., {Gallart}, C.,
  {Monelli}, M., {Groenewegen}, M.~A.~T., {Holtzman}, J.~A., {Olsen}, K.~A.~G.,
  {Seth}, A.~C., {Weisz}, D.~R., \& {the ANGST/ANGRRR Collaboration}. 2008,
  \pasp, 120, 583

\bibitem[{{Girardi} {et~al.}(2005){Girardi}, {Groenewegen}, {Hatziminaoglou},
  \& {da Costa}}]{Girardi05}
{Girardi}, L., {Groenewegen}, M.~A.~T., {Hatziminaoglou}, E., \& {da Costa}, L.
  2005, \aap, 436, 895

\bibitem[{{Girardi} \& {Marigo}(2007)}]{Girardi07}
{Girardi}, L., \& {Marigo}, P. 2007, \aap, 462, 237

\bibitem[{{Girardi} {et~al.}(2010){Girardi}, {Williams}, {Gilbert},
  {Rosenfield}, {Dalcanton}, {Marigo}, {Boyer}, {Dolphin}, {Weisz},
  {Melbourne}, {Olsen}, {Seth}, \& {Skillman}}]{Girardi10}
{Girardi}, L., {Williams}, B.~F., {Gilbert}, K.~M., {Rosenfield}, P.,
  {Dalcanton}, J.~J., {Marigo}, P., {Boyer}, M.~L., {Dolphin}, A., {Weisz},
  D.~R., {Melbourne}, J., {Olsen}, K.~A.~G., {Seth}, A.~C., \& {Skillman}, E.
  2010, \apj, 724, 1030

\bibitem[{{Gullieuszik} {et~al.}(2008{\natexlab{a}}){Gullieuszik}, {Greggio},
  {Held}, {Moretti}, {Arcidiacono}, {Bagnara}, {Baruffolo}, {Diolaiti},
  {Falomo}, {Farinato}, {Lombini}, {Ragazzoni}, {Brast}, {Donaldson}, {Kolb},
  {Marchetti}, \& {Tordo}}]{Gullieuszik08a}
{Gullieuszik}, M., {Greggio}, L., {Held}, E.~V., {Moretti}, A., {Arcidiacono},
  C., {Bagnara}, P., {Baruffolo}, A., {Diolaiti}, E., {Falomo}, R., {Farinato},
  J., {Lombini}, M., {Ragazzoni}, R., {Brast}, R., {Donaldson}, R., {Kolb}, J.,
  {Marchetti}, E., \& {Tordo}, S. 2008{\natexlab{a}}, \aap, 483, L5

\bibitem[{{Gullieuszik} {et~al.}(2008{\natexlab{b}}){Gullieuszik}, {Held},
  {Rizzi}, {Girardi}, {Marigo}, \& {Momany}}]{Gullieuszik08}
{Gullieuszik}, M., {Held}, E.~V., {Rizzi}, L., {Girardi}, L., {Marigo}, P., \&
  {Momany}, Y. 2008{\natexlab{b}}, \mnras, 388, 1185

\bibitem[{{Holtzman} {et~al.}(2006){Holtzman}, {Afonso}, \&
  {Dolphin}}]{Holtzman06}
{Holtzman}, J.~A., {Afonso}, C., \& {Dolphin}, A. 2006, \apjs, 166, 534

\bibitem[{{Iben} \& {Renzini}(1983)}]{Iben83}
{Iben}, Jr., I., \& {Renzini}, A. 1983, \araa, 21, 271

\bibitem[{{Ilbert} {et~al.}(2010){Ilbert}, {Salvato}, {Le Floc'h}, {Aussel},
  {Capak}, {McCracken}, {Mobasher}, {Kartaltepe}, {Scoville}, {Sanders},
  {Arnouts}, {Bundy}, {Cassata}, {Kneib}, {Koekemoer}, {Le F{\`e}vre}, {Lilly},
  {Surace}, {Taniguchi}, {Tasca}, {Thompson}, {Tresse}, {Zamojski}, {Zamorani},
  \& {Zucca}}]{Ilbert10}
{Ilbert}, O., {Salvato}, M., {Le Floc'h}, E., {Aussel}, H., {Capak}, P.,
  {McCracken}, H.~J., {Mobasher}, B., {Kartaltepe}, J., {Scoville}, N.,
  {Sanders}, D.~B., {Arnouts}, S., {Bundy}, K., {Cassata}, P., {Kneib}, J.,
  {Koekemoer}, A., {Le F{\`e}vre}, O., {Lilly}, S., {Surace}, J., {Taniguchi},
  Y., {Tasca}, L., {Thompson}, D., {Tresse}, L., {Zamojski}, M., {Zamorani},
  G., \& {Zucca}, E. 2010, \apj, 709, 644

\bibitem[{{Jerjen} \& {Rejkuba}(2001)}]{Jerjen01}
{Jerjen}, H., \& {Rejkuba}, M. 2001, \aap, 371, 487

\bibitem[{{Kennicutt} {et~al.}(1994){Kennicutt}, {Tamblyn}, \&
  {Congdon}}]{Kennicutt94}
{Kennicutt}, Jr., R.~C., {Tamblyn}, P., \& {Congdon}, C.~E. 1994, \apj, 435, 22

\bibitem[{{Kriek} {et~al.}(2010){Kriek}, {Labb{\'e}}, {Conroy}, {Whitaker},
  {van Dokkum}, {Brammer}, {Franx}, {Illingworth}, {Marchesini}, {Muzzin},
  {Quadri}, \& {Rudnick}}]{Kriek10}
{Kriek}, M., {Labb{\'e}}, I., {Conroy}, C., {Whitaker}, K.~E., {van Dokkum},
  P.~G., {Brammer}, G.~B., {Franx}, M., {Illingworth}, G.~D., {Marchesini}, D.,
  {Muzzin}, A., {Quadri}, R.~F., \& {Rudnick}, G. 2010, \apjl, 722, L64

\bibitem[{{Kroupa}(2001)}]{Kroupa01}
{Kroupa}, P. 2001, \mnras, 322, 231

\bibitem[{{Laskar} {et~al.}(2011){Laskar}, {Berger}, \& {Chary}}]{Laskar11}
{Laskar}, T., {Berger}, E., \& {Chary}, R.-R. 2011, \apj, 739, 1

\bibitem[{{Mannucci} {et~al.}(2009){Mannucci}, {Cresci}, {Maiolino}, {Marconi},
  {Pastorini}, {Pozzetti}, {Gnerucci}, {Risaliti}, {Schneider}, {Lehnert}, \&
  {Salvati}}]{Mannucci09}
{Mannucci}, F., {Cresci}, G., {Maiolino}, R., {Marconi}, A., {Pastorini}, G.,
  {Pozzetti}, L., {Gnerucci}, A., {Risaliti}, G., {Schneider}, R., {Lehnert},
  M., \& {Salvati}, M. 2009, \mnras, 398, 1915

\bibitem[{{Maraston}(2005)}]{Maraston05}
{Maraston}, C. 2005, \mnras, 362, 799

\bibitem[{{Maraston} {et~al.}(2006){Maraston}, {Daddi}, {Renzini}, {Cimatti},
  {Dickinson}, {Papovich}, {Pasquali}, \& {Pirzkal}}]{Maraston06}
{Maraston}, C., {Daddi}, E., {Renzini}, A., {Cimatti}, A., {Dickinson}, M.,
  {Papovich}, C., {Pasquali}, A., \& {Pirzkal}, N. 2006, \apj, 652, 85

\bibitem[{{Marigo} \& {Girardi}(2007)}]{Marigo07}
{Marigo}, P., \& {Girardi}, L. 2007, \aap, 469, 239

\bibitem[{{Marigo} {et~al.}(2008){Marigo}, {Girardi}, {Bressan}, {Groenewegen},
  {Silva}, \& {Granato}}]{Marigo08}
{Marigo}, P., {Girardi}, L., {Bressan}, A., {Groenewegen}, M.~A.~T., {Silva},
  L., \& {Granato}, G.~L. 2008, \aap, 482, 883

\bibitem[{{McQuinn} {et~al.}(2011){McQuinn}, {Skillman}, {Dalcanton},
  {Dolphin}, {Holtzman}, {Weisz}, \& {Williams}}]{McQuinn11}
{McQuinn}, K.~B.~W., {Skillman}, E.~D., {Dalcanton}, J.~J., {Dolphin}, A.~E.,
  {Holtzman}, J., {Weisz}, D.~R., \& {Williams}, B.~F. 2011, \apj, 740, 48

\bibitem[{{Melbourne} {et~al.}(2010){Melbourne}, {Williams}, {Dalcanton},
  {Ammons}, {Max}, {Koo}, {Girardi}, \& {Dolphin}}]{Melbourne10}
{Melbourne}, J., {Williams}, B., {Dalcanton}, J., {Ammons}, S.~M., {Max}, C.,
  {Koo}, D.~C., {Girardi}, L., \& {Dolphin}, A. 2010, \apj, 712, 469

\bibitem[{{Moustakas} {et~al.}(2011){Moustakas}, {Zaritsky}, {Brown}, {Cool},
  {Dey}, {Eisenstein}, {Gonzalez}, {Jannuzi}, {Jones}, {Kochanek}, {Murray}, \&
  {Wild}}]{Moustakas11}
{Moustakas}, J., {Zaritsky}, D., {Brown}, M., {Cool}, R., {Dey}, A.,
  {Eisenstein}, D.~J., {Gonzalez}, A.~H., {Jannuzi}, B., {Jones}, C.,
  {Kochanek}, C.~S., {Murray}, S.~S., \& {Wild}, V. 2011, ArXiv e-prints

\bibitem[{{Muzzin} {et~al.}(2009){Muzzin}, {van Dokkum}, {Franx}, {Marchesini},
  {Kriek}, \& {Labb{\'e}}}]{Muzzin09}
{Muzzin}, A., {van Dokkum}, P., {Franx}, M., {Marchesini}, D., {Kriek}, M., \&
  {Labb{\'e}}, I. 2009, \apjl, 706, L188

\bibitem[{{Nikolaev} \& {Weinberg}(2000)}]{Nikolaev00}
{Nikolaev}, S., \& {Weinberg}, M.~D. 2000, \apj, 542, 804

\bibitem[{{Onodera} {et~al.}(2010){Onodera}, {Arimoto}, {Daddi}, {Renzini},
  {Kong}, {Cimatti}, {Broadhurst}, \& {Alexander}}]{Onodera10}
{Onodera}, M., {Arimoto}, N., {Daddi}, E., {Renzini}, A., {Kong}, X.,
  {Cimatti}, A., {Broadhurst}, T., \& {Alexander}, D.~M. 2010, \apj, 715, 385

\bibitem[{{Persson} {et~al.}(1983){Persson}, {Aaronson}, {Cohen}, {Frogel}, \&
  {Matthews}}]{Persson83}
{Persson}, S.~E., {Aaronson}, M., {Cohen}, J.~G., {Frogel}, J.~A., \&
  {Matthews}, K. 1983, \apj, 266, 105

\bibitem[{{Pozzetti} {et~al.}(2010){Pozzetti}, {Bolzonella}, {Zucca},
  {Zamorani}, {Lilly}, {Renzini}, {Moresco}, {Mignoli}, {Cassata}, {Tasca},
  {Lamareille}, {Maier}, {Meneux}, {Halliday}, {Oesch}, {Vergani}, {Caputi},
  {Kova{\v c}}, {Cimatti}, {Cucciati}, {Iovino}, {Peng}, {Carollo}, {Contini},
  {Kneib}, {Le F{\'e}vre}, {Mainieri}, {Scodeggio}, {Bardelli}, {Bongiorno},
  {Coppa}, {de la Torre}, {de Ravel}, {Franzetti}, {Garilli}, {Kampczyk},
  {Knobel}, {Le Borgne}, {Le Brun}, {Pell{\`o}}, {Perez Montero},
  {Ricciardelli}, {Silverman}, {Tanaka}, {Tresse}, {Abbas}, {Bottini}, {Cappi},
  {Guzzo}, {Koekemoer}, {Leauthaud}, {Maccagni}, {Marinoni}, {McCracken},
  {Memeo}, {Porciani}, {Scaramella}, {Scarlata}, \& {Scoville}}]{Pozzetti10}
{Pozzetti}, L., {Bolzonella}, M., {Zucca}, E., {Zamorani}, G., {Lilly}, S.,
  {Renzini}, A., {Moresco}, M., {Mignoli}, M., {Cassata}, P., {Tasca}, L.,
  {Lamareille}, F., {Maier}, C., {Meneux}, B., {Halliday}, C., {Oesch}, P.,
  {Vergani}, D., {Caputi}, K., {Kova{\v c}}, K., {Cimatti}, A., {Cucciati}, O.,
  {Iovino}, A., {Peng}, Y., {Carollo}, M., {Contini}, T., {Kneib}, J., {Le
  F{\'e}vre}, O., {Mainieri}, V., {Scodeggio}, M., {Bardelli}, S., {Bongiorno},
  A., {Coppa}, G., {de la Torre}, S., {de Ravel}, L., {Franzetti}, P.,
  {Garilli}, B., {Kampczyk}, P., {Knobel}, C., {Le Borgne}, J., {Le Brun}, V.,
  {Pell{\`o}}, R., {Perez Montero}, E., {Ricciardelli}, E., {Silverman}, J.~D.,
  {Tanaka}, M., {Tresse}, L., {Abbas}, U., {Bottini}, D., {Cappi}, A., {Guzzo},
  L., {Koekemoer}, A.~M., {Leauthaud}, A., {Maccagni}, D., {Marinoni}, C.,
  {McCracken}, H.~J., {Memeo}, P., {Porciani}, C., {Scaramella}, R.,
  {Scarlata}, C., \& {Scoville}, N. 2010, \aap, 523, A13+

\bibitem[{{Raimondo} {et~al.}(2005){Raimondo}, {Brocato}, {Cantiello}, \&
  {Capaccioli}}]{Raimondo05}
{Raimondo}, G., {Brocato}, E., {Cantiello}, M., \& {Capaccioli}, M. 2005, \aj,
  130, 2625

\bibitem[{{Salzer} {et~al.}(2005){Salzer}, {Lee}, {Melbourne}, {Hinz},
  {Alonso-Herrero}, \& {Jangren}}]{Salzer05}
{Salzer}, J.~J., {Lee}, J.~C., {Melbourne}, J., {Hinz}, J.~L.,
  {Alonso-Herrero}, A., \& {Jangren}, A. 2005, \apj, 624, 661

\bibitem[{{Sanders} {et~al.}(2007){Sanders}, {Salvato}, {Aussel}, {Ilbert},
  {Scoville}, {Surace}, {Frayer}, {Sheth}, {Helou}, {Brooke}, {Bhattacharya},
  {Yan}, {Kartaltepe}, {Barnes}, {Blain}, {Calzetti}, {Capak}, {Carilli},
  {Carollo}, {Comastri}, {Daddi}, {Ellis}, {Elvis}, {Fall}, {Franceschini},
  {Giavalisco}, {Hasinger}, {Impey}, {Koekemoer}, {Le F{\`e}vre}, {Lilly},
  {Liu}, {McCracken}, {Mobasher}, {Renzini}, {Rich}, {Schinnerer}, {Shopbell},
  {Taniguchi}, {Thompson}, {Urry}, \& {Williams}}]{Sanders07}
{Sanders}, D.~B., {Salvato}, M., {Aussel}, H., {Ilbert}, O., {Scoville}, N.,
  {Surace}, J.~A., {Frayer}, D.~T., {Sheth}, K., {Helou}, G., {Brooke}, T.,
  {Bhattacharya}, B., {Yan}, L., {Kartaltepe}, J.~S., {Barnes}, J.~E., {Blain},
  A.~W., {Calzetti}, D., {Capak}, P., {Carilli}, C., {Carollo}, C.~M.,
  {Comastri}, A., {Daddi}, E., {Ellis}, R.~S., {Elvis}, M., {Fall}, S.~M.,
  {Franceschini}, A., {Giavalisco}, M., {Hasinger}, G., {Impey}, C.,
  {Koekemoer}, A., {Le F{\`e}vre}, O., {Lilly}, S., {Liu}, M.~C., {McCracken},
  H.~J., {Mobasher}, B., {Renzini}, A., {Rich}, M., {Schinnerer}, E.,
  {Shopbell}, P.~L., {Taniguchi}, Y., {Thompson}, D.~J., {Urry}, C.~M., \&
  {Williams}, J.~P. 2007, \apjs, 172, 86

\bibitem[{{Sommariva} {et~al.}(2011){Sommariva}, {Mannucci}, {Cresci},
  {Maiolino}, {Marconi}, {Nagao}, {Baroni}, \& {Grazian}}]{Sommariva11}
{Sommariva}, V., {Mannucci}, F., {Cresci}, G., {Maiolino}, R., {Marconi}, A.,
  {Nagao}, T., {Baroni}, A., \& {Grazian}, A. 2011, ArXiv e-prints

\bibitem[{{Spergel} {et~al.}(2007){Spergel}, {Bean}, {Dor{\'e}}, {Nolta},
  {Bennett}, {Dunkley}, {Hinshaw}, {Jarosik}, {Komatsu}, {Page}, {Peiris},
  {Verde}, {Halpern}, {Hill}, {Kogut}, {Limon}, {Meyer}, {Odegard}, {Tucker},
  {Weiland}, {Wollack}, \& {Wright}}]{Spergel07}
{Spergel}, D.~N., {Bean}, R., {Dor{\'e}}, O., {Nolta}, M.~R., {Bennett}, C.~L.,
  {Dunkley}, J., {Hinshaw}, G., {Jarosik}, N., {Komatsu}, E., {Page}, L.,
  {Peiris}, H.~V., {Verde}, L., {Halpern}, M., {Hill}, R.~S., {Kogut}, A.,
  {Limon}, M., {Meyer}, S.~S., {Odegard}, N., {Tucker}, G.~S., {Weiland},
  J.~L., {Wollack}, E., \& {Wright}, E.~L. 2007, \apjs, 170, 377

\bibitem[{{Swinbank} {et~al.}(2004){Swinbank}, {Smail}, {Chapman}, {Blain},
  {Ivison}, \& {Keel}}]{Swinbank04}
{Swinbank}, A.~M., {Smail}, I., {Chapman}, S.~C., {Blain}, A.~W., {Ivison},
  R.~J., \& {Keel}, W.~C. 2004, \apj, 617, 64

\bibitem[{{Valdes} {et~al.}(1995){Valdes}, {Campusano}, {Velasquez}, \&
  {Stetson}}]{Valdes95}
{Valdes}, F.~G., {Campusano}, L.~E., {Velasquez}, J.~D., \& {Stetson}, P.~B.
  1995, \pasp, 107, 1119

\bibitem[{{Vassiliadis} \& {Wood}(1993)}]{Vassiliadis93}
{Vassiliadis}, E., \& {Wood}, P.~R. 1993, \apj, 413, 641

\bibitem[{{V{\'a}zquez} \& {Leitherer}(2005)}]{Vazquez05}
{V{\'a}zquez}, G.~A., \& {Leitherer}, C. 2005, \apj, 621, 695

\bibitem[{{Vulcani} {et~al.}(2010){Vulcani}, {Poggianti},
  {Arag{\'o}n-Salamanca}, {Fasano}, {Rudnick}, {Valentinuzzi}, {Dressler},
  {Bettoni}, {Cava}, {D'Onofrio}, {Fritz}, {Moretti}, {Omizzolo}, \&
  {Varela}}]{Vulcani10}
{Vulcani}, B., {Poggianti}, B.~M., {Arag{\'o}n-Salamanca}, A., {Fasano}, G.,
  {Rudnick}, G., {Valentinuzzi}, T., {Dressler}, A., {Bettoni}, D., {Cava}, A.,
  {D'Onofrio}, M., {Fritz}, J., {Moretti}, A., {Omizzolo}, A., \& {Varela}, J.
  2010, ArXiv e-prints

\bibitem[{{Wagenhuber} \& {Groenewegen}(1998)}]{Wagenhuber98}
{Wagenhuber}, J., \& {Groenewegen}, M.~A.~T. 1998, \aap, 340, 183

\bibitem[{{Weisz} {et~al.}(2011{\natexlab{a}}){Weisz}, {Dalcanton}, {Williams},
  {Gilbert}, {Skillman}, {Seth}, {Dolphin}, {McQuinn}, {Gogarten}, {Holtzman},
  {Rosema}, {Cole}, {Karachentsev}, \& {Zaritsky}}]{Weisz11}
{Weisz}, D.~R., {Dalcanton}, J.~J., {Williams}, B.~F., {Gilbert}, K.~M.,
  {Skillman}, E.~D., {Seth}, A.~C., {Dolphin}, A.~E., {McQuinn}, K.~B.~W.,
  {Gogarten}, S.~M., {Holtzman}, J., {Rosema}, K., {Cole}, A., {Karachentsev},
  I.~D., \& {Zaritsky}, D. 2011{\natexlab{a}}, \apj, 739, 5

\bibitem[{{Weisz} {et~al.}(2011{\natexlab{b}}){Weisz}, {Dolphin}, {Dalcanton},
  {Skillman}, {Holtzman}, {Williams}, {Gilbert}, {Seth}, {Cole}, {Gogarten},
  {Rosema}, {Karachentsev}, {McQuinn}, \& {Zaritsky}}]{Weisz11b}
{Weisz}, D.~R., {Dolphin}, A.~E., {Dalcanton}, J.~J., {Skillman}, E.~D.,
  {Holtzman}, J., {Williams}, B.~F., {Gilbert}, K.~M., {Seth}, A.~C., {Cole},
  A., {Gogarten}, S.~M., {Rosema}, K., {Karachentsev}, I.~D., {McQuinn},
  K.~B.~W., \& {Zaritsky}, D. 2011{\natexlab{b}}, \apj, 743, 8

\bibitem[{{Williams} {et~al.}(2007){Williams}, {Ciardullo}, {Durrell},
  {Vinciguerra}, {Feldmeier}, {Jacoby}, {Sigurdsson}, {von Hippel}, {Ferguson},
  {Tanvir}, {Arnaboldi}, {Gerhard}, {Aguerri}, \& {Freeman}}]{Williams07}
{Williams}, B.~F., {Ciardullo}, R., {Durrell}, P.~R., {Vinciguerra}, M.,
  {Feldmeier}, J.~J., {Jacoby}, G.~H., {Sigurdsson}, S., {von Hippel}, T.,
  {Ferguson}, H.~C., {Tanvir}, N.~R., {Arnaboldi}, M., {Gerhard}, O.,
  {Aguerri}, J.~A.~L., \& {Freeman}, K. 2007, \apj, 656, 756

\bibitem[{{Williams} {et~al.}(2009{\natexlab{a}}){Williams}, {Dalcanton},
  {Dolphin}, {Holtzman}, \& {Sarajedini}}]{Williams09b}
{Williams}, B.~F., {Dalcanton}, J.~J., {Dolphin}, A.~E., {Holtzman}, J., \&
  {Sarajedini}, A. 2009{\natexlab{a}}, \apjl, 695, L15

\bibitem[{{Williams} {et~al.}(2009{\natexlab{b}}){Williams}, {Dalcanton},
  {Seth}, {Weisz}, {Dolphin}, {Skillman}, {Harris}, {Holtzman}, {Girardi}, {de
  Jong}, {Olsen}, {Cole}, {Gallart}, {Gogarten}, {Hidalgo}, {Mateo}, {Rosema},
  {Stetson}, \& {Quinn}}]{Williams09a}
{Williams}, B.~F., {Dalcanton}, J.~J., {Seth}, A.~C., {Weisz}, D., {Dolphin},
  A., {Skillman}, E., {Harris}, J., {Holtzman}, J., {Girardi}, L., {de Jong},
  R.~S., {Olsen}, K., {Cole}, A., {Gallart}, C., {Gogarten}, S.~M., {Hidalgo},
  S.~L., {Mateo}, M., {Rosema}, K., {Stetson}, P.~B., \& {Quinn}, T.
  2009{\natexlab{b}}, \aj, 137, 419

\bibitem[{{Williams} {et~al.}(2010){Williams}, {Dalcanton}, {Stilp}, {Gilbert},
  {Ro{\v s}kar}, {Seth}, {Weisz}, {Dolphin}, {Gogarten}, {Skillman}, \&
  {Holtzman}}]{Williams10}
{Williams}, B.~F., {Dalcanton}, J.~J., {Stilp}, A., {Gilbert}, K.~M., {Ro{\v
  s}kar}, R., {Seth}, A.~C., {Weisz}, D., {Dolphin}, A., {Gogarten}, S.~M.,
  {Skillman}, E., \& {Holtzman}, J. 2010, \apj, 709, 135

\bibitem[{{Winters} {et~al.}(2000){Winters}, {Le Bertre}, {Jeong}, {Helling},
  \& {Sedlmayr}}]{Winters00}
{Winters}, J.~M., {Le Bertre}, T., {Jeong}, K.~S., {Helling}, C., \&
  {Sedlmayr}, E. 2000, \aap, 361, 641

\bibitem[{{Winters} {et~al.}(2003){Winters}, {Le Bertre}, {Jeong}, {Nyman}, \&
  {Epchtein}}]{Winters03}
{Winters}, J.~M., {Le Bertre}, T., {Jeong}, K.~S., {Nyman}, L.-{\AA}., \&
  {Epchtein}, N. 2003, \aap, 409, 715

\end{thebibliography}

\clearpage

\end{document}